\documentclass[10pt]{article}
\usepackage{amsmath}
\usepackage{latexsym}
\usepackage{amssymb}
\usepackage{amsfonts}
\usepackage{amsthm}
\title{ FROM KALMAN TO EINSTEIN AND MAXWELL: \\ THE STRUCTURAL CONTROLLABILITY REVISITED   }
\author{J.-F. POMMARET \\ CERMICS, Ecole des Ponts ParisTech, France \\ 
 jean-francois.pommaret@wanadoo.fr \\
ORCID: 0000-0003-0907-2601 }
\date{  }
\begin{document}
\maketitle

\noindent
{\bf ABSTRACT}  \\

\noindent
In the Special Relativity paper of Einstein (1905), only a footnote provides a reference to the conformal group of space-time for the Minkowski metric 
$\omega$. In this paper, we prove that General Relativity (1915) will depend on the following {\it cornerstone} result of differential homological algebra (1990). Let $K$ be a differential field and $D=K[d_1,...,d_n]$ be the ring of differential operators with coefficients in $K$. If $M$ is the differential module over $D$ defined by the Killing operator ${\cal{D}} :T \rightarrow S_2T^*: \xi \rightarrow \Omega = {\cal{L}}(\xi) \omega$ and $N$ is the differential module over $D$ defined by the $Cauchy = ad(Killing)$ adjoint operator with torsion submodule $t(N)$, then $t(N) \simeq {ext}^1_D(M) = 0$ and the Cauchy operator can be thus parametrized by stress functions having strictly nothing to do with $\Omega$. This result is largely superseding the Kalman controllability test in classical OD control theory and is showing that controllability is a structural "{\it built-in}" property of an OD/PD control system not depending on the choice of inputs and outputs, contrary to the engineering tradition. It also points out the {\it terrible confusion} done by Einstein (1915) while following Beltrami (1892), both of them using the Einstein operator but ignoring that it was self-adjoint in the framework of differential double duality (1995). We finally prove that the structure of electromagnetism and gravitation only depends on the nonlinear {\it elations} of the conformal group of space-time, showing thus that {\it nothing is left from the mathematical foundations of both general relativity and gauge theory}.   \\

\vspace{2cm}

Controllability; Einstein equations; Maxwell equations; \\

\vspace{1cm}

\noindent
{\bf COMMENTS} \\
This paper is the natural continuation of the 3 recent arXiv preprints now published as paper or book chapters with open access:\\
https://doi.org/10.5772/intechopen.1000851    \\
https://doi.org/10.4236/apm.2024.142004    \\
NOVA SCIENCE PUBLISHERS, ISBN: 979-8-89113-607-6   \\

\newpage

\noindent
{\bf 1) INTRODUCTION}  \\

Being a specialist of systems of PDE in group theory and control theory, it has been a challenge 
for me to apply the new methods of {\it Differential Homological Algebra} introduced around 1990 in order to 
study gravitational waves. The three last papers published in 2024 (37 p., 51 p., 67 p.):  \\
\noindent
https://doi.org/10.5772/intechopen.1000851    \\
https://doi.org/10.4236/apm.2024.142004    \\
NOVA SCIENCE PUBLISHERS, ISBN: 979-8-89113-607-6    \\
\noindent
could be roughly summarized by the single formula: \\
\[    \fbox{  $  t(N) \simeq  {ext}^1(M) = 0    $  }  \] 
where $M$ is the differential module defined by the Killing operator and $N$ is the differential module 
defined by the $Cauchy=ad(Killing)$ operator with torsion submodule $t(N)$, as extension modules are torsion modules 
that do not depend on the resolution of $M$ that MUST be used, namely the differential sequence in which the order of an operator is under its arrow:    \\
\[  \fbox{  $  0 \rightarrow \Theta \rightarrow n \underset 1{\stackrel{Killing}{\longrightarrow}}  \frac{n(n+1)}{2}
            \underset 2{\stackrel{Riemann}{\longrightarrow}}  \frac{n^2 (n^2 - 1)}{12} 
            \underset 1{\stackrel{Bianchi}{\longrightarrow}}  \frac{n^2(n^2 - 1)(n -2) }{24}   $  }   \]
This result, a cornerstone of homological algebra, points out the {\it terrible confusion} done by Einstein (1915) while following Beltrami (1892), 
both using the Einstein operator but ignoring that it was was self-adjoint in the framework of {\it differential double duality}. The Cauchy operator can be thus parametrized ({\it backwards} !) by $n(n+1)/2$ stress functions {\it having strictly nothing to do with the metric}, exactly like in the case of 
the single Airy stress function for plane elasticity, because {\it the Airy parametrization is only the adjoint of the Riemann operator} when $n=2$ !.  \\
Though unpleasant it is, {\it  nothing is left from Einstein general relativity, that is to say both Einstein equations and gravitational waves}, because 
we shall prove that the Einstein equations are not compatible with differential double duality and that {\it the Cauchy operator can indeed be 
parametrized by the adjoint of the Ricci operator}, with no reference to the Einstein operator !.  \\
 
However, one of he most striking facts of this paper is also provided by the next example (Pommaret, 2005):  \\

\noindent
{\bf Example 1.1}: ({\it Kalman system}): {\it THE PREVIOUS FORMULA IS THE KALMAN TEST IN CLASSICAL CONTROL THEORY}. As any operator is the adjoint of its own adjoint because $ad(ad({\cal{D}}))={\cal{D}}$, one can exchange $M$ and $N$ in the formula, that is $t(N) \simeq {ext}^1(M) \Leftrightarrow  t(M) \simeq {ext}^1(N)$. Hence, if $M$ is the differential module defined by the formally surjective Kalman operator $(y^k, u^r) \rightarrow ( - d y^k + A^k_l y^l + B^k_r u^r)$ with inputs $u$ and outputs $y$ while $N$ is the differential module defined by its adjoint operator with torsion submodule $t(N)$, then the Kalman controllability test amounts to say that the given control system is controllable if and only if $N=0$. Introducing Lagrange multipliers $\lambda = ({\lambda}_k)$, the kernel of the adjoint operator is defined by the OD equations $(y^k \rightarrow d {\lambda}_k + {\lambda}_l A^l_k = 0  , u^r \rightarrow {\lambda}_k B^k_r = 0)$ with {\it all} their differential consequences, namely: \\
\[  d\lambda + \lambda A= 0, \lambda B = 0  \Rightarrow d(\lambda B)=(d \lambda)B= - ( \lambda A )B =0  \Rightarrow {\lambda} A B = 0, ... \]
and so on, as a way to recover the well known controllability matrix $ (B, AB, A^2B, ...)$. It follows that $t(N)=N$ is already a torsion module and that the Kalman 
system is controllable if and only if $N=t(N)=0$ as claimed. Moreover, a control system is controllable if and only if it is parametrizable, that is $M$ can be embedded into a free differential module. In fact, when $n=1$, $D=K[d]$ is a principal ideal domain, that is any ideal can be generated by a single element, and it is well known that any torsion-free module over $D$ is indeed free. Accordingly, the kernel of the projection of $Dy + Du$ onto $M$ is free too and there is no loss of generality by supposing that the control system is made by differentially independent equations. The controllability of an OD control system is thus the purely structural property 
$t(M) = 0$ independently of the presentation, a fact amounting to the impossibility to find any torsion element, that is any linear combination of the the control variables that could be a solution of an autonomous OD equation for itself. \\

\noindent
{\bf Example 1.2}: With $m=3, n=1$ and a parameter $a = a(x) \in K$, let us consider the formally surjective  first order operator:   \\
\[ \fbox{  $   {\cal{D}}_1: ( {\eta}^1, {\eta}^2, {\eta}^3 ) \longrightarrow (d {\eta}^1 - a \, {\eta}^2 - d {\eta}^3= {\zeta}^1, \,\,  {\eta}^1 - d {\eta}^2 + d {\eta}^3 = {\zeta}^2 )$ } \]
Multiplying on the left by two test functions $({\lambda}^1, {\lambda}^2)$ and integrating by parts , we obtain:  \\
\[  \fbox{  $  ad({\cal{D}}_1): ({\lambda}^1, {\lambda}^2) \longrightarrow ( - d {\lambda}^1 + {\lambda}^2 = {\mu}^1, \,\, - a {\lambda}^1 + d {\lambda}^2 = {\mu}^2, \,\, 
d {\lambda}^1 - d {\lambda}^2 = {\mu}^3 )   $  }  \]
In order to look for the CC of this operator, we obtain:  \\
 \[ {\lambda}^2 - a {\lambda}^1 = {\mu}^1 + {\mu}^2 + {\mu}^3 \Rightarrow  - (\partial a + a^2 - a)  {\lambda}^1 =d ({\mu}^1 + {\mu}^2  + {\mu}^3) +  
 (a - 1) {\mu}^2 + a {\mu}^3    \]
 Introducing the notation $j_q(\mu) $ for all the derivatives of $\mu$ up to order $q$, we obtain therefore:  \\
 \[  \fbox{  $   (\partial a + a^2 - a) \lambda \in j_1( \mu) $  }  \]
 When the structural controllability condition is satisfied, that is when $a$ is not a solution of the {\it Riccati} equation in the bracket, we may obtain a second order CC operator of the form:,   \\
  \[   \fbox{  $ ad({\cal{D}}) : ( {\mu}^1, {\mu}^2, {\mu}^3) \longrightarrow  d^2 {\mu}^1 + d^2 {\mu}^2 + d^2 {\mu}^3 + ... = \nu  $  }  \]
Multiplying on the left by a test function $\xi$ and integrating by parts, we obtain the second order injective parametrization, provided that $\partial a + a^2 - a \neq 0$: \\
\[  \fbox{  $   {\cal{D}}: \xi  \rightarrow  ( d^2 \xi + ... = {\eta}^1, \,\, d^2 \xi +... = {\eta}^2, \,\, d^2 \xi - ... = {\eta}^3 )    $  }  \]
We have the long exact (splitting) sequence and its adjoint (splitting) sequence which is also exact:
\[   \fbox{  $         \begin{array}{rcccccccl}
 0  & \longrightarrow  & \xi & \underset 2{\stackrel{{\cal{D}}}{\longrightarrow}} & \eta & \underset 1{\stackrel{{\cal{D}}_1}{\longrightarrow} }  &  \zeta  & \longrightarrow   & 0  \\
&  &  &  &  &  &  &  &    \\
 0  & \longleftarrow   & \nu & \underset 2{\stackrel{ad({\cal{D}})}{\longleftarrow}} & \mu & \underset 1{\stackrel{ad({\cal{D}}_1)}{\longleftarrow} }  &  \lambda  & \longleftarrow &  0
 \end{array}     $  }       \]  
At no moment one has to decide about the choice of inputs and outputs and we advise the reader to effect {\it any choice} for applying the Kalman test when $a$ is a constant parameter. Of course $ a = cst \Rightarrow a(a -1) \neq 0 $ in a coherent way.  \\

\noindent
{\bf Example 1.3}: ({\it Double pendulum}): Many examples can be found in classical ordinary differential control theory because it is known that a linear control system is controllable if and only if it is parametrizable. In our opinion, the best and simplest one is the so-called double pendulum in which a rigid bar is able to move horizontally with reference position $x$ and we attach two pendulums with respective length $l_1$ and $l_2$ making the (small) angles ${\theta}_1$ and ${\theta}_2$ with the vertical, the corresponding control system does not depend on the mass of each pendulum and the two equations easily follow by projection from the Newton laws:  \\
\[  \fbox{  $ {\cal{D}}_1 \eta = 0 \,\, \Leftrightarrow \,\,\,  d^2x +l_1d^2{\theta}^1 +g {\theta}^1=0, \hspace{1cm}  d^2x + l_2 d^2{\theta}^2 + g{\theta}^2=0  $  }   \]
where $g$ is the gravity. A first result, still not acknowledged by the control community, is to prove that this control system is controllable if and only if $l_1 \neq  l_2$ without using a tedious computation through the standard Kalman test but, {\it equivalently}, to prove that the corresponding second order operator $ad({\cal{D}}_1)$ is injective. Though this is not evident, such a result comes from the fact $D$ is a principal ideal ring when $n = 1$ and thus, if the differential module $M_1$ is torsion-free, then $M_1$ is also free and has a basis allowing to split the short exact resolution $0 \rightarrow D^2 \stackrel{{\cal{D}}_1}{\longrightarrow} D^3 \rightarrow M_1 \rightarrow 0$ with $M_1 \simeq D$ in this case. When learning control theory, it has also been a surprise to be unable to find examples in which the controllability was explicitly shown not to depend on the choice of inputs and outputs among the system variables, like in such an example as we shall see.\\
Hence, multiplying on the left the first OD equation by ${\lambda}^1$, the second by ${\lambda}^2$, then adding and integrating by parts, we get:
 \[  \fbox{  $  ad({\cal{D}}_1) \lambda= \mu  \,\,\, \Leftrightarrow  \,\,\, 
  \left\{  \begin{array}{lcccl}
 x & \longrightarrow  & d^2 {\lambda}^1 + d^2 {\lambda}^2 & = &    {\mu}^1  \\
 {\theta}^1 & \longrightarrow &  l_1 d^2 {\lambda}^1   + g {\lambda}^1 & = & {\mu}^2 \\
 {\theta}^2  & \longrightarrow  &  l_2 d^2 {\lambda}^2  + g {\lambda}^2 &  =  & {\mu}^3
\end{array}  \right.  $   }    \]
The main problem is that the operator $ad({\cal{D}}_1)$ is {\it not} formally integrable because we have: 
\[   g(l_2 {\lambda}^1 + l_1 {\lambda}^2) = l_2 {\mu}^2 + l_1 {\mu}^3 - l_1 l_2 {\mu}^1  \]
and is thus injective if and only if $l_1 \neq l_2$ because, differentiating twice this equation, we also get:
\[       (l_2 / l_1) {\lambda}^1 + (l_1 / l_2) {\lambda}^2   \in j_2 (\mu)      \]
Hence, if $l_1 \neq l_2$, we finally obtain $ \lambda \in j_2 (\mu)$ and, after substitution, a single fourth order CC for $\mu$ showing that $ad({\cal{D}})$ is indeed a fourth order operator, {\it a result not evident indeed at first sight}. It follows that we have thus been able to work out the parametrizing operator ${\cal{D}}$ of order 4, namely:   \\
\[  \fbox{  $   {\cal{D}} \phi=  \eta \,\,\,  \Leftrightarrow  \,\,\,    \left\{  \begin{array}{rcl}
  - l_1 l_2 d^4\phi - g(l_1+l_2)d^2 \phi - g^2 \phi & = & x  \\
  l_2 d^4\phi +g d^2 \phi & = & {\theta}_1  \\
  l_1 d^4 \phi + gd^2 \phi  & = & {\theta}_2
  \end{array} \right.  $   }   \]  
 This parametrization is injective iff $l_1 \neq l_2$ because we have successively with $g\neq 0$:  \\
 \[     l_2 d^2 \phi + g \phi = 0, \Rightarrow   l_1 d^2 \phi +g \phi = 0 \Rightarrow    g(l_1 - l_2) \phi =0 \Rightarrow  \phi = 0  \]  
We have the long exact splitting sequence and its adjoint splitting sequence which is also exact:
\[   \fbox{  $         \begin{array}{rcccccccl}
0  & \longrightarrow  & 1 & \underset 4{\stackrel{{\cal{D}}}{\longrightarrow}} & 3 & \underset 2{\stackrel{{\cal{D}}_1}{\longrightarrow} }  &  2  & \longrightarrow   &  0  \\
&  &  &  &  &  &  &  &    \\
 0  & \longleftarrow   & 1 & \underset 4{\stackrel{ad({\cal{D}})}{\longleftarrow}} & 3 & \underset 2{\stackrel{ad({\cal{D}}_1)}{\longleftarrow} }  &  2  & \longleftarrow  &  0   
 \end{array}     $  }       \]
 We now study the way to split these sequences. As any operator is the adjoint of its own adjoint, we define the lift $ad({\cal{P}}_1): \mu \rightarrow \lambda$ of the lower sequence as follows:\\
 \[  \fbox{ $   g^2 (l_1 - l_2) {\lambda}^1 = g (l_1 - l_2) {\mu}^2 - g (l_1)^2 {\mu}^1 + l_1 l_2 d^2 {\mu}^2 + (l_1)^2 d^2 {\mu}^3 - (l_1)^2 l_2 d^ 2 {\mu}^1  $ }  \]
 \[  \fbox{ $  g^2(l_1 - l_2) {\lambda}^2 =  g (l_1 - l_2) {\mu}^3 + g(l_2)^2 {\mu}^1 - (l_2)^2 d^2 {\mu}^2 - l_1 l_2 d^2 {\mu}^3 + l_1 (l_2)^2 d^2 {\mu}^1   $  }   \]
 obtain the lift ${\cal{P}}_1: ({\zeta}^1, {\zeta}^2) \rightarrow  (x, {\theta}^1, {\theta}^2) $ of the upper sequence, up to a factor $g^2 (l_1 - l_2)$, namely:  \\
  \[ \fbox{ $ - g (l_1)^2 {\zeta}^1 + g (l_2)^2 {\zeta}^2 - (l_1)^2 l_2 d^2 {\zeta}^1+ l_1 (l_2)^2 {\zeta}^2 = x  $ }  \]
  \[  \fbox{ $ g(l_1 - l_2) {\zeta}^1 + l_1 l_2 d^2 {\zeta}^1 - (l_2)^2 d^2 {\zeta}^2 = {\theta}^1 $ }  \]
  \[  \fbox{  $  g (l_1 - l_2) {\zeta}^2 + (l_1)^2 d^2 {\zeta}^1 - l_1 l_2 d^2{\zeta}^2 = {\theta}^2 $ }  \] 
  We finally consider the case $l_1= l_2 = l$. Subtracting the two OD equations, we discover that $ z = {\theta}^1 - {\theta}^2$ is an observable quantity that satisfies the autonomous system $ l d^2 z + g z = 0$ existing for a single pendulum. It follows that $ z $ is a torsion element and the system cannot be controllable. When $ z = 0 \Rightarrow {\theta}^1 = {\theta}^2 = \theta$ we let the reader prove that the remaining OD equation $d^2 x + l d^2 {\theta} + g {\theta} = 0$ can be parametrized by $l d^2 \xi + g \xi = x, - d^2 \xi = {\theta}$. \\
At this stage of the reading, we invite the reader to realize this experiment with a few dollars, check how the controllability depends on the lengths and wonder how this example may have {\it anything} to do with the Cosserat, Einstein or Maxwell equations !.  \\

\vspace{1cm}

This paper is also a kind of Summary Note sketching in a rather self-contained but condensed way the results  presented through a series of lectures at the Albert Einstein Institute (AEI, Berlin/Potsdam), October, 23-27, 2017. The initial motivation for studying the methods used in this paper has been a $1000 \$ $ challenge proposed in $1970$ by J. Wheeler in the physics department of Princeton University while the author of this paper was a visiting student of D.C. Spencer in the close-by mathematics department, namely:  \\ 
{\it Is it possible to express the generic solutions of Einstein equations in vacuum by means of the derivatives of a certain number of arbitrary functions, like the potentials for Maxwell equations ?}.\\
After recalling the negative answer we already provided in 1995 (Pommaret, 1995), the main purpose of this paper is to use again these new techniques of {\it differential double duality} in order to revisit the mathematical foundations of the concepts and equations involved in {\it general relativity} and {\it gauge theory} that are leading to gravitational waves. At the same time, we point out the fact that the above parametrization problem is equivalent to the controllability property of a control system, such a result showing for the first time that it is a {\it structural} property, that is a property that does not depend on the choice of inputs and outputs or even on the {\it presentation} of the system, that is on a change of {\it all} the independent variables used to describe the system, contrary to the commonly accepted point of view of the control community. Many explicit examples are illustrating the paper, ranging from ordinary differential (OD) or partial differential (PD) control theory to mathematical physics, explaining in particular why the mathematical foundations of both gravitation {\it and} electromagnetism only depend on the structure of the conformal group of space-time. 
Accordingly, the foundations of control theory, engineering and mathematical physics must be revisited within this new framework which has been initiated by (Oberst, 1990) but only for systems with constant coefficients, though striking it may sometimes look like. Of course, it is rather easy to study systems involving OD equations as we saw and we shall need new tools for studying systems of PD equations, though these new methods can also be used for OD equations. An additional difficulty will be met when dealing with operators having variable coefficients.  \\

\noindent
{\bf Example 1.4}: While using Kalman test in control theory, it is often useful to transform a second order system $d^2y=0$ to a first order system 
$d z^1 - z^2=0, d z^2=0$ by setting $y=z^1, d y= z^2$, transforming one OD equation for one unknown to two OD equations for two unknown. However, the mathematical community is not aware that, more generally, this has been exactly the procedure followed by Spencer from transforming ANY system of PD equations of order $q$ with $n$ independent variables and $m$ unknowns to a new system of PD equations of order one. One of the best examples to be met in the literature has been provided by (Macaulay, 1916). With $n=3, m= 1, q= 2$ and using the jet notation $d_{23} y=y_{23}$, let us consider the second order homogeneous system 
$P y \equiv y_{33} = 0, Q y \equiv y_{23} - y_{11} =0, R y \equiv y_{22} = 0$. With a reference to (Janet, 1920), differentiating once, we notice that all the derivatives of order $3$ vanish but $y_{123} - y_{111}=0$ and that all derivatives of order $4$ do vanish. We obtain therefore eight arbitrary parametric jets:
\[   \{ z^1 = y, z^2 = y_1, z^3 = y_2, z^4 = y_3, z^5= y_{11}, z^6 = y_{12}, z^7 = y_{13}, z^8 = y_{111}  \}  \]
satisfying the non-homogeneous first order "equivalent system", called "First Spencer operator":  \\
\[ d_3 z^1 - z^4=0, d_3 z^2 -  z^7=0, d_3 z^3 - z^5=0, d_3 z^4 = 0, d_3 z^5 - z^8 = 0, .... , d_1 z^1 - z^2= 0, ..., d_1 z^8=0  \]
with the eight new parametric jets  $ \{ z^1, z^2, z^3, ... , z^8 \} $. In the present situation, we can integrate the system explicitly. We have indeed at once a basis of eight solutions, namely:  \\
\[  \{ f_{\tau}(x) \mid 1 \leq \tau \leq  8  \}  =   \{ 1, x^1, x^2, x^3, x^1 x^2, x^1 x^3, x^2 x^3 + \frac{1}{2} (x^1)^2, x^1 x^2 x^3 + \frac{1}{6} (x^1)^3 \}   \]
and the space of solutions is a vector space ${\cal{V}}$ of dimension $8$ over the field $\mathbb{Q}$ of constants of $K$.  \\

\noindent
{\bf Example 1.5}: With $n=1, m=3, q=1, a= a(x) \in K= \mathbb{Q}(x)$, let us consider again the system of OD equations met in Example 1.2:  \\
\[    d y^1 - d y^2 - a(x) y^3 = 0, \hspace{2cm} y^1  + d y^2 - dy^3 = 0  \]
In classical control theory, the Kalman controllability test can only be applied whenever $a$ is a constant parameter and we let the reader check that, {\it independently of the choice of one input and two outputs among the three unknowns}, the test is providing the condition $a (a-1) \neq 0$, that is both $a\neq 0, a \neq 1$. When $a$ is no longera constant coefficient, not a lot of things can be said independently of the choice of input and output. \\
Let us multiply the first equation by a test function ${\lambda}^1$, the second by a test function ${\lambda}^2$, add and integrate by parts in order to look for the kernel of the adjoint operator, namely:
\[  y^1 \rightarrow - d {\lambda}^1  + {\lambda}^2 = 0, \,\,\,  y^2 \rightarrow  d {\lambda}^1 - d {\lambda}^2 = 0, \,\,\, y^3 \rightarrow -a {\lambda}^1 + d {\lambda}^2 = 0  \]
We get ${\lambda}^2 = d {\lambda}^1$ and thus $d {\lambda}^1 - a {\lambda}^1 = 0, d^2 {\lambda}^1 - a {\lambda}^1 = 0$, a result leading to the striking Riccati inequality condition: 
\[        {\lambda}^1 = 0 \Leftrightarrow  {\lambda}^2 = 0  \Leftrightarrow    \partial a + a (a - 1) \neq 0   \]
in a coherent way with the result of the Kalman test when $a$ is a constant parameter. We notice that the result obtained is a structural property of the system.  \\

\noindent
{\bf Example 1.6}: (Pommaret, 2023 a) With two independent variables $(x^1,x^2)$ 
and one unknown $y$, let us consider the following second order system with constant coefficients:
\[ \left\{
\begin{array}{rll}
Py\equiv &d_{22}y   &=u\\
Qy\equiv &d_{12}y-y &=v
\end{array}
\right.   \]
where now $P$ and $Q$ are PD operators with coefficients in the subfield $k=\mathbb{Q}$ of constants of the differential field 
$K= k(x^1,x^2)$. We obtain at once from a first use of crossed derivatives:
\[  d_2 y = d_1u -d_2v  \]
and from a second use:
\[  y=d_{11}u-d_{12}v-v   \]
and could hope to obtain the $4^{th}$-order generating compatibility conditions (CC) by substitution, that is to say:
\[
\left\{
\begin{array}{rll}
A\equiv &d_{1122}u-d_{1222}v-d_{22}v-u&=0\\
B\equiv &d_{1112}u-d_{11}u-d_{1122}v &=0
\end{array}
\right.  \]
with the only generating CC : $ w \equiv d_{11} A - d_{12} B - B = 0 $ .  \\
However, {\it in this particular case}, there is an unexpected {\it unique second order} generating CC:
\[  C\equiv d_{12}u-u-d_{22}v=0 \]
as we now have indeed $PQ-QP=0$ both with $A\equiv d_{12}C+C$ and 
$B\equiv d_{11}C$, a result leading to $C\equiv d_{22}B-d_{12}A+A$. 
Accordingly, the systems $A=0, B=0$ on one side and $C=0$ on the other side 
are completely different though they have the same solutions in $u, v$ which can be parametrized injectively by $y$.\\
Finally, setting $u=0, v=0$, we notice that the preceding homogeneous 
system can be written in the form ${\cal D}y=0$ and admits the only 
solution $y=0$. More precisely, if a linear system $R_q \subset J_q(E)$ of order $q$ on $E$ is given we may find two integers $(r,s)$ such that, prolonging $r+s$ times to obtain $R_{q+r+s} $ and keeping only the equations of order $q+r$, we obtain a system $R^{(s)}_{q+r}$ providing all the informations on the solutions up to any order ( {\it prolongation / projection} (PP) procedure)(Pommare, 1978, 1994. In the present case, we get successively:   \\
\[    0 = R^{(4)}_2 \subset  R^{(3)}_2 \subset R^{(2)}_2 \subset R^{(1)}_2 \subset R_2 \subset J_2(E) \]
with strict inclusions and respective dimensions: $ 0 < 1 < 2 < 3 < 4 < 6 $.  \\

\noindent
{\bf Example 1.7}: Denoting by $y^k_i=d_iy^k $ for $i=1,2 $ and $k=1,2,3 $ the 
formal derivatives of the three differential indeterminates 
$y^1,y^2,y^3$, we consider the system of three PD equations for 3 unknowns 
and 2 independent variables $(x^1, x^2)$ which is defining a differential module $M$ over the non-commutative ring $D= \mathbb{Q}(a) (x^1, x^2) [d_1, d_2]$ of differential operators with coefficients in $\mathbb{Q}(a)(x^1, x^2)$ when $a$ is a constant parameter:  \\
\[ \left\{  \begin{array}{rll}
{\Phi}^1 \equiv  & y^3_2 -  y^2_2  & = 0 \\
{\Phi}^2 \equiv  & y^2_2  + y^1_1 - a x^2 y^1 & = 0 \\
{\Phi}^3 \equiv  & y^3_1  - y^2_1 & = 0 
\end{array} \right.  \]
No one among $(y^1,y^2)$ can be given arbitrarily  and that there is a unique generating CC, namely:  \\
\[ \Psi \equiv d_2 {\Phi}^3 - d_1 {\Phi}^1 = 0  \]  
Also, setting $z=y^3 - y^2$, we get both $z_1=0, z_2=0$ and $z$ is an autonomous element. Then one can easily prove that
any other autonomous element can be expressible by means of a differential operator acting on $z$ which is therefore a generator of the torsion module $t(M) \subset M$. Accordingly, in the present situation, any autonomous element is a constant multiple of $z$.\\
Finally, setting $z=0$ and thus $ y^3 =  y^2$, we obtain for $(y^1, y^2)$, after substitution:  \\
\[     {\Phi}'\equiv    y^2_2+ y^1_1 - a x^2 y^1 = 0  \]
which is defining an operator ${\cal{D}}' : (y^1, y^2) \rightarrow {\Phi}'$ and a torsion-free module $M' \simeq M / t(M)$ in the short exact sequences:  \\
\[     0 \rightarrow D \rightarrow D^2  \rightarrow M' \rightarrow 0 , \hspace{15mm}  0  \rightarrow t(M) \rightarrow M \rightarrow M' \rightarrow 0  \]
Multiplying the previous operator by a test function $\lambda$ and integrating by parts, the kernel of the adjoint operator 
$ad({\cal{D}}') : \lambda \rightarrow ({\mu}^1, {\mu}^2)$ is defined by: \\
\[   y^2 \rightarrow - d_2 \lambda = {\mu}^2, y^1 \rightarrow  - d_1 \lambda  - a x^2 \lambda = {\mu}^1 \Rightarrow     
 d_1 {\mu}^2 - d_2 {\mu}^1  - a x^2 {\mu}^2   =  a \lambda              \]
We have thus two quite different situations:   \\
\noindent
$\bullet \,\,  a = 0 $ :  The adjoint operator is not injective and we are in the situation of the $div $ operator when $n=2$ that can be parametrized by the $curl$ operator in such a way that $M'$ is neither free nor projective but $M' \subset D$ with a strict inclusion.       \\
\noindent
$ \bullet \,\,  a \neq 0 $, say $a=1$: The adjoint operator is injective and, using the fact that any operator can be written as the adjoint of an operator, we have obtained a lifting operator $ad({\cal{P}}) : ({\mu}^1, {\mu}^2) \rightarrow \lambda$ such that $ ad({\cal{D}}') \circ ad({\cal{P}}) = id_{\lambda} \Rightarrow {\cal{P}} \circ {\cal{D}}'= id_{{\Phi}'}$. We shall prove later on that $M'$ is not free but projective, thus torsion-free,  because this lift provides an isomorphism $D^2 \simeq M' \oplus D$ and an isomorphism $M \simeq t(M) \oplus M'$ which may not exist in general.  \\

We now sketch the main result that will be proved and illustrated through this paper. In particular, its application to Einstein general relativity and Maxwell electromagnetism will prove that the mathematical foundations of these two apparently well established theories
will have to be entirely revisited but for quite different reasons. \\

Roughly, if a differential module $M$ is defined y a linear differential operator ${\cal{D}}$ and we denote by $N$ the differential module defined by the (formal) adjoint operator $ad({\cal{D}})$, we shall prove:  \\

\noindent
{\bf Theorem 1.8}: (Pommaret, 2001, 2005) The differential module ${ext}^1(M) = ext^1_D(M,D)=t(N)$ does not depend on the presentation of $M$.\\

\noindent
{\bf Example 1.9 }: As a first striking result, that does not seem to have been noticed by mechanicians up till now, let us consider the situation of classical elasticity theory where $\cal{D}$ is the Killing operator for the euclidean metric, namely $\Omega \equiv  \cal{D}\xi=\cal{L}(\xi)\omega \in S_2 T^*$ and ${\cal{D}}_1$ the corresponding CC, namely the linearized Riemann curvature with $n^2(n^2-1)/12$ components. In that case, as it is well known that the Poincar\'{e} sequence for the exterior derivative is self-adjoint {\it up to sign} (for $n=3$ the adjoints of $grad,curl,div$ are respectively $- div,curl,- grad$) then {\it the first extension module does not depend on the differential sequence used} and therefore vanishes. Accordingly, $ad(\cal{D})$ generates the CC of $ad({\cal{D}}_1)$. Hence, in order to parametrize the Cauchy stress equations, that is $ad(\cal{D})$, namely:  \\
\[  \fbox{ $  {\sigma}^{12} = {\sigma}^{21}, \,\, d_1 {\sigma}^{11} + d_2 {\sigma}^{21} = 0, \,\,  d_1 {\sigma}^{12} + d_2 {\sigma}^{22} = 0   $ }  \]
one just needs to compute $ad({\cal{D}}_1)$. For $n=2$, we get:\\
\[    \phi (d_{11}{\Omega}_{22}+d_{22}{\Omega}_{11}-2 d_{12}{\Omega}_{12})= (d_{22}\phi )\, {\Omega}_{11} -2 (d_{12}\phi)\,  {\Omega}_{12}+(d_{11}\phi) \, {\Omega}_{22}+ d_i(...)^i   \]
and recover the parametrization by means of the Airy function in a rather unexpected way:  \\
\[   \fbox{ $ {\sigma}^{11} = d_{22} \phi, \,\, {\sigma}^{12} = {\sigma}^{21} = - d_{12} \phi, \,\, {\sigma}^{22} = d_{11} \phi  $  }  \]
Exhibiting a parametrization for $n\geq 3$ thus becomes an exercise of computer algebra, the number of (pseudo)-potentials being the number $n^2(n^2-1)/12$ of components of the Riemann tensor.\\
We now treat the case of Cosserat equations with zero second members, namely (Cosserat, 1909):  \\
\[  \fbox{ $   d_1 {\sigma}^{1,1} + d_2 {\sigma}^{2,1} = 0,  d_1 {\sigma}^{1,2} + d_2 {\sigma}^{2,2} = 0, 
d_i {\mu}^{i,12} + {\sigma}^{1,2} -{\sigma}^{2,1}= 0   $ }   \]
For this, instead of using the Janet sequence as before, we now use the Spencer sequence which is isomorphic to the gauge sequence though with quite different operators. However, according to the general theorems of homological algebra, the existence of a parametrization does not depend on the differential sequence used and therefore follows again, like in the previous example, from the fact that the Poincar\' {e} sequence is self-adjoint up to the sign. In the present situation, we have $C_r={\wedge}^rT^*\otimes R_1\simeq {\wedge}^rT^*\otimes \cal{G}$ with ${dim(\cal{G})}=n(n+1)/2$. We have shown that the Cosserat equations were just $ad(D_1)$, their {\it first order} parametrization is thus described by $ad(D_2)$ and needs $dim(C_2)=n^2(n^2-1)/4$ (pseudo)-potentials. We provide the details when $n=2$ but we know at once that we must use $3$ (pseudo)-potentials only. The case $n=3$ could be treated similarly and is left as an exercise.\\
   The Spencer operator $D_1$ is described by the equations:\\
\[  {\partial}_1{\xi}_1=A_{11}, {\partial}_1{\xi}_2-{\xi}_{1,2}=A_{12}, {\partial}_2{\xi}_1-{\xi}_{2,1}=A_{21}, {\partial}_2{\xi}_2=A_{22}, {\partial}_1{\xi}_{1,2}=B_1, {\partial}_2{\xi}_{1,2}=B_2  \]
because $R_1$ is defined by the equations  ${\xi}_{1,1}=0,{\xi}_{1,2}+{\xi}_{2,1}=0, {\xi}_{2,2}=0$. \\
   Accordingly the 3 CC describing the Spencer operator $D_2$ are:\\
\[  {\partial}_2A_{11}-{\partial}_1A_{21}+B_1=0, {\partial}_2A_{12}-{\partial}_1A_{22}+B_2=0, {\partial}_2B_1-{\partial}_1B_2=0  \]
Multiplying these equations respectively by ${\phi}^1, {\phi}^2, {\phi}^3$, then summing and integrating by part, we get $ad(D_2)$ and the desired first order parametrization in the form (Pommaret, 1994, 2010):\\
\[ \fbox{  $   {\sigma}^{1,1}=-{\partial}_2{\phi}^1, {\sigma}^{1,2}=-{\partial}_2{\phi}^2, {\sigma}^{2,1}={\partial}_1{\phi}^1, {\sigma}^{2,2}={\partial}_1{\phi}^2, {\mu}^{1,12}=-{\partial}_2{\phi}^3+{\phi}^1, {\mu}^{2,12}={\partial}_1{\phi}^3+{\phi}^2  $  }  \]
as announced previously. As we are dealing with PD equations with constant coefficients, it is important to notice that such a parametrization could also have been obtained by localization (exercise). When the stress is symmetric, that is ${\sigma}^{1,2} = {\sigma}^{2,1}$, the Airy parametrization can be recovered if we cancel the couple-stress with ${\phi}_1 = {\partial}_2 {\phi}_3, {\phi}_2 = - {\partial}_1 {\phi}_3$ and set ${\phi}_3 = - \phi$.\\
Changing the presentation will be studied later on as we shall need additional tools.  \\

We end this Introduction with one of the best examples we know in order to understand that working out differential sequences is not an easy task, even on rather elementary examples.   \\

\noindent
{\bf Example 1.10}: (Macaulay, 1916) Let us revisit Example 1.4 using more advanced methods. With $m = 1, n = 3, q = 2, K = \mathbb{Q}$, let us consider the linear second order system $R_2 \subset J_2(E)$ with $dim(E) = 1$ while using jet notations (Pommaret, 1978, 1994, Spencer, 1965): 
\[  y_{33} = 0, \hspace{1cm}  y_{23} - y_{11} = 0, \hspace{1cm} y_{22} = 0  \]
We let the reader check easily that $dim(g_2)= 3, dim(g_3) = 1$ with only parametric jet $y_{111}$, $g_4 = 0 $ and thus $dim( R_2)= 8 = 2^3$, a result leading to s $dim(R_{3 + r}) = 8$ that is $R_{3+r}\simeq R_3, \forall r\geq 0$.  We recall the dimensions of the following jet bundles:  
\[  \begin{array}{cccccccccc}
q & \rightarrow & 0 & 1 & 2 & 3 & 4 & 5 & 6 & 7   \\
S_qT^* & \rightarrow & 1 & 3 & 6 & 10 & 15 & 21 & 28 & 36  \\
J_q(E) & \rightarrow & 1 & 4 & 10 & 20 & 35 & 56 & 84 & 120
\end{array}  \]
both with the commutative and exact diagram allowing to construct {\it inductively} the Spencer bundles $C_r \subset C_r (E)$ and the Janet bundles $F_r$ for $r= 0, 1, ... ,n$ with $F_0=J_q(E)/R_q$ and $C_0=R_q \subset J_q(E) = C_1(E)$ while replacing the system $R_q \subset J_q(E)$  of order $q$ on $E$ by the system $R_{q+1} \subset J_1(R_q)$ of order $1$ on $R_q$ when $q$ is large enough, that is $q=4$ in the present example because $g_4=0$.  \\
\[ \fbox{  $  \begin{array}{ccccccccccc}
 & &  &  &   &  & 0 & & 0 & & \\
& &  &  &   & & \downarrow & & \downarrow & & \\
   &   & 0 & \longrightarrow  &  R_{q+1}    &  \longrightarrow & J_1(R_q) & \longrightarrow & C_1 & \longrightarrow & 0  \\
&  &  &  &   &  &  \downarrow &  &  \downarrow &  &  \\
  &  & 0 & \longrightarrow  &   J_{q+1}(E)  &  \longrightarrow   & J_1(J_q(E))   & \longrightarrow  &  C_1(E) & \longrightarrow &  0  \\
  & &   &  &  \parallel &  &  \downarrow &   &  \downarrow  &    &   \\
 0  &  \longrightarrow &   R_{q+1}  & \longrightarrow   &  J_{q+1}(E) & \longrightarrow & J_1(F_0 )& \longrightarrow & F_1 & \rightarrow & 0   \\
& &   &  &   &  &  \downarrow &  &  \downarrow   &  &  \\
 &  &  &  &    &  &  0  &  &  0  &  &  
\end{array}   $  }  \]
showing that we have indeed:   \\
\[    C_r = {\wedge}^r T^* \otimes R_q / \delta ({\wedge}^{r-1} T^* \otimes g_{q+1}) \]
\[    C_r(E) = {\wedge}^r T^* \otimes J_q(E) / \delta ({\wedge}^{r-1} T^* \otimes S_{q+1}T^* \otimes E)    \]
\[   F_r = {\wedge}^r T^* \otimes J_q(E) / ( {\wedge}^r T^* \otimes R_q + \delta ({\wedge}^{r-1} T^* \otimes S_{q+1}T^ \otimes E))  \]
When $R_q\subset J_q(E)$ is involutive, that is formally integrable (FI) with an involutive symbol $g_q$, then these three differential sequences are formally exact on the jet level and, in the Spencer sequence: \\
\[ \fbox{  $  0 \longrightarrow  \Theta \underset q{\stackrel{j_q}{\longrightarrow}} C_0 \underset 1{\stackrel{D_1}{\longrightarrow}} C_1 \underset 1{\stackrel{D_2}{\longrightarrow}} ... 
\underset 1{\stackrel{D_n}{\longrightarrow}} C_n  \longrightarrow 0   $  }   \]
the first order involutive operators $D_1, D_2, ..., D_n$ are induced by the standard Spencer operator $d:R_{q+1} \longrightarrow T^* \otimes R_q$ already defineded that can be extended to $d: {\wedge}^r T^* \otimes R_{q+1}\longrightarrow {\wedge}^{r+1} \otimes R_q$. A similar condition is also valid for the Janet sequence:  \\
\[  \fbox{  $  0 \longrightarrow \Theta \longrightarrow E \underset q{\stackrel{{\cal{D}}}{\longrightarrow}} F_0 \underset 1{\stackrel{{\cal{D}}_1}{\longrightarrow}} F_1 
\underset 1{\stackrel{{\cal{D}}_2}{\longrightarrow}} ... \underset 1{\stackrel{{\cal{D}}_n}{\longrightarrow}} F_n \longrightarrow 0  $  }  \]
which can be thus constructed "{\it as a whole}" from the previous extension of the Spencer operator (See (Pommaret, 1978), p 183 + 185 + 391 for the main diagrams, (Pommaret, 1994) for other explicit computations on the Macaulay example and (Pommaret, 2005) for its application to group theory). However, such a result is still not known and not even acknowledged today in mathematical physics, particularly in general relativity which is {\it never} using the Spencer $\delta$-cohomology in order to define the Riemann or Bianchi operators (Pommaret, 1988). The study of the present Macaulay example will be sufficient in order to justify our comment.  \\
First of all, as $g_2$ is {\it not} $2$-acyclic and the coeficients are constant, the CC are of order two as follows:\\
\[ Q w - R v = 0,  \,  R u - P w = 0, \, P v - Q u =  0 \, \Rightarrow \,  P (Q w - R v ) + Q (R u - P w ) + R (P v - Q u) \equiv  0    \]  
The simplest formally exact resolution, {\it which is quite far from being a Janet sequence}, is thus: \\
\[   \fbox{  $  0  \longrightarrow \Theta \longrightarrow 1 \underset 2{\stackrel{\cal{D}}{\longrightarrow}}  3  
\underset 2{\stackrel{{\cal{D}}_1}{ \longrightarrow }} 3 \underset 2{\stackrel{{\cal{D}}_2}{\longrightarrow }}
1 \longrightarrow 0  $  }  \]                                       
Secondly, as the first prolongation of $R_2$ becoming involutive is $R_4$, an idea could be to start with the system $R_3 \subset J_3(E)$ but we have proved in (Pommaret, 2016) that the simplest formally exact sequence that could be obtained, {\it which is also quite far from being a Janet sequence}, is:  \\
\[  \fbox{ $    0 \longrightarrow \Theta \longrightarrow 1\underset 3{\longrightarrow} 12 \underset 1{\longrightarrow} 21 \underset 2{\longrightarrow} 46 \underset 1{\longrightarrow} 72 
\underset 1{\longrightarrow} 48 \underset 1{\longrightarrow} 12 \longrightarrow 0   $  }  \]
Indeed, the Euler-Poincar\'{e} characteristic is $1 - 12 + 21 - 46 + 72 - 48 + 12 = 0 $ but  we notice that the orders of the successive operators may vary up and down. \\
Then, decomposing any solution on the basis already exhibited, we may set $f_{\mu}(x)= {\lambda}^{\tau}(x) {\partial}_{\mu} f_{\tau}(x) $ for any section of $R_4$ by inverting a $8 \times 8 $ matrix. With standard notations for multi-index notation, we obtain for the Spencer operator $d:R_4 \rightarrow T^* \otimes R_4: {\wedge}^0 T^* \otimes {\cal{V}} \rightarrow  {\wedge}^1T^* \otimes {\cal{V}}$ the formula:  \\
\[  \fbox{  $   {\partial}_i f_{\mu}(x) - f_{\mu + 1_i}(x) = {\partial}_i {\lambda}^{\tau}(x) {\partial}_{\mu} f_{\tau}(x)   $  }  \]
showing that the Spencer sequence is isomorphic to a tensor product of the Poincar\'{e} sequence.  \\
We finally let the reader discover that the {\it Fundamental Diagram I} relating the upper Spencer sequence to the lower Janet sequence is (See (Pommaret, 1978), p 19-22 for details):  \\
\[  \fbox{  $    \begin{array}{ccccccccccccccc}
 &  &  &  &  &  &  0 &  &  0  &  &  0  &  &  0  &  &   \\
&  &  &  &  &  & \downarrow &  &  \downarrow  &  &  \downarrow &  & \downarrow &  &  \\
  &  &  0  & \longrightarrow & \Theta & \underset 4{\stackrel{j_4}{\longrightarrow}} & 8 & \underset 1{\stackrel{D_1}{\longrightarrow}} & 24 & \underset 1{\stackrel{D_2}{\longrightarrow}} & 24 & \underset 1{\stackrel{D_3}{\longrightarrow}} & 8 & \longrightarrow & 0  \\
&  &  &  &  &  & \downarrow &  &  \downarrow  &  &  \downarrow &  & \downarrow &  &  \\
 &  &  0  & \longrightarrow & 1 & \underset 4{\stackrel{j_4}{\longrightarrow}} & 35 & \underset 1{\stackrel{D_1}{\longrightarrow}} & 84 & \underset 1{\stackrel{D_2}{\longrightarrow}} & 70 & \underset 1{\stackrel{D_3}{\longrightarrow}} & 20 & \longrightarrow & 0  \\
&  &  &  &  \parallel &  & \downarrow &  &  \downarrow  &  &  \downarrow &  & \downarrow &  &  \\
0 & \longrightarrow & \Theta & \longrightarrow & 1 & \underset 4{\stackrel{{\cal{D}}}{\longrightarrow}} & 27 & \underset 1{\stackrel{{\cal{D}}_1}{\longrightarrow}} & 60 & \underset 1{\stackrel{{\cal{D}}_2}{\longrightarrow}} & 46 & \underset 1{\stackrel{{\cal{D}}_3}{\longrightarrow}} &  12 & \longrightarrow & 0\\ 
&  &  &  &  &  & \downarrow &  &  \downarrow  &  &  \downarrow &  & \downarrow &  &  \\
 &  &  &  &  &  &  0 &  &  0  &  &  0  &  &  0  &  &
\end{array}   $  }   \]
In the present example, the Spencer bundles are $C_r ={\wedge}^r T^* \otimes R_4$ and their dimensions are quite lower that the dimensions of the Janet bundles. Among the long exact sequences that {\it must} be used the following involves a $540 \times 600$ matrix and we wish good luck to anybody using computer algebra:  \\
\[   0 \rightarrow R_7 \rightarrow J_7(1) \rightarrow J_3(27) \rightarrow J_2(60) \rightarrow J_1(46) \rightarrow F_3 \rightarrow 0  \Rightarrow   
8 - 120 + 540 - 600 + 184 - 12 = 0 \]   \\    

The content of the paper will follow this Introduction. In section 2 we shall recall, in the most self-contained and elementary way as possible, the concepts and main results of homological algebra before extending them to the differential framework (See Zentralblatt review Zbl 1079.93001). In section 3 we shall apply them in order to revisit the mathematical foundations of general relativity. In section 4 we shall prove that the structure of the conformal group {\it must} also be carefully revisited because, contrary to Riemannian geometry, the corresponding differential sequence will drastically depend on the dimension of the ground manifold. In section 6 we present a modern version of a few results found by H. Poincar\'{e} in 1901 but rarely quoted in this conformal background. In section 6 we shall apply {\it all} the previous results obtained for the conformal group in order to revisit the mathematical foundations of both electromagnetism and gravitation by chasing in the {\it fundamental diagram II}, before concluding in section 7. \\

We invite the reader to keep constantly in mind the motivating examples presented in the Introduction as these new methods, found by pure mathematicians, have {\it never} been applied to OD/PD control theory with variable coefficients or mathematical physics (general relativity and gauge theory), a fact explaining why we have not been able to find other references.   \\

\noindent
{\bf 2) DIFFERENTIAL HOMOLOGICAL ALGEBRA} \\

It becomes clear from the previous motivating examples that there is a need for classifying the properties of systems of OD (classical control theory) or PD (mathematical physics) equations in a way that does not depend on their presentations and {\it this is the purpose of differential homological algebra}. The crucial idea will be indeed to obtain such a classification from the families of modules they allow to define over integral domains in the following way (See (Pommaret, 2001) or Zbl 1079.93001) but a much more advanced "{\it purity}" classification in which {\it torsion-free} amounts to {\it 0-pure} (Pommaret, 2001, 2015):  \\
\[  \fbox{  $  FREE \,\, \subset \, \, PROJECTIVE \,\, \subset \,\, REFLEXIVE \,\, \subset \, \, TORSION-FREE  $  }   \]
pointing out the fact that such a classification just disappears when $n=1$. \\

\noindent
{\bf 2.1) MODULE THEORY} \\

Before entering the heart of this section, we need a few technical definitions and results from commutative algebra, in particular for {\it localization}. The reader may look at (Rotman, 1979, Kashiwara, 1995, Pommaret, 2001) for most of the proofs as we are using quite standard notations. \\

\noindent
{\bf Definition 2.1.1}: A {\it ring} $A$ is a non-empty set with two associative binary operations respectively called {\it addition} and {\it multiplication},
respectively sending $a,b\in A$ to $a+b\in A$ and $ab\in A$ in such a waythat $A$ becomes an abelian group for the multiplication, 
so that $A$ has a zero element denoted by $0$, every $a\in A$ has an additive inversedenoted by $-a$ and the multiplication is 
distributive over the addition, that is to say $a(b+c)=ab+ac, (a+b)c=ac+bc, \forall a,b,c\in A$.\\
A ring $A$ is said to be {\it unitary} if it has a (unique) element $ab=ba, \forall a,b\in A$. \\
A non-zero element $a\in A$ is called a {\it zero-divisor} if one can find a non-zero $b\in A$ such that $ab=0$ and 
a ring is called an {\it integral domain} if it has no zero-divisor.\\

\noindent
{\bf Definition 2.1.2}: A ring $K$ is called a {\it field} if every non-zero element
$a\in K$ is a {\it unit}, that is one can find an element $b\in K$ such that 
$ab=1\in K$.\\

\noindent
{\bf Definition 2.1.3}: A {\it module} $M$ over a ring $A$ or simply an 
$A$-{\it module} is a set of elements $x,y,z,...$ which is an abelian group
for an addition $(x,y)\rightarrow x+y$ with an action $A\times M\rightarrow
M:(a,x)\rightarrow ax$ satisfying:\\
$\bullet$ \hspace{1cm} $a(x+y)=ax+ay, \forall a\in A, \forall x,y\in M$\\
$\bullet$ \hspace{1cm} $a(bx)=(ab)x, \forall a,b\in A, \forall x\in M$\\
$\bullet$ \hspace{1cm} $(a+b)x=ax+bx, \forall a,b\in A, \forall x\in M$\\
$\bullet$ \hspace{2cm} $1x=x, \forall x\in M$ \\
The set of modules over a ring $A$ will be denoted by $mod(A)$. A module 
over a field is called a {\it vector space}.\\

\noindent
{\bf Definition 2.1.4}: A map $f:M\rightarrow N$ between two $A$-modules is
called a {\it homomorphism} over $A$ if $f(x+y)=f(x)+f(y), \forall x,y\in
M$ and $f(ax)=af(x), \forall a\in A, \forall x\in M$. We successively
define:\\
$\bullet$ \hspace{2cm} $ker(f)=\{x\in M{\mid} f(x)=0\}$\\
$\bullet$ \hspace{2cm} $im(f)=\{y\in N{\mid} \exists x\in M, f(x)=y\}$ \\
$\bullet$ \hspace{2cm} $coker(f)=N/im(f)$ \\

\noindent
{\bf Definition 2.1.5}: We say that a chain of modules and homomorphisms is a 
{\it sequence} if the composition of two successive such homomorphisms is
zero. A sequence is said to be {\it exact} if the kernel of each map is
equal to the image of the map preceding it. An injective homomorphism is
called a {\it monomorphism}, a surjective homomorphism is called an 
{\it epimorphism} and a bijective homomorphism is called an 
{\it isomorphism}. A short exact sequence is an exact sequence made by a
monomorphism followed by an epimorphism.\\

\noindent
{\bf Proposition 2.1.6}: If one has a short exact sequence:
\[0\longrightarrow
M'\stackrel{f}{\longrightarrow}M\stackrel{g}{\longrightarrow}M''
\longrightarrow 0  \]
then the following conditions are equivalent:\\
$\bullet$ There exists a monomorphism $v:M''\rightarrow M$ such that 
$g\circ v=id_{M''}$.\\
$\bullet$ There exists an epimorphism $u:M\rightarrow M'$ such that 
$u\circ f=id_{M'}$.\\
$\bullet$ There exist isomorphisms $\varphi=(u,g):M\rightarrow M'\oplus M''$
and $\psi=f+v:M'\oplus M''\rightarrow M$ that are inverse to each other and
provide an isomorphism $M\simeq M'\oplus M''$\\

\noindent
{\bf Definition 2.1.7}: In the above situation, we say that the short exact
sequence {\it splits} and $u(v)$ is called a {\it lift} for $f(g)$. In
particular we have the relation: $f\circ u+v\circ g=id_M$.\\

\noindent
{\bf Definition 2.1.8}: A left (right) {\it ideal} $\mathfrak{a}$ in a ring $A$ is a 
submodule of $A$ considered as a left (right) module over itself. When the
inclusion $\mathfrak{a}\subset A$ is strict, we say that $\mathfrak{a}$ is
a {\it proper ideal} of $A$.\\

\noindent
{\bf Lemma 2.1.9}: If $\mathfrak{a}$ is an ideal in a ring $A$, the set of
elements $rad(\mathfrak{a})=\{a\in A{\mid} \exists n\in \mathbb{N}, 
a^n\in \mathfrak{a}\}$ is an ideal of $A$ containing $\mathfrak{a}$ and
called the {\it radical} of $\mathfrak{a}$. An ideal is called {\it perfect} 
or {\it radical} if it is equal to its radical. \\

\noindent
{\bf Definition 2.1.10}: For any subset $S\subset A$, the smallest ideal
containing $S$ is called the ideal {\it generated} by $S$. An ideal generated 
by a single element is called a {\it principal ideal} and a ring is called
a {\it principal ideal ring} if any ideal is principal. The simplest
example is that of polynomial rings in one indeterminate over a field. When
$\mathfrak{a}$ and $\mathfrak{b}$ are two ideals of $A$, we shall denote by
$\mathfrak{a}+\mathfrak{b}$ ($\mathfrak{a}\mathfrak{b}$) the ideal
generated by all the sums $a+b$ (products $ab$) with 
$a\in\mathfrak{a}, b\in\mathfrak{b}$.\\

\noindent
{\bf Definition 2.1.11}: An ideal $\mathfrak{p}$ of a ring $A$ is called a 
{\it prime ideal} if, whenever $ab\in \mathfrak{p}$ ($aAb\in \mathfrak{p}$ in
the non-commutative case) then either $a\in \mathfrak{p}$ or
$b\in\mathfrak{p}$. The set of proper prime ideals of $A$ is denoted by 
$spec(A)$ and called the {\it spectrum} of $A$.\\

\noindent
{\bf Definition 2.1.12}: The {\it annihilator} of a module $M$ in $A$
is the ideal $ann_A(M)$ of $A$ made by all the elements $a\in A$ such that 
$ax=0, \forall x\in M$.\\

From now on, all rings considered will be unitary integral domains, that is
rings containing 1 and having no zero-divisor. For the sake of clarity, as
a few results will also be valid for modules over non-commutative rings, we
shall denote by ${}_AM_B$ a module $M$ which is a left module for $A$ with
operation $(a,x)\rightarrow ax$ and a right module for $B$ with operation
$(x,b)\rightarrow xb$. In the commutative case, lower indices are not
needed. If $M={}_AM$ and $N={}_AN$ are two left $A$-modules, the set of
$A$-linear maps $f:M\rightarrow N$ will be denoted by $hom_A(M,N)$ or
simply $hom(M,N)$ when there will be no confusion and there is a canonical
isomorphism $hom(A,M)\simeq M:f\rightarrow f(1)$ with inverse $x\rightarrow
(a\rightarrow ax)$. When $A$ is commutative, $hom(M,N)$ is again an
$A$-module for the law $(bf)(x)=f(bx)$ as we have indeed:
\[ (bf)(ax)=f(bax)=f(abx)=af(bx)=a(bf)(x).\]
In the non-commutative case, things are much more complicate and we have:\\

\noindent
{\bf Lemma 2.1.13}: Given ${}_AM_B$ and ${}_AN$, then $hom_A(M,N)$ becomes a left
module over $B$ for the law $(bf)(x)=f(xb)$.\\

\noindent
{\it Proof}: We just need to check the two relations:
\[ (bf)(ax)=f(axb)=af(xb)=a(bf)(x),\]
\[ (b'(b''f))(x)=(b''f)(xb')=f(xb'b'')=((b'b'')f)(x).\]
\hspace*{12cm}            $ \Box $  \\

A similar result can be obtained (exercise) with ${ }_AM$ and ${ }_AN_B$, 
where $hom_A(M,N)$ now becomes a right $B$-module for the law 
$(fb)(x)=f(x)b$.\\

Now we recall that a sequence of modules and maps is exact if the kernel 
of any map is equal to the image of the map preceding it and we have:\\

\noindent
{\bf Theorem 2.1.14}: If $M,M',M''$ are $A$-modules, the sequence:
\[ M'\stackrel{f}{\rightarrow}M\stackrel{g}{\rightarrow} M''\rightarrow 0 \]
is exact if and only if the sequence:
\[ 0\rightarrow hom(M'',N)\rightarrow hom(M,N)\rightarrow hom(M',N) \]
is exact for any $A$-module $N$.\\

\noindent
{\it Proof}: Let us consider homomorphisms $h:M\rightarrow N$,
$h':M'\rightarrow N$, $h'':M''\rightarrow N$ such that $h''\circ g=h$,
$h\circ f=h'$. If $h=0$, then $h''\circ g=0$ implies $h''(x'')=0, \forall
x''\in M''$ because $g$ is surjective and we can find $x\in M$ such that
$x''=g(x)$. Then $h''(x'')=h''(g(x))=h''\circ g(x)=0$. Now, if $h'=0$, we
have $h\circ f=0$ and $h$ factors through $g$ because the initial sequence
is exact. Hence there exists $h'':M''\rightarrow N$ such that $h=h''\circ
g$ and the second sequence is exact.\\
We let the reader prove the converse as an exercise.\\
\hspace*{12cm}  $ \Box $   \\

\noindent
{\bf Corollary 2.1.15}: The short exact sequence:
\[  0\rightarrow M'\rightarrow M\rightarrow M''\rightarrow 0  \]
splits if and only if the short exact sequence:
\[ 0\rightarrow hom(M'',N)\rightarrow hom(M,N)\rightarrow hom(M',N)\rightarrow 0 \]
is exact for any module $N$.\\

\noindent
{\bf Definition 2.1.16}: If $M$ is a module over a ring $A$, a {\it system of
  generators} of $M$ over $A$ is a family $\{x_i\}_{i\in I}$ of elements of
  $M$ such that any element of $M$ can be written $x=\sum_{i\in I}a_ix_i$
  with only a finite number of nonzero $a_i$.\\

\noindent
{\bf Definition 2.1.17}: An $A$-module is called {\it noetherian} if every
submodule of $M$ (and thus $M$ itself) is finitely generated.\\

One has the following technical lemma:\\

\noindent
{\bf Lemma 2.1.18}: In a short exact sequence of modules, the central module is
noetherian if and only if the two other modules are noetherian.\\

We obtain in particular:\\

\noindent
{\bf Proposition 2.1.19}: If $A$ is a noetherian ring and $M$ is a finitely
generated module over $A$, then $M$ is noetherian.\\

\noindent
{\it Proof}: Applying the lemma to the short exact sequence $0\rightarrow A^{r-1}\rightarrow A^r\rightarrow A\rightarrow 0$ 
where the epimorphism on the right is the projection onto the first factor, we deduce by induction that $A^r$ is noetherian. 
Now, if $M$ is generated by $\{x_1,...,x_r\}$, there is an epimorphism $A^r\rightarrow M:(1,0,...,0)\rightarrow x_1,...,\{0,...,0,1\}\rightarrow x_r$ 
and $M$ is noetherian because of the lemma. \\
\hspace*{12cm}  $ \Box $   \\

In the preceding situation, the kernel of the epimorphism $A^r \rightarrow  M$ is also finitely generated, say by $\{y_1,...,y_s\}$ 
and we therefore obtain the exact sequence $A^s \rightarrow A^r\rightarrow M\rightarrow 0$
that can be extended inductively to the left.\\

\noindent
{\bf Definition 2.1.20}: In this case, we say that $M$ is {\it finitely presented}.\\

We now present the basic elements of the technique of {\it localization} in the non-commutative case 
as it will be needed later on in a few proofs. We start with a basic definition:\\

\noindent
{\bf Definition 2.1.21}: A subset $S$ of a ring $A$ is said to be {\it
  multiplicatively closed} if $\forall s,t\in S\Rightarrow st\in S$ and
  $1\in S$. For simplicit, we shal suppose from now that $A$ is an integral domain and consider $S=A-\{0\}$.

In a general way, whenever $A$ is a non-commutative ring, that is $ab\neq
ba$ when $a,b\in A$, we shall set the following definition:\\

\noindent
{\bf Definition 2.1.22}: By a {\it left ring of fractions} or 
{\it left localization} of a noncommutative ring $A$ with respect to a 
multiplicatively closed subset $S$ of $A$, we mean a ring denoted by 
$S^{-1}A$ and a homomorphism $\theta={\theta}_S:A\rightarrow S^{-1}A$ 
such that:\\
1) $\theta(s)$ is invertible in $S^{-1}A, \forall s\in S$.\\
2) Each element of $S^{-1}A$ or {\it fraction} has the form 
${\theta(s)}^{-1}\theta(a)$ for some $s\in S, a\in A$.\\
3) $ker(\theta)=\{a\in A{\mid} \exists s\in S, sa=0\}$.\\
A {\it right ring of fractions} or {\it right localization} can be
similarly defined.\\

In actual practice, a fraction will be simply written $s^{-1}a$ and we have
to distinguish carefully $s^{-1}a$ from $as^{-1}$. We shall meet four problems (Pommaret, 2021 d):  \\    
$\bullet$  \,\, How to compare $s^{-1}a$ with $as^{-1}$ ?.  \\
$\bullet$ \,\, How to decide when we shall say that $ s^{-1}a=t^{-1}b$ ?.  \\
$\bullet$ \,\, How to multiply $s^{-1}a$ by $t^{-1}b$ ?.  \\
$\bullet$ \,\, How to find a common denominator for $s^{-1}a + t^{-1}b$ ?.   \\
The following proposition is essential and will be completed by two technical lemmas that
will be used for constructing localizations.\\

The following proposition is essential for constructing localizations:\\

\noindent
{\bf Proposition 2.1.23}: If there exists a left localization of $A$ with respect
to $S$, then we must have:\\
1) $Sa\cap As\neq 0, \forall a\in A, \forall s\in S$.\\
2) If $s\in S$ and $a\in A$ are such that $as=0$, then there exists $t\in
S$ such that $ta=0$.\\

\noindent
{\it Proof}: The element $\theta(a)\theta(s)^{-1}$ in $S^{-1}A$ must be of
the form $\theta(t)^{-1}\theta(b)$ for some $t\in S, b\in A$. Accordingly, 
$\theta(a)\theta(s)^{-1}=\theta(t)^{-1}\theta(b)\Rightarrow
\theta(t)\theta(a)=\theta(b)\theta(s)$ and thus $\theta(ta-bs)=0\Rightarrow
\exists u\in S, uta=ubs$ with $ut\in S, ub\in A$. Finally, $as=0\Rightarrow
\theta(a)\theta(s)=0\Rightarrow \theta(a)=0$ because $\theta(s)$ is
invertible in $S^{-1}A$. Hence $\exists t\in S$ such that $ta=0$.\\
\hspace*{12cm}                 $\Box $ \\

\noindent
{\bf Definition 2.1.24}: A set $S$ satisfying the condition 1) is called a {\it
  left Ore set}.\\

\noindent
{\bf Lemma 2.1.25}: If $S$ is a left Ore set in a noetherian ring, then $S$ also
satisfies the condition 2) of the preceding lemma.\\

\noindent
{\bf Lemma 2.1.26}: If $S$ is a left Ore set in a ring $A$, then $As\cap At\cap
S\neq 0, \forall s,t \in S$ and two fractions can be brought to the same
denominator.\\

Let $K$ be a {\it differential field} with $n$ commuting derivations $({\partial}_1,...,{\partial}_n)$ and consider the ring $D=K[d_1,...,d_n]=K[d]$ of differential operators with coefficients in $K$ with $n$ commuting formal derivatives satisfying $d_ia=ad_i + {\partial}_ia$ in the operator sense. If $P=a^{\mu}d_{\mu}\in D=K[d]$, the highest value of ${\mid}\mu {\mid}$ with $a^{\mu}\neq 0$ is called the {\it order} of the {\it operator} $P$ and the ring $D$ with multiplication $(P,Q)\longrightarrow P\circ Q=PQ$ is filtred by the order $q$ of the operators. We have the {\it filtration} $0\subset K=D_0\subset D_1\subset  ... \subset D_q \subset ... \subset D_{\infty}=D$. As an algebra, $D$ is generated by $K=D_0$ and $T=D_1/D_0$ with $D_1=K\oplus T$ if we identify an element $\xi={\xi}^id_i\in T$ with the vector field $\xi={\xi}^i(x){\partial}_i$ of differential geometry, but with ${\xi}^i\in K$ now. It follows that $D={ }_DD_D$ is a {\it bimodule} over itself, being at the same time a left $D$-module by the composition $P \longrightarrow QP$ and a right $D$-module by the composition $P \longrightarrow PQ$. We define the {\it adjoint} functor $ad:D \longrightarrow D^{op}:P=a^{\mu}d_{\mu} \longrightarrow  ad(P)=(-1)^{\mid \mu \mid}d_{\mu}a^{\mu}$ and we have $ad(ad(P))=P$ both with $ad(PQ)=ad(Q)ad(P), \forall P,Q\in D$. Such a definition can be extended to any matrix of operators by using the transposed matrix of adjoint operators (See (Pommaret, 2021) for more details and applications to control theory or mathematical physics). \\  

\noindent
{\bf Proposition 2.1.27}: $D$ is an Ore domain and $S=D - \{ 0 \} \Rightarrow S^{-1}D = D S^{-1}$. \\

\noindent
{\it Proof}: For this, if $P, Q \in  D $, let us consider the inhomogeneous system $ad(P)y = u, ad(Q) y = v$. As the number of derivative of $(u,v)$ is quite larger 
than the number of derivatives of the single y, there is at least one compatibility condition (CC) for $(u,v)$ of the form $U u = V v$ leading to the identity 
$ U P = V Q $ and $D$ is an Ore domain. Conversely, if $U, V \in D$, we may repeat the same procedure with $ad(U), ad(V)$ in order to get $ad(P), ad(Q)$ such that $ad(P)ad(U)=ad(Q)ad(V)$ and thus to get $P, Q \in D$ such that $UP = V Q$ and thus $U^{-1}V= P Q^{-1}$, a result showing the importance of the adjoint (Compare to (Kashiwara, 1995), p 27).\\
\hspace*{12cm}           $\Box  $   \\

Accordingly, if $y=(y^1, ... ,y^m)$ are differential indeterminates, then $D$ acts on $y^k$ by setting $d_iy^k=y^k_i \longrightarrow d_{\mu}y^k=y^k_{\mu}$ with $d_iy^k_{\mu}=y^k_{\mu+1_i}$ and $y^k_0=y^k$. We may therefore use the jet coordinates in a formal way as in the previous section. Therefore, if a system of OD/PD equations is written in the form ${\Phi}^{\tau}\equiv a^{\tau\mu}_ky^k_{\mu}=0$ with coefficients $a\in K$, we may introduce the free differential module $Dy=Dy^1+ ... +Dy^m\simeq D^m$ and consider the differential {\it module of equations} $I=D\Phi\subset Dy$, both with the residual {\it differential module} $M=Dy/D\Phi$ or $D$-module and we may set $M={ }_DM$ if we want to specify the ring of differential operators. We may introduce the formal {\it prolongation} with respect to $d_i$ by setting $d_i{\Phi}^{\tau}\equiv a^{\tau\mu}_ky^k_{\mu+1_i}+({\partial}_ia^{\tau\mu}_k)y^k_{\mu}$ in order to induce maps $d_i:M \longrightarrow M:{\bar{y} }^k_{\mu} \longrightarrow {\bar{y}}^k_{\mu+1_i}$ by residue with respect to $I$ if we use to denote the residue $Dy \longrightarrow M: y^k \longrightarrow {\bar{y}}^k$ by a bar like in algebraic geometry. However, for simplicity, we shall not write down the bar when the background will indicate clearly if we are in $Dy$ or in $M$. As a byproduct, the differential modules we shall consider will always be {\it finitely generated} ($k=1,...,m<\infty$) and {\it finitely presented} ($\tau=1, ... ,p<\infty$). Equivalently, introducing the {\it matrix of operators} ${\cal{D}}=(a^{\tau\mu}_kd_{\mu})$ with $m$ columns and $p$ rows, we may introduce the morphism $D^p \stackrel{{\cal{D}}}{\longrightarrow} D^m:(P_{\tau}) \longrightarrow (P_{\tau}{\Phi}^{\tau})$ over $D$ by acting with $D$ {\it on the left of these row vectors} while acting with ${\cal{D}}$ {\it on the right of these row vectors} by composition of operators with $im({\cal{D}})=I$. The {\it presentation} of $M$ is defined by the exact cokernel sequence $D^p \stackrel{{\cal{D}}}{\longrightarrow} D^m \longrightarrow M \longrightarrow 0 $. We notice that the presentation only depends on $K, D$ and $\Phi$ or $ \cal{D}$, that is to say never refers to the concept of (explicit local or formal) solutions. It follows from its definition that $M$ can be endowed with a {\it quotient filtration} obtained from that of $D^m$ which is defined by the order of the jet coordinates $y_q$ in $D_qy$. We have therefore the {\it inductive limit} $0 \subseteq M_0 \subseteq M_1 \subseteq ... \subseteq M_q \subseteq ... \subseteq M_{\infty}=M$ with $d_iM_q\subseteq M_{q+1}$ and $M=DM_q$ for $q\gg 0$ with prolongations $D_rM_q\subseteq M_{q+r}, \forall q,r\geq 0$. It is important to notice that it may be sometimes quite difficult to work out $I_q$ or $M_q$ from a given presentation which is not involutive (Pommaret, 1978, 1994, 2023 a, 2024 b). \\

We are now in position to construct the ring of fractions $S^{-1}A$ whenever
$S$ is a left Ore set. For this, usingthe preceding lemmas, let us define an equivalence relation on $S\times A$ 
by saying that $(s,a)\sim (t,b)$ if one can find $u,v\in S$ such that
$us=vt\in S$ and $ua=vb$. Such a relation is clearly reflexive and
symmetric, thus we only need to prove that it is transitive. So let
$(s_1,a_1)\sim (s_2,a_2)$ and $(s_2,a_2)\sim (s_3,a_3)$. Then we can find 
$u_1,u_2\in A$ such that $u_1s_1=u_2s_2\in S$ and $u_1a_1=u_2a_2$. Also we
can find $v_2,v_3\in A$ such that $v_2s_2=v_3s_3\in S$ and
$v_2a_2=v_3a_3$. Now, from the Ore condition, one can find $w_1,w_3\in A$
such that $w_1u_1s_1=w_3v_3s_3\in S$ and thus $w_1u_2s_2=w_3v_2s_2\in S$,
that is to say $(w_1u_2-w_3v_2)s_2=0$. Hence, unless $A$ is an integral
domain, using the second condition of the last proposition, we can find
$t\in S$ such that $t(w_1u_2-w_3v_2)=0 \Rightarrow tw_1u_2=tw_3v_2$. Changing 
$w_1$ and $w_3$ if necessary, we may assume that
$w_1u_2=w_3v_2 \Rightarrow w_1u_1a_1=w_1u_2a_2=w_3v_2a_2=w_3v_3a_3$ as
wished. We finally define $S^{-1}A$ to be the quotient of $S\times A$ by
the above equivalence relation with $\theta:A\rightarrow
S^{-1}A:a\rightarrow 1^{-1}a$.\\
The sum $(s,a)+(t,b)$ will be defined to be $(us=vt,ua+vb)$ and the product
$(s,a)\times (t,b)$ will be defined to be $(st,ab)$.\\
A similar approach can be used in order to define and construct modules of
fractions whenever $S$ satifies the two conditions of the last
proposition. For this we need a preliminary lemma:\\

\noindent
{\bf Lemma 2.1.28}: If $S$ is a left Ore set in a ring $A$ and $M$ is a left
module over $A$, the set:
\[t_S(M)=\{x\in M{\mid} \exists s\in S, sx=0\}  \]
is a submodule of $M$ called the $S$-{\it torsion submodule} of $M$.\\

\noindent
{\it Proof}: If $x,y\in t_S(M)$, we may find $s,t\in S$ such that $sx=0,
ty=0$. Now, we can find $u,v\in A$ such that $us=vt\in S$ and we
successively get $us(x+y)=usx+vty=0\Rightarrow x+y\in t_S(M)$. Also,
$\forall a\in A$, using the Ore condition for $S$, we can find $b\in A,
t\in S$ such that $ta=bs$ and we get $tax=bsx=0\Rightarrow ax\in t_S(M)$.\\
\hspace*{12cm}   $ \Box $    \\

\noindent
{\bf Definition 2.1.29}: By a {\it left module of fractions} or 
{\it left localization} of $M$ with respect to $S$, we mean a left module 
$S^{-1}M$ over $S^{-1}A$ both with a homomorphism
$\theta={\theta}_S:M\rightarrow S^{-1}M:x\rightarrow 1^{-1}x$ such that:\\
1) Each element of $S^{-1}M$ has the form $s^{-1}\theta(x)$ for 
$s\in S,x\in M$.\\
2) $ker({\theta}_S)=t_S(M)$.\\

In order to construct $S^{-1}M$, we shall define an equivalence relation on
$S\times M$ by saying that $(s,x)\sim (t,y)$ if there exists $u,v\in A$
such that $us=vt\in S$ and $ux=vy$. Checking that this relation is
reflexive, symmetric and transitive can be done as before (exercise) and we
define $S^{-1}M$ to be the quotient of $S\times M$ by this equivalence 
relation. \\
The main property of localization is expressed by the following theorem:\\

\noindent
{\bf Theorem 2.1.30}: If one has an exact sequence:
\[    M'\stackrel{f}{\longrightarrow} M \stackrel{g}{\longrightarrow} M''
\]
then one also has the exact sequence:
\[  S^{-1}M'\stackrel{S^{-1}f}{\longrightarrow} S^{-1}M
\stackrel{S^{-1}g}{\longrightarrow} S^{-1}M''  \]
where $S^{-1}f(s^{-1}x)=s^{-1}f(x)$.\\

We now turn to the definition and brief study of tensor products of modules
over rings that will not be necessarily commutative unless stated
explicitly.\\
Let $M=M_A$ be a right $A$-module and $N={}_AN$ be a left $A$-module. We
may introduce the free $\mathbb{Z}$-module made by finite formal linear
combinations of elements of $M\times N$ with coefficients in
$\mathbb{Z}$.\\

\noindent
{\bf Definition 2.1.31}: The tensor product of $M$ and $N$ over $A$ is the
$\mathbb{Z}$-module $M{\otimes}_AN$ obtained by quotienting the above 
$\mathbb{Z}$-module by the submodule generated by the elements of the
form:
\[  (x+x',y)-(x,y)-(x',y), (x,y+y')-(x,y)-(x,y'), (xa,y)-(x,ay) \]
and the image of $(x,y)$ will be denoted by $x\otimes y$.\\

It follows from the definition that we have the relations:
\[  (x+x')\otimes y=x\otimes y+x'\otimes y, x\otimes (y+y')=x\otimes
y+x\otimes y', xa\otimes y=x\otimes ay \]
and there is a canonical isomorphism $M{\otimes}_AA\simeq M,
A{\otimes}_AN\simeq N$. When $A$ is commutative, we may use left modules
only and $M{\otimes}_AN$ becomes a left $A$-module.\\

\noindent
{\bf Example 2.1.32}: If $A=\mathbb{Z}, M=\mathbb{Z}/2\mathbb{Z}$ and
$N=\mathbb{Z}/3\mathbb{Z}$, we have 
$(\mathbb{Z}/2\mathbb{Z}){\otimes}_{\mathbb{Z}}(\mathbb{Z}/3\mathbb{Z})=0$
because $x\otimes y=3(x\otimes y)-2(x\otimes y)=x\otimes 3y-
2x\otimes y=0-0=0$.\\

As a link with localization, we let the reader prove that the multiplication 
map $S^{-1}A\times M\rightarrow S^{-1}M$ given by $(s^{-1}a,x)\rightarrow
s^{-1}ax$ induces an isomorphism $S^{-1}A{\otimes}_AM\rightarrow S^{-1}M$
of modules over $S^{-1}A$ when $S^{-1}A$ is considered as a right module over
$A$ and $M$ as a left module over $A$. \\
When $A$ is a commutative integral domain and $S=A-\{0\}$, the field 
$K=Q(A)=S^{-1}A$ is called the field of fractions of $A$ and we have the
short exact sequence:
\[   0\longrightarrow A \longrightarrow K \longrightarrow K/A
\longrightarrow 0  \]
If now $M$ is a left $A$-module, we may tensor this sequence by $M$ on the
right with $A\otimes M=M$ but we do not get in general an exact sequence. 
The defect of exactness {\it on the left} is nothing else but the 
{\it torsion submodule} $t(M)=\{m\in M{\mid} \exists 0\neq a\in A, am=0\} 
\subseteq M$ and we have the long exact sequence:
\[ 0 \longrightarrow t(M) \longrightarrow M \longrightarrow K{\otimes}_AM 
\longrightarrow K/A{\otimes}_AM \longrightarrow 0 \]
as we may describe the central map as follows:
\[  m \longrightarrow  1\otimes m=\frac{a}{a}\otimes m=
\frac{1}{a}\otimes am  \hspace{5mm},\hspace{5mm} \forall 0\neq a\in A\]
Such a result, based on the localization technique, allows to
understand why controllability has to do with the so-called 
``simplification'' of the {\it transfer matrix}. In particular, a module 
$M$ is said to be a {\it torsion module} if $t(M)=M$ and a 
{\it torsion-free module} if $t(M)=0$.\\

\noindent
{\bf Definition 2.1.33}: A module in $mod(A)$ is called a {\it free module} if it
has a {\it basis}, that is a system of generators linearly independent over
$A$. When a module $F$ is free, the number of generators in a basis, and
thus in any basis (exercise), is called the {\it rank} of $F$ over $A$ and
is denoted by $rank_A(F)$. In particular, if $F$ is free of finite rank $r$,
then $F\simeq A^r$.\\

More generally, if $M$ is any module over a ring $A$ and $F$ is a maximum
free submodule of $M$, then $M/F=T$ is a torsion module. Indeed, if $x\in
M, x\notin F$, then one can find $a\in A$ such that $ax\in F$ because,
otherwise, $F\subset\{F,x\}$ should be free submodules of $M$ with a strict 
inclusion. In that case, the {\it rank} of $M$ is by definition the rank 
of $F$ over $A$ and one has equivalently :\\

\noindent
{\bf Lemma 2.1.34}: $rk_A(M)=dim_K(K{\otimes}_AM)$.\\

\noindent
{\it Proof}: Taking the tensor product by $K$ over $A$ of the short exact
sequence $0\rightarrow F\rightarrow M \rightarrow T \rightarrow 0$, we get
an isomorphism $K{\otimes}_AF\simeq K{\otimes}_AM$ because
$K{\otimes}_AT=0$ (exercise) and the lemma follows from the definition of
the rank.\\
\hspace*{12cm}  $  \Box $   \\

We now provide two proofs of the {\it additivity property of the rank}, 
the second one being also valid for non-commutative rings.

\noindent
{\bf Proposition 2.1.35}: If $0\rightarrow M'\stackrel{f}{\rightarrow}
M\stackrel{g}{\rightarrow} M''\rightarrow 0$ is a short exact sequence of 
modules over a ring $A$, then we have $rk_A(M)=rk_A(M')+rk_A(M'')$.\\

\noindent
{\it Proof 1}: Using localization with respect to the multiplicatively 
closed subset $S=A-\{0\}$, this proposition is just a straight consequence
of the definition of rank and the fact that localization preserves 
exactness.\\
{\it Proof 2}: Let us consider the following diagram with exact left/right
columns and central row:\\
\[
\begin{array}{rcccccl}
  & 0 & & 0 & & 0 & \\
  & \downarrow & & \downarrow & & \downarrow & \\
0\rightarrow & F'& \rightarrow & F'\oplus F''& \rightarrow &
  F''&\rightarrow 0\\
  & \;\;\;\downarrow i'& & \;\;\downarrow i & & \;\;\;\;
\downarrow i''&  \\
0\rightarrow & M'& \stackrel{f}{\rightarrow} & M &\stackrel{g}{\rightarrow}
  & M'' &\rightarrow 0 \\
  & \;\;\;\downarrow p'& & \;\;\;\downarrow p & & \;\;\;\;\downarrow p''
&   \\
0\rightarrow & T'& \rightarrow & T &\rightarrow & T''& \rightarrow 0 \\
 & \downarrow & & \downarrow & & \downarrow &   \\
 & 0 & & 0 & & 0 &  
\end{array}
\]
where $F'(F'')$ is a maximum free submodule of $M'(M'')$ and 
$T'=M'/F'(T''=M''/F'')$ is a torsion module. Pulling back by $g$ the image
under $i''$ of a basis of $F''$, we may obtain by linearity a map 
$\sigma:F''\rightarrow M$ and we define $i=f\circ i'\circ{\pi}'+
\sigma\circ{\pi}''$ where ${\pi}':F'\oplus F''\rightarrow F'$ and 
${\pi}'':F'\oplus F''\rightarrow F''$ are the canonical projections on 
each factor of the direct sum. We have $i{\mid}_{F'}=f\circ i'$ and 
$g\circ i=g\circ \sigma\circ {\pi}''=i''\circ {\pi}''$. Hence, the
diagram is commutative and thus exact with $rk_A(F'\oplus F'')=
rk_A(F')+rk_A(F'')$ trivially. Finally, if $T'$ and $T''$ are torsion
modules, it is easy to check that $T$ is a torsion module too and $F'\oplus
F''$ is thus a maximum free submodule of $M$.\\
\hspace*{12cm}                       $\Box $  \\

\noindent
{\bf Definition 2.1.36}: If $f:M\rightarrow N$ is any morphism, the {\it rank} of
$f$ will be defined to be $rk_A(f)=rk_A(im(f))$.\\

We provide a few additional properties of the rank that will be used in the
sequel. For this we shall set $M^*=hom_A(M,A)$ and, for any moprphism
$f:M\rightarrow N$ we shall denote by $f^*:N^*\rightarrow M^*$ the
corresponding morphism which is such that 
$f^*(h)=h\circ f, \forall h\in hom_A(N,A)$.\\

\noindent
{\bf Proposition 2.1.37}: When $A$ is a commutative integral domain and $M$ is a
finitely presented module over $A$, then $rk_A(M)=rk_A(M^*)$.\\

\noindent
{\it Proof}: Applying $hom_A(\bullet,A)$ to the short exact sequence in the
proof of the preceding lemma while taking into account $T^*=0$, we get a
monomorphism $0\rightarrow M^*\rightarrow F^*$ and obtain therefore 
$rk_A(M^*)\leq rk_A(F^*)$. However, as $F\simeq A^r$ with $r<\infty$ 
because $M$ is finitely generated, we get $F^*\simeq A^r$ too because 
$A^*\simeq A$. It follows that $rk_A(M^*)\leq rk_A(F^*)=rk_A(F)=rk_A(M)$
and thus $rk_A(M^*)\leq rk_A(M)$.\\
Now, if $F_1\stackrel{d}{\rightarrow} F_0\rightarrow M\rightarrow 0$ is a
finite presentation of $M$, applying $hom_A(\bullet,A)$ to this
presentation, we get the ker/coker exact sequence:\\
\[  0 \leftarrow N \leftarrow F_1^* \stackrel{d^*}{\leftarrow} F_0^*
\leftarrow M^* \leftarrow 0  \]
Applying $hom_A(\bullet,A)$ to this sequence while taking into account the two useful 
isomorphisms $F_0^{**}\simeq F_0,F_1^{**}\simeq F_1$, we get the ker/coker
exact sequence:\\
\[ 0 \rightarrow N^* \rightarrow F_1 \stackrel{d}{\rightarrow} F_0
\rightarrow M \rightarrow 0  \]
Counting the ranks, we obtain:\\
\[  rk_A(N)-rk_A(M^*)=rk_A(F_1^*)-rk_A(F_0^*)=rk_A(F_1)-rk_A(F_0)=
rk_A(N^*)-rk_A(M) \]
and thus:\\
\[  (rk_A(M)-rk_A(M^*))+(rk_A(N)-rk_A(N^*))=0  \]
As both two numbers in this sum are non-negative, they must be zero and we
finally get the very important formulas $rk_A(M)=rk_A(M^*), rk_A(N)=rk_A(N^*)$.\\
\hspace*{12cm}   $ \Box $  \\

\noindent
{\bf Corollary 2.1.38}: Under the condition of the proposition, we have 
$rk_A(f)=rk_A(f^*)$.\\

\noindent
{\it Proof}: Introducing the $ker/coker$ exact sequence:
\[ 0 \rightarrow K \rightarrow M \stackrel{f}{\rightarrow} N 
\rightarrow Q \rightarrow 0\]
we have: $rk_A(f)+rk_A(Q)=rk_A(N)$. Applying $hom_A(\bullet,A)$ and taking
into account Theorem 2.A.14, we have the exact sequence:
\[ 0\rightarrow Q^*\rightarrow N^*\stackrel{f^*}{\rightarrow} M^*  \]
and thus : $rk_A(f^*)+rk_A(Q^*)=rk_A(N^*)$. Using the preceding
proposition, we get $rk_A(Q)=rk_A(Q^*)$ and $rk_A(N)=rk_A(N^*)$, that is to
say $rk_A(f)=rk_A(f^*)$.\\
\hspace*{12cm}  $ \Box $   \\   \\   \\

\noindent
{\bf 2.2) HOMOLOGICAL ALGEBRA}\\

Having in mind the previous section, we now need a few definittions and
results from homological algebra (Rotman, 1979). In all that follows, $A,B,C,...$ are
modules over a ring $A$ or vector spaces over a field $k$ and the linear 
maps are making the diagrams commutative.\\
We start recalling the well known Cramer's rule for linear systems through
the exactness of the ker/coker sequence for modules. We introduce the
notations $rk=rank$, $nb=number$, $dim=dimension$, $ker=kernel$, $im=image$, 
$coker=cokernel$. When $\Phi :A\rightarrow B$ is a linear map (homomorphism),
we introduce the so-called ker/coker exact sequence:
\[0\longrightarrow ker(\Phi) \longrightarrow A
\stackrel{\Phi}{\longrightarrow} B \longrightarrow coker(\Phi) 
\longrightarrow 0 \]
\noindent
where $coker(\Phi)=B/im(\Phi) $.\\
In the case of vector spaces over a field $k$, we successively have 
$rk(\Phi)=dim(im(\Phi))$, $dim(ker(\Phi))=dim(A)-rk(\Phi)$, 
$dim(coker(\Phi))=dim(B)-rk(\Phi)=nb$ of
compatibility conditions, and obtain by substraction:
\[ dim(ker(\Phi))-dim(A)+dim(B)-dim(coker(\Phi))=0 \]
In the case of modules, using localization, we may replace the dimension by
the rank and obtain the same relations because of the additive property of
the rank. The following theorem is essential:\\

\noindent
{\bf Snake Theorem 2.2.1}: When one has the following commutative diagram
resulting from the the two central vertical short exact sequences by
exhibiting the three corresponding horizontal ker/coker exact sequences:
\[    \fbox{ $ \begin{array}{ccccccccccc}
 & & 0 & & 0 & & 0 & & & & \\
 & &\downarrow & & \downarrow & & \downarrow & & & & \\
0&\longrightarrow &K&\longrightarrow &A&\longrightarrow
&A'&\longrightarrow &Q&\longrightarrow &0\\
 & &\downarrow & &\;\;\;\downarrow \! \Phi&
&\;\;\;\;\downarrow \! {\Phi}'& &\downarrow & & \\
0&\longrightarrow &L&\longrightarrow &B&\longrightarrow
&B'&\longrightarrow &R& \longrightarrow &0 \\
 & &\downarrow & &\;\;\;\downarrow \! \Psi & &\;\;\;\;
\downarrow \! {\Psi}'& & \downarrow & & \\
0 & \longrightarrow &M& \longrightarrow &C&
\longrightarrow &C'& \longrightarrow &S& \longrightarrow
&0 \\
 & & & & \downarrow & & \downarrow & & \downarrow & & \\
 & & & &0& &0& &0& &
\end{array}     $   }\]
then there exists a connecting map $M \longrightarrow Q$ both with a long
exact sequence:
\[   \fbox{  $0 \longrightarrow K \longrightarrow L \longrightarrow M  \longrightarrow Q \longrightarrow R \longrightarrow S  \longrightarrow 0.  $  }  \]

We may now introduce {\it cohomology theory} through the following definition:\\

\noindent
{\bf Definition 2.2.2}: If one has a sequence $A \stackrel{\Phi}{\longrightarrow} B \stackrel{\Psi}{\longrightarrow} C $,
then one may introduce with $coboundary=im(\Phi)\subseteq ker(\Psi)=cocycle \subseteq B$ and define 
the cohomology at $B$ to be the quotient $cocycle/coboundary $.\\

\noindent
{\bf Theorem 2.2.3}: The following commutative diagram where the two central 
vertical sequences are long exact sequences and the horizontal lines are
ker/coker exact sequences:
\[   \fbox{ $      \begin{array}{ccccccccccccc}
 & &0& &0& &0& & & & & & \\
 & &\downarrow & & \downarrow & & \downarrow & & & & & &   \\
0&\longrightarrow &K&\longrightarrow &A&\longrightarrow  &A'&\longrightarrow &Q&\longrightarrow &0& & \\
 & &\downarrow & &\;\;\;\downarrow \! \Phi &&\;\;\;\;\downarrow \! \Phi' & &\downarrow & & & & \\
0&\longrightarrow &L&\longrightarrow&B&\longrightarrow &B'&\longrightarrow &R&\longrightarrow &0& & \\
   & &\downarrow & &\;\;\;\downarrow \! \Psi & &\;\;\;\;\downarrow \! \Psi'&  &\downarrow &  & & &  \\
0&\longrightarrow &\fbox{M}&\longrightarrow &C&\longrightarrow &C'&\longrightarrow &S&\longrightarrow &0& & \\
 & &\downarrow & &\;\;\;\downarrow \! \Omega & &\;\;\;\;\downarrow \! \Omega'& &\downarrow & & & & \\
0&\longrightarrow &N&\longrightarrow &D&\longrightarrow &D'&\longrightarrow &T&\longrightarrow &0& & \\
 & & & &\downarrow & &\downarrow & &\downarrow & & & & \\
 & & & &0& &0& &0 & & & & 
\end{array} $  }  \]
induces an isomorphism between the cohomology at $M$ in the left vertical
column and the kernel of the morphism $Q\rightarrow R$ in the right
vertical column.\\

We now introduce the {\it extension functor} in an elementary manner, using
the standard notation $hom_A(M,A)=M^*$. First of all, by a {\it free 
resolution} of an $A$-module $M$, we understand a long exact sequence:
\[  ...\stackrel{d_2}{\longrightarrow} F_1 \stackrel{d_1}{\longrightarrow}
F_0 \longrightarrow M \longrightarrow 0  \]
where $F_0, F_1, ... $are free modules, that is to say modules isomorphic to
powers of $A$ and $M=coker(d_1)=F_0/im(d_1)$. We may {\it take out} $M$ and obtain the
{\it deleted sequence}:
\[  ...\stackrel{d_2}{\longrightarrow} F_1 \stackrel{d_1}{\longrightarrow}
F_0 \longrightarrow 0  \]
which is of course no longer exact. If $N$ is any other $A$-module, we may 
apply the functor $hom_A(\bullet,N)$ and obtain the sequence:
\[ ...\stackrel{d_2^*}{\longleftarrow} hom_A(F_1,N) \stackrel{d_1^*}
{\longleftarrow} hom_A(F_0,N) \longleftarrow 0  \]

\noindent
{\bf Definition 2.2.4}: One may define:   \\
\[    {ext}^0_A(M,N)=ker(d_1^*)=hom_A(M,N), \hspace{1cm}  {ext}^i_A(M,N)=ker(d_{i+1}^*)/im(d_i^*), \forall i\geq 1 \]

One can prove that the extension modules do not depend on the resolution of $M$ chosen and have the following two main properties, 
the first of which only is classical [19,23].\\

\noindent
{\bf Proposition 2.2.5}: If $0\rightarrow M'\rightarrow M \rightarrow M''\rightarrow 0$ is a short exact sequence of $A$-modules, then we have
the following {\it connecting long exact sequence}:
\[ \fbox{ $  0\rightarrow hom_A(M'',N) \rightarrow hom_A(M,N) \rightarrow hom_A(M',N) \rightarrow ext_A^1(M'',N) \rightarrow ... $  }   \]
of extension modules.\\

We provide two different proofs of the following proposition:\\

\noindent
{\bf Proposition 2.2.6}: $ext^i(M) = ext^i_A(M,A)$ is a torsion module, $\forall i\geq 1$.\\

\noindent
{\it Proof} 1: Let $F$ be a maximal free submodule of $M$. From the short 
exact sequence: 
\[0\longrightarrow F\longrightarrow M\longrightarrow M/F\longrightarrow 0\]
where $M/F$ is a torsion module, we obtain the long exact sequence:
\[    ...\rightarrow ext^{i-1}(F)\rightarrow ext^i(M/F) \rightarrow  ext^i(M)\rightarrow ext^i(F)\rightarrow ...  \]
As $ext^i(F)=0, \forall i\geq 1$ from the definitions, we get $ext^i(M)\simeq ext^i(M/F), \forall i \geq 2$. 
Now it is known that the tensor by the field $K$ of any exact sequence is again an exact sequence. 
Accordingly, we have from the definition:
\[   K{\otimes}_A ext^i(M/F, A)\simeq ext^i_A(M/F,K) \simeq ext^i_K(K{\otimes}_AM/F,K)=0 , \forall i \geq 1 \]
We finally obtain from the above sequence $ K{\otimes}_Aext^i(M)=0  \Rightarrow ext^i(M)$ torsion, $\forall i\geq 1$.\\

\noindent
{\it Proof} 2: Having in mind that $B_i=im(d_i^*)$ and
$Z_i=ker(d_{i+1}^*)$, we obtain $rk(B_i)=rk(d_i^*)=rk(d_i)$ and
$rk(Z_i)=rk(F_i^*)-rk(d_{i+1}^*)=rk(F_i)-rk(d_{i+1})$. However, we started
from a resolution, that is an exact sequence in which
$rk(d_i)+rk(d_{i+1})=rk(F_i)$. It follows that $rk(B_i)=rk(Z_i)$ and thus
$rk(H_i)=rk(Z_i)-rk(B_i)=0$, that is to say $ext^i(M)$ is a torsion
module for $i\geq 1$, $\forall M\in mod(A)$.\\
  \hspace*{12cm}  $ \Box $  \\

As we have seen in the Motivating Examples of the Introduction, the same module may have many 
very different presentations. In particular, we have the {\it Schanuel lemma} (Pommare, 2001):  \\

\noindent
{\bf Lemma 2.2.7}: If $F'_1\stackrel{d'_1}{\longrightarrow}F'_0\rightarrow M\rightarrow 0$ 
and $F''_1\stackrel{d''_1}{\longrightarrow}F''_0\rightarrow M\rightarrow 0$ are two {\it presentations} of $M$, 
there exists a presentation $F_1\stackrel{d_1}{\longrightarrow}F_0\rightarrow M\rightarrow 0$ of $M$ 
projecting onto the preceding ones.\\

\noindent
{\bf Definition 2.2.8}: An $A$-module $P$ is {\it projective} if there exists a
free module $F$ and another (thus projective) module $Q$ such that $P\oplus
Q\simeq F$. Any free module is projective.\\ 

\noindent
{\bf Proposition 2.2.9}: The short exact sequence:\\
\[  0\longrightarrow M'\stackrel{f}{\longrightarrow}M 
\stackrel{g}{\longrightarrow} M''\longrightarrow 0 \]
splits whenever $M''$ is projective.\\

\noindent
{\bf Proposition 2.2.10}: When $P$ is a projective module and $N$ is any module, we have: 
\[  {ext}^i_A(P,N)=0,\forall i\geq 1   \]

\noindent
{\bf Proposition 2.2.11}: When $P$ is a projective module, applying
$hom_A(P,\bullet)$ to any short exact sequence gives a short exact sequence.\\

\noindent
{\bf 2.3) DIFFERENTIAL DUALITY}  \\

The {\it main but highly not evident} trick will be to introduce the {\it adjoint operator} $\tilde{\cal D}=ad(\cal D)$ by the formula of
integration by part:
\[ <\lambda,{\cal D}\xi>=<\tilde{\cal D}\lambda,\xi>+div(\hspace{5mm}) \]
where $\lambda$ is a test row vector and $< >$ denotes the usual contraction. The adjoint can also be defined formally, as in computer
algebra packages, by setting:  \\
\[   ad(a) = a, , \forall a \in K, ad(d_i) = - d_i, ad(P Q) = ad(Q) ad(P), \forall P, Q \in D  \] 
Another way is to define the adjoint of an operator directly on $D$ by setting:  \\
\[  P=\sum_{0\leq {\mid}\mu{\mid}\leq p}a^{\mu}d_{\mu}\longrightarrow  ad(P)=\sum_{0\leq{\mid}\mu{\mid}\leq p}(-1)^{{\mid}\mu{\mid}}d_{\mu}a^{\mu} \] 
for any $P\in D$ with $ord(P)=p$ and to extend such a definition by linearity.\\
We shall denote by $N$ the differential module defined from $ad({\cal D})$ exactly like $M$ was defined from $\cal D$ and we have the following fundamental theorem which is not easily accessible to intuition [ K, K2  ]:\\

\noindent
{\bf Theorem 2.3.1}: There is a long exact sequence: \\
\[  \fbox{ $ 0 \longrightarrow  ext^1(N) \longrightarrow M \stackrel{\epsilon}{\longrightarrow}  M^{**}  \longrightarrow ext^2(N) \longrightarrow  0  $  }  \]
and the two following statements are equivalent:\\
$\bullet$ The corresponding operator is simply (doubly) parametrizable.\\
$\bullet$ The corresponding module is torsion-free (reflexive).\\

\noindent
{\it Proof}: Let us start with a free presentation of $M$:
\[  F_1\stackrel{d_1}{\longrightarrow} F_0 \longrightarrow M \longrightarrow 0 \]
By definition, we have $M=coker(d_1)\Longrightarrow N=coker(d^*_1)$ and we may exhibit the following free resolution of $N$ where 
$M^*=ker(d^*_1)=im(d^*_0)\simeq coker (d^*_{-1})$:  \\
\[  \fbox{  $  \begin{array}{rcccl}
0 \longleftarrow N \longleftarrow F^*_1\stackrel{d^*_1}{\longleftarrow} &F^*_0 &\stackrel{d^*_0}{\longleftarrow}& F^*_{-1} & \stackrel{d^*_{-1}}{\longleftarrow} F^*_{-2} \\ 
                                                                                                        & \uparrow &                                         &  \downarrow &   \\
                                                                                                        &     M^* &        =               &  M^* &   \\
                                                                                                         &     \uparrow   &             &   \downarrow &    \\
                                                                                                        &     0    &                       &   0    &
\end{array}  $  }   \]
The deleted sequence is:
\[ 0 \longleftarrow F^*_1\stackrel{d^*_1}{\longleftarrow}F^*_0  \stackrel{d^*_0}{\longleftarrow}F^*_{-1}  \stackrel{d^*_{-1}}{\longleftarrow} F^*_{-2}  \]
Applying $hom_A(\bullet,A)$ and using the canonical isomorphism $F^{**}\simeq F$ for any free module $F$, we get the sequence:  \\
\[   \fbox{ $ \begin{array}{rcccl}
0\longrightarrow F_1\stackrel{d_1}{\longrightarrow} &F_0 &
\stackrel{d_0}{\longrightarrow}
&F_{-1}&\stackrel{d_{-1}}{\longrightarrow} F_{-2}\\
  & \downarrow &  & \uparrow &   \\
  &    M  &\stackrel{\epsilon}{\longrightarrow} &M^{**} &   \\
  & \downarrow &  & \uparrow &   \\
  &   0        &  &    0     &   
\end{array}  $ } \]
in which $ \epsilon : M \rightarrow  M^{**}$ is defined by $(\epsilon (m))(f) = f(m), \forall f \in hom_A (M, A)$. \\
Denoting as usual a coboundary space by $B$, a cocycle space by $Z$ and the cohomology by $H=Z/B$, we get the commutative and exact diagram:
\[
\begin{array}{rcccccl}
0\longrightarrow & B_0 & \longrightarrow & F_0 &\longrightarrow & M &
\longrightarrow 0 \\
  & \downarrow &  &  \parallel &  & \downarrow \epsilon &  \\
0\longrightarrow & Z_0 & \longrightarrow & F_0 &\longrightarrow & M^{**} &  
\end{array}  \]
An easy snake chase provides at once $H_0=Z_0/B_0=ext^1(N)\simeq  ker(\epsilon)$ and it follows that $ker(\epsilon) \subseteq  M$ is a torsion module, 
that is $ ker(\epsilon) \subseteq t(M)$. \\
Now, if $m \in t(M)$, then we can find $0 \neq a \in A$ such that $a m =0$. Hence, $\forall f \in hom_A(M, A)$, we have $a (f(m)) = f(am) = f(0) = 0$ 
and thus $f(m)=0$ because $A$ is an integral domain. We obtain therefore $t(M) \subseteq  ker(\epsilon) \subseteq M$ and thus $t(M)=ker(\epsilon)$.   \\
Finally, as $B_{-1}=im(\epsilon)$ and $Z_{-1}\simeq M^{**}$, we finally obtain:  \\
 \[  H_{-1}=Z_{-1}/B_{-1}=ext^2 (N)\simeq coker(\epsilon) \]
 Accordingly, a torsion-free (reflexive) module is described by an operator that admits a single (double) step parametrization.\\
 As $ ad(ad({\cal{D}}))= {\cal{D}}$, it is important to notice that one can exchange $M$ and $N$ in any case.  \\
\hspace*{12cm}  $ \Box $ \\

The same proof also provides an effective test for applications by using $D$ and $ad$ instead of $A$ and $*$ in the differential framework. 
In particular, a control system is controllable if it does not admit any ``{\it autonomous element}'', that is to say any finite linear combination 
of the control variables and their derivatives that satisfies,  {\it for itself}, at least one OD or PD equation. More precisely, starting with 
the control system described by an operator ${\cal{D}}_1$, one MUST construct ${\tilde{\cal{D}}}_1$ and then $\cal{D}$ such that 
$\tilde{\cal{D}}$ generates all the compatibility conditions of ${\tilde{\cal{D}}}_1$. Finally, $M$ is torsion-free if and only if ${\cal{D}}_1$ 
generates all the compatibility conditions of $\cal{D}$. Though striking it could be, {\it this is the true generalization of the standard Kalman test} 
as we already claimed in the Introduction.\\

\noindent
{\bf Corollary 2.3.2}: The constructive test in order to know if an operator ${\cal{D}}_1$ can be parametrized by an operator ${\cal{D}}$ has five successive steps along with the following diagram in operator language:  \\

 \[  \begin{array}{rcccccccl}
   &  &  &  & & & &  {\zeta}' &    \hspace{1cm}   \fbox{5}   \\
   &  &  &  & & &  \stackrel{{ {\cal{D}}_1}'}{\nearrow}  & &  \\
  \fbox{4} \hspace{7mm}  &  &  & \xi  & \stackrel{{\cal{D}}}{\longrightarrow} &  \eta & \stackrel{{\cal{D}}_1}{\longrightarrow} & \zeta         &\hspace{1cm}   \fbox{1}  \\
   &  &  &  &  &  &  &  &  \\
   & &  &  &  &  &  &   &  \\
  \fbox{3} \hspace{7mm}  &  &  & \nu & \stackrel{ad({\cal{D}})}{\longleftarrow} & \mu & \stackrel{ad({\cal{D}}_1)}{\longleftarrow} & \lambda &\hspace{1cm} \fbox{2}    \\ 
  & & & & & & & &     
  \end{array}    \] 
  \noindent
\[{\cal{D}}_1 \,\,\,parametrized \,\,\,by \,\,\,{\cal{D}} \Leftrightarrow {\cal{D}}_1={\cal{D}}'_1  \Leftrightarrow ext^1(N)=0  \Leftrightarrow  \epsilon \,\,\, injective  \Leftrightarrow t(M)=0\] 
\noindent
Any new CC brought by ${\cal{D}}'_1$ is a torsion element of the differential module defined by 
${\cal{D}}_1$.    \\
\hspace*{12cm}  $\Box$  \\

\noindent
{\it Proof}: We have used the fact that $ad(ad({\cal{D}}))={\cal{D}}$ and the parametrization is existing if and only if we may have ${{\cal{D}}_1}' = {\cal{D}}_1$ whenever ${{\cal{D}}_1}'$ generates the CC of ${\cal{D}} $ as $ad({\cal{D}}) \circ ad({\cal{D}}_1)=0  \Rightarrow {\cal{D}}_1 \circ {\cal{D}}=0$, that is ${\cal{D}}_1$ is surely among the CC of ${\cal{D}}$ but other CC may also exist. In addition, denoting by $M_1$ the differentia module determined by ${\cal{D}}_1$ and using the fact that 
$rk_D({\cal{D}}'_1)= rk_D({\cal{D}}_1)= p - rk_D({\cal{D}})$ because $rk_D(ad({\cal{D}}))= p - rk_D(ad({\cal{D}}_1))$, then any new CC provides an element of 
$t(M_1)$.  \\

\noindent
{\bf Corollary 2.3.3}: The constructive test in order to know if an operator ${\cal{D}}_1$ can be parametrized by an operator ${\cal{D}}$ which can be itself parametrized by an operator ${\cal{D}}_{-1}$ has 5 steps which are drawn in the following diagram where $ad({\cal{D}})$ generates the CC of $ad({\cal{D}}_1)$ and ${\cal{D}}_1'$ generates the CC of ${\cal{D}}=ad(ad({\cal{D}}))$ while $ad({\cal{D}}_{-1})$ generates the CC of $ad({\cal{D}})$ and ${\cal{D}}'$ generates the CC of ${\cal{D}}_{-1}$:  \\
\[  \begin{array}{rcccccccl}
 & & & & & {\eta}'     & &  {\zeta}' &\hspace{15mm} \fbox{5}  \\
 & & & &  \stackrel{{\cal{D}}'}{\nearrow}   & & \stackrel{{\cal{D}}'_1}{\nearrow} &  &  \\
\fbox{4} \hspace{15mm}&\phi & \stackrel{{\cal{D}}_{-1}}{\longrightarrow}& \xi  & \stackrel{{\cal{D}}}{\longrightarrow} &  \eta & \stackrel{{\cal{D}}_1}{\longrightarrow} & \zeta &\hspace{15mm}   \fbox{1}  \\
 &  &  &  &  &  &  &  &  \\
 &  &  &  &  &  &  &  &  \\
 \fbox{3} \hspace{15mm}& \theta &\stackrel{ad({\cal{D}}_{-1})}{\longleftarrow}& \nu & \stackrel{ad({\cal{D}})}{\longleftarrow} & \mu & \stackrel{ad({\cal{D}}_1)}{\longleftarrow} & \lambda &\hspace{15mm}  \fbox{2}
  \end{array}  \]
 \[{\cal{D}}_1 \,\,\,parametrized \,\,\,by \,\,\,{\cal{D}} \Leftrightarrow {\cal{D}}_1={\cal{D}}'_1  \Leftrightarrow ext^1(N)=0  \Leftrightarrow  \epsilon \,\,\, injective  \Leftrightarrow t(M)=0\] 
\[{\cal{D}} \,\,\,parametrized \,\,\,by \,\,\,{\cal{D}}_{-1} \Leftrightarrow {\cal{D}}={\cal{D}}' \Leftrightarrow ext^2(N)=0 \Leftrightarrow \epsilon \,\,\, surjective  \hspace{17mm}  \]

\noindent
{\bf Corollary 2.3.4}: In the differential module framework, if $F_1 \stackrel{{\cal{D}}_1}{\longrightarrow} F_0 \stackrel{p}{\longrightarrow} M \rightarrow 0$ is a finite free presentation of $M=coker({\cal{D}}_1)$ with $t(M)=0$, then we may obtain an exact sequence $F_1 \stackrel{{\cal{D}}_1}{\longrightarrow} F_0 \stackrel{{\cal{D}}}{\longrightarrow} E $ of free differential modules where ${\cal{D}}$ is the parametrizing operator. However, there may exist other parametrizations $F_1 \stackrel{{\cal{D}}_1}{\longrightarrow} F_0 \stackrel{{\cal{D}}'}{\longrightarrow} E' $ called {\it minimal parametrizations} such that $coker({\cal{D}}')$ is a torsion module and we have thus $rk_D(M)=rk_D(E')$.  \\

\noindent
{\bf Example 2.3.5 }: When $n\geq3$, the existence of the Poincar\'{e} differential sequence:    \\
\[   0 \rightarrow \Theta \rightarrow {\wedge}^0 T^* \stackrel{d}{\rightarrow} {\wedge}^1 T^* \stackrel{d}{\rightarrow} ...  \stackrel{d}{\rightarrow} {\wedge}^{n-1} T^* 
\stackrel{d}{\rightarrow} {\wedge}^n T^* \rightarrow 0  \]
for the exterior derivative $"d"$, proves that the differential module defined by the last operator is surely reflexive. However, when $n=3$, the operators involved, 
namely $(grad, curl, div)$, are such that the $div$ may be parametrized by an operator defining a torsion module as follows by considering the involutive system:  \\
\[  \left  \{   \begin{array}{lcc}
     d_3 y^3 & = & z^1  \\
     d_3 y^2 & = & z^2  \\
     d_2 y^1 - d_1 y^2 & = & z^3 
\end{array}  \right. \fbox{ $\begin{array}{ccc}
1 & 2 & 3 \\
1 & 2 & 3 \\
1 & 2 & \bullet
\end{array} $  } \Rightarrow  d_1 z^1 + d_2 z^2 + d_3 z^3 =0     \]  \\

Now, in order to have a full picture of the correspondence existing between differential modules and differential operators, it just remains to explain
why and how we can pass from left to right modules and conversely. By this way, we shall be able to take into account the behaviour of the adjoint of
an operator under changes of coordinates. We start with a technical but quite useful  lemma (Pommaret, 2001, 2023 b):\\

\noindent
{\bf Lemma 2.3.6}: If $f\in aut(X)$ is a local diffeomorphism of $X$, we may set
$y=f(x) \Rightarrow x=f^{-1}(y)=g(y)$ and introduce the {\it jacobian}
$\Delta (x)=det({\partial}_if^k(x))\neq 0$. Then, we have the identity:
\[   \frac{\partial}{\partial y^k}(\frac{1}{\Delta (g(y))}
{\partial}_if^k(g(y))\equiv 0.   \]

Accordingly, we notice that, if ${\cal{D}}:E\rightarrow F$ is an operator,
the way to obtain the adjoint through an integration by parts proves that
the test function is indeed a section of the {\it adjoint bundle}
$\tilde{F}=F^*\otimes {\Lambda}^nT^*$ and that we get an operator
$ad({\cal{D}}):\tilde{F}\rightarrow \tilde{E}$. This is in particular the 
reason why, in elasticity, the deformation is a covariant tensor but the 
stress is a contravariant tensor density and, in electromagnetism, the EM 
field is a covariant tensor (in fact a 2-form) but the induction is a 
contravariant tensor density.\\
Also, if we define the adjoint formally, we get, in the operator sense:\\
\[ ad(\frac{1}{\Delta}{\partial}_if^k\frac{\partial}{\partial
    y^k})=-\frac{\partial}{\partial y^k}\circ
    (\frac{1}{\Delta}{\partial}_if^k)=-\frac{1}{\Delta}{\partial}_if^k
\frac{\partial}{\partial y^k}=
-\frac{1}{\Delta}\frac{\partial}{\partial x^i}  \]
and obtain therefore:
\[\frac{\partial}{\partial x^i}=
{\partial}_if^k(x)\frac{\partial}{\partial y^k} \Rightarrow 
ad(\frac{\partial}{\partial x^i})=-\frac{\partial}{\partial x^i}=\Delta
    ad(\frac{1}{\Delta}{\partial}_if^k(x)\frac{\partial}{\partial y^k})  \]
a result showing that the adjoint of the gradient operator 
$d:{\Lambda}^0T^*\rightarrow {\Lambda}^1T^*$ is minus the exterior
derivative $d:{\Lambda}^{n-1}T^*\rightarrow {\Lambda}^nT^*$.\\

If $A$ is a differential ring and $D=A[d]$ as usual, we may introduce the
ideal $I=\{P\in D{\mid}P(1)=0\}$ and obtain $A\simeq D/I$ both with the
direct sum decomposition $D\simeq A\oplus I$. In fact, denoting by $D_q$
the submodule over $A$ of operators of order $q$, $A$ can be identified 
with the subring $D_0\subset D$ of zero order operators and we may
consider any differential module over $D$ as a module over $A$, just 
`` {\it forgetting}'' about its differential structure. Caring about the
notation, we shall set $T=D_1/D_0=\{\xi=a^id_i{\mid}a^i\in A\}$ with
$\xi(a)={\xi}^i{\partial}_ia,\forall a\in A$, so that $D$ can be generated by
$A$ and $T$. \\             

The module counterpart is more tricky and is based on the following theorem (Pommaret, 2001):\\

\noindent
{\bf Theorem 2.3.7}: If $M$ and $N$ are right $D$-modules, then $hom_A(M,N)$
becomes a left $D$-module.\\

\noindent
{\it Proof}: We just need to define the action of $\xi\in T$ by the
formula:
\[  (\xi f)(m)=f(m\xi)-f(m)\xi ,\hspace{1cm} \forall m\in M  \]
Indeed, setting $(af)(m)=f(m)a=f(ma)$ and introducing the bracket
$(\xi,\eta)\rightarrow [\xi,\eta]$ of vector fields, we let the reader 
check that $a(bf)=(ab)f, \forall a,b\in A$ and that we have the formulas:
\[ \xi(af)=(\xi(a)+a\xi)f,\hspace{1cm}(\xi\eta-\eta\xi)f=[\xi,\eta]f,
\forall a\in A,\forall \xi,\eta\in T  \]
in the operator sense.\\
\hspace*{12cm}  $ \Box $  \\

Finally, if $M$ is a left $D$-module, according to the comment following lemma 3.1.13, then $M^*=hom_D(M,D)$ is a right $D$-module 
and thus $N=N_r$ is a right $D$-module. However, we have the following technical proposition:\\

\noindent
{\bf Proposition 2.3.8}: ${\Lambda}^nT^*$ has a natural right module structure
over $D$. \\

\noindent
{\it Proof}: If $\alpha=adx^1\wedge ...\wedge dx^n\in T^*$ is a volume form
with coefficient $a\in A$, we may set 
$\alpha.P=ad(P)(a)dx^1\wedge...\wedge dx^n$. As $D$ is generated by $A$ and
$T$, we just need to check that the above formula has an intrinsic meaning
for any $\xi\in T$. In that case, we check at once:
\[  \alpha.\xi=-{\partial}_i(a{\xi}^i)dx^1\wedge...\wedge dx^n=
-\cal{L}(\xi)\alpha \]
by introducing the Lie derivative of $\alpha$ with respect to $\xi$, along
the intrinsic formula ${\cal{L}}(\xi)=i(\xi)d+di(\xi)$ where $i( )$ is the
interior multiplication and $d$ is the exterior derivative of exterior
forms. According to well known properties of the Lie derivative, we get :
\[\alpha.(a\xi)=(\alpha.\xi).a-\alpha.\xi(a), \hspace{5mm}
\alpha.(\xi\eta-\eta\xi)=-[\cal{L}(\xi),\cal{L}(\eta)]\alpha=
-\cal{L}([\xi,\eta])\alpha=\alpha.[\xi,\eta].  \]
\hspace*{12cm}  $ \Box $ \\

According to the preceding theorem and proposition, the left differential module corresponding to $ad(\cal{D})$ is not $N_r$ but rather 
$N_l=hom_A({\Lambda}^nT^*,N_r)$. When $D$ is a commutative ring, this side changing procedure is no longer needed.\\

Of course, keeping the same module $M$ but changing its presentation or 
even using an isomorphic module $M'$ (2 OD equations of order 2 or 4 OD 
equations of order 1 as in the case of the double pendulum), then $N$ may 
change to $N'$. The following result, {\it totally unaccessible to intuition},
justifies ``{\it a posteriori}'' the use of the extension functor by proving
that the above results are unchanged and are thus ``{\it intrinsic}'' (Pommaret, 2001, 2005):   //

\noindent
{\bf Theorem 2.3.9}: $N$ and $N'$ are {\it projectively equivalent}, that is to say one can find projective modules $P$ and $P'$ 
such that $N \oplus P\simeq N' \oplus P'$.\\

\noindent
{\it Proof}: According to Schanuel lemma, we can always suppose, with no
loss of generality, that the resolution of $M$ projects onto the
resolution of $M'$. The kernel sequence is a splitting sequence made up with
projective modules because the kernel of the projection of $F_i$ onto
$F'_i$ is a projective module $P_i$ for $i=0,1$. Such a property still holds 
when applying duality. Hence, if $C$ is the kernel of the epimorphism from
$P_1$ to $P_0$ induced by $d_1$, then $C$ is a projective module and the
top short exact sequence splits in the following commutative and exact 
diagram:\\
\[ \begin{array}{lllllllllll}
 & & 0 & & 0 & & 0 & & & & \\
 & & \downarrow & & \downarrow & & \downarrow & & & & \\
0 & \rightarrow & C & \rightarrow & P_1 & \rightarrow & P_0 & \rightarrow &
 0 & & \\
 & & \downarrow & & \downarrow & & \downarrow & & \downarrow & & \\
0 & \rightarrow & K & \rightarrow & F_1 &\stackrel{d_1}{\rightarrow} & F_0
 & \rightarrow & M & \rightarrow & 0 \\ 
 & & \downarrow & & \downarrow & & \downarrow & & \downarrow & & \\
0 & \rightarrow & K' & \rightarrow & F'_1 & \stackrel{d'_1}{\rightarrow} &
 F'_0 & \rightarrow & M' & \rightarrow & 0 \\
 & & \downarrow & & \downarrow & & \downarrow & & \downarrow & & \\
 & & 0 & & 0 & & 0 & & 0 & &
\end{array}  \]
Applying $hom_A(\bullet,A)$ to this siagram while taking into account
Corollary 3.1.15, we get the following commutative and exact diagram:\\
\[ \begin{array}{lllllllllll}
 & & 0 & & 0 & & 0 & & & & \\
 & & \uparrow & & \uparrow & & \uparrow & & & & \\
0 & \leftarrow & C^* & \leftarrow & P_1^* & \leftarrow & P_0^* & \leftarrow &
 0 & & \\
 & & \uparrow & & \uparrow & & \uparrow & & \uparrow & & \\
0 & \leftarrow & N & \leftarrow & F_1^* &\stackrel{d_1^*}{\leftarrow} & F_0^*
 & \leftarrow & M^* & \leftarrow & 0 \\ 
 & & \uparrow & & \uparrow & & \uparrow & & \uparrow & & \\
0 & \leftarrow & N'& \leftarrow & {F'_1}^* & \stackrel{{d'_1}^*}{\leftarrow} &
 {F'_0}^* & \leftarrow & {M'}^* & \leftarrow & 0 \\
 & & \uparrow & & \uparrow & & \uparrow & & \uparrow & & \\
 & & 0 & & 0 & & 0 & & 0 & &
\end{array}  \]
In this diagram $C^*$ is also a projective module, the upper and left 
short exact sequences split and we obtain $N\simeq N'\oplus C^*$.\\
  \hspace*{12cm}  $ \Box $   \\ 

Accordingly, using the properties of the extension functor, we get:\\

\noindent
{\bf Corollary 2.3.10}: $ext^i(N) \simeq ext^i(N'), \,\,\, \forall i \geq 1$.  \\

\noindent
{\bf Remark 2.3.11}: When $A$ is a principal ideal ring, it is well known (See (Pommaret, 2001, Rotman, 1979) for more details) that any torsion-free module over $A$ is free and thus projective. Accordingly, the kernel of the projection of $F_0$ onto $M$ is free and we can always suppose, 
with no loss of generality, that $d_1$ and $d'_1$ are monomorphisms [8]. In that case, there is an isomorphism  $P_0\simeq P_1$ in the proof of the preceding theorem and $C=0\Rightarrow C^*=0$, that is to say $N\simeq N'$. This is the very specific situation only considered by OD control theory 
where the OD equations defining the control systems are always supposed to be differentially 
independent (linearly independent over $D$).\\

 \noindent
 {\bf Example 2.3.12}: Revisiting the introductory example 1.6, we discover that, the only solution of the given system being $y=0$, the differential modules 
 defined by the systems $(A=0, B=0)$ or $(C=0)$ are isomorphic to $M=Dy$ and we have the following commutative and exact diagram of operators:
 \[   \begin{array}{rcccccccccl}
 & & 0 & & 0 & & 0 & & &   \\
 & & \downarrow & & \downarrow & & \downarrow && & & \\
 0 & \rightarrow & y & \underset 2 {\stackrel{{\cal{D}}}{ \rightarrow}} & (u,v) & \underset 2{\stackrel{{\cal{D}}'_1}{\rightarrow }}& C & \rightarrow & 0 & &  \\
  & & \parallel & & \parallel & & \hspace{3mm} \downarrow 2 & & \downarrow & &  \\
 0 & \rightarrow & y & \underset 2{\stackrel{{\cal{D}}}{ \rightarrow}}&(u,v)&\underset 4{\stackrel{{\cal{D}}_1}{\rightarrow }}&(A, B)&\rightarrow & w & \rightarrow &0 \\ 
  & & \downarrow & & \downarrow & & \hspace{3mm} \downarrow 2 & & \parallel & &  \\
& & 0 & &0 &  \rightarrow & w & =  & w &\rightarrow & 0  \\
& & & & & & \downarrow & & \downarrow & &  \\
& & & & & & 0 & & 0 & &  
 \end{array}  \]
 Translating this result in the language of differential modules, we obtain the commutative and exact diagram showing that $C \simeq D$:  \\
\[ \begin{array}{rcccccccccl}
 & & 0 & & 0 & &  & & & & \\
 & & \downarrow & & \downarrow & & & & & & \\
0 & \rightarrow & C & = & D & \rightarrow & 0 &  & 0 & & \\
 & & \parallel & & \downarrow & & \downarrow & & \downarrow & & \\
0 & \rightarrow & D & \rightarrow & D^2 &\stackrel{d_1}{\rightarrow} & D^2& \rightarrow & D & \rightarrow & 0 \\ 
& & \downarrow & & \downarrow & & \downarrow & & \parallel & & \\
 & &0 &\rightarrow & D &\stackrel{d'_1}{\rightarrow} & D^2 & \rightarrow & D & \rightarrow & 0 \\
 & &  & & \downarrow & & \downarrow  & & \downarrow & & \\
& &  & & 0 & & 0 & & 0 & &
\end{array}  \]
Applying $hom_D(\bullet, D)$ we obtain the commutative and exact diagram:  
\[ \begin{array}{lllllllllll}
 & & 0 & & 0 & & 0 & & & & \\
 & & \uparrow & & \uparrow & & \uparrow & & & & \\
0 & \leftarrow & D & = & D & \leftarrow & 0 & & 0 & & \\
 & & \uparrow & & \uparrow & & \uparrow & & \uparrow & & \\
0 & \leftarrow & N & \leftarrow & D^2 &\stackrel{d_1^*}{\leftarrow} & D^2 & \leftarrow & D & \leftarrow & 0 \\ 
 & & \uparrow & & \uparrow & & \uparrow & & \parallel & & \\
0 & \leftarrow & N'& \leftarrow & D & \stackrel{{d'_1}^*}{\leftarrow} & D^2 & \leftarrow & D & \leftarrow & 0 \\
 & & \uparrow & & \uparrow & & \uparrow & & \uparrow & & \\
 & & 0 & & 0 & & 0 & & 0 & &
\end{array}  \]
We obtain $N \simeq N' \oplus D $ and ${ext}^1(N) = {ext}^1(N') =0$ because $ad({\cal{D}}'_1) $ is an injective operator with $N' = 0$, exactly like 
${\cal{D}}$ is an injective operator, and the bottom horizontal sequence splits.  \\

We are now ready for exhibiting the final desired link with operator theory. \\

\noindent
{\bf Theorem 2.3.13}: One has \fbox{ $ad(\overline{\cal{D}}) = \overline{ad(\cal{D})}$} for any change $\bar{x}=\varphi(x)$ of independent variables.   \\

\noindent
{\it Proof}: As the proof is rather technical, we shall divide it into three steps:\\
\noindent
{Step 1}: We start providing the tricky computation for a change ${\bar{x}}^j= {\varphi}^j(x)$ on any $\xi = {\xi}^i(x){\partial}_i \in T$. Dealing with operators and no longer with vector fields, we may set $\xi = {\xi}^i d_i \in D$, writing ${\bar{x}}^j = {\partial}_i {\varphi}^j {\xi}^i = \frac{\partial {\bar{x}}^j}{\partial x^i} {\xi}^i$ in order to keep the duality existing between $x$ and $\bar{x}$. Using crucially Lemma ...  with now $\Delta = det( {\partial}_i {\varphi}^j)$, we obtain successively in the framework of operators:  \\
\[   \xi = {\xi}^i d_i \in D  \Rightarrow  ad(\xi) = - d_i {\xi}^i = - \xi - {\partial}_i {\xi}^i \in D  \]
\[ ad(\bar{\xi}) = - \frac{d}{d {\bar{x}}^j} {\bar{\xi}}^j= - \frac{d}{d {\bar{x}}^j} \frac{ \partial {\bar{x}}^j}{\partial x^i} {\xi}^i =
 - \xi - {\partial}_i{\xi}^i - {\xi}^i \frac{\partial }{\partial {\bar{x}}^j} ((\frac{1}{\Delta} \frac{\partial {\bar{x}}^j}{\partial x^i} )\Delta) = 
 - \xi - {\partial}_i {\xi}^i - \frac{1}{\Delta} {\xi}^i {\partial}_i \Delta  \]
and thus $ \Delta ad(\bar{\xi}) = - {\Delta} \xi  \Delta - {\partial}_i (\Delta \xi) = - d_i {\xi}^i \Delta = ad(\xi) \Delta$, that is \fbox{$ ad(\bar{\xi}) = \overline{ad(\xi)}$}.  \\
   
\noindent
{Step 2}: As any operator $P \in D$ can be written as $P = {\xi}_1 ... {\xi}_r$ with ${\xi}_1, ..., {\xi}_r \in D$, we obtain from the first step:
\[  \Delta ad(\bar{P}) = \Delta ad({\bar{\xi}}_1... {\bar{\xi}}_r)= \Delta ad({\bar{\xi}}_r) ... ad({\bar{\xi}}_1= ad({\xi}_r) \Delta ad({\bar{\xi}}_{r-1}) ... ad({\bar{\xi}}_1) =
     ad({\xi}_r) ... ad({\xi}_1) \Delta  \]
and thus the formula $ \Delta ad(\bar{P}) = ad(P) \Delta $, that is \fbox{ $ ad(\bar{P}) = \overline {ad(P)} $ }.  \\

\noindent
{Step 3}: With $\cal{D}= \xi \rightarrow \eta$ and $ad(\cal{D}): \mu \rightarrow \nu$, using an integration by parts with contraction $ <   >   $, we get:  \\
\[   < \mu, {\cal{D}} \xi > - < ad({\cal{D}}) \mu, \xi > = \frac{\partial}{\partial x^i} ( )^i \]
As any contraction is a $n$-form, we obtain in the new coordinate system:  \\
\[  < \bar{\mu}, \overline{\cal{D}}\bar{\xi} > = \frac{1}{\Delta} < \mu, {\cal{D}} \xi > , \hspace{2cm} < \overline{ad({\cal{D}})} \bar{\mu} , \bar{\xi} > = 
\frac{1}{\Delta} < ad ({\cal{D}}) \mu , \xi >   \]
and thus:  \\
\[  < \bar{\mu} , \overline{\cal{D}} \bar{\xi} > - < \overline{ad(\cal{D})} \bar{\mu}, \bar{\xi} > = \frac{1}{\Delta} \frac{\partial}{\partial x^i} ( )^i =
 \frac{1}{\Delta} \frac{\partial {\bar{x}}^j}{\partial x^i} \frac{\partial}{\partial {\bar{x}}^j}( )^i = 
 \frac{\partial }{\partial {\bar{x}}^j} (\frac{1}{\Delta} \frac{\partial {\bar{x}}^j }{\partial x^i} ( )^i ) \]
and thus $ad(\overline{\cal{D}}) = \overline{ad(\cal{D})}$ as the adjoint of an operator is uniquely defined by such an identity.  \\
\hspace*{12cm}  $\Box$  \\

\noindent
{\bf Corollary 2.3.14}: To any linear differential operator $E \underset q{\stackrel{\cal{D}}{\longrightarrow}} F $ of order $q$ we may associate another linear 
differential operator ${\wedge}^nT^*\otimes E^* \underset q{\stackrel{ad(\cal{D})}{\longleftarrow}} {\wedge}^nT^* \otimes F^*$ of order $q$, in such a way that $ad(ad(\cal{D})) = \cal{D}$ but it is important to notice that its arrow is now going {\it backwards}, that is {\it from right to left}. We shall use to set 
$ad(E) = {\wedge}^n T^* \otimes  E^*$ in order to simplify the notations for applications while keeping the same dimension.  \\

\noindent
{\bf Important Remark 2.3.15}: In actual practice and in the operator framework, we may consider an operator matrix acting on the left of column vectors (sections of vector bundles). \\
Similarly, in the framework of left $D$-modules, we may use now row vectors and write:   \\
\[    D {\otimes }_A F^* \stackrel{{\cal{D}}}{\longrightarrow} D {\otimes}_A E^* \stackrel{p}{\rightarrow}  M  \rightarrow 0 \]
with ${\cal{D}}$ acting now by composition on the right of row vectors while $D$ is acting on the left by usual composition of operators. We shall set 
$ D(E) = D {\otimes}_A E^*$ with $E^* = hom_A(E, A)$ and obtain therefore $hom_D(D(E), A)= E$. Applying $ hom_D(\bullet, D)$ and 
using right $D$-modules or using the {\it side changing functor}  $ hom_A( {\wedge}^nT^* , \bullet )$ and using left $D$-modules, we get:. \\
\[    D(F)  \rightarrow D(E) \rightarrow M  \rightarrow 0    \]
In the dual situation, we shall obtain:  \\
\[     0 \leftarrow N \leftarrow  D {\otimes}_A {\wedge}^n T {\otimes}_A F \stackrel{  ad({\cal{D}})}{\longleftarrow }  D {\otimes}_A {\wedge}^n T {\otimes}_A E  \]
in order to keep on going with left differential modules. Such a difficulty is explaining why adjoint operators have {\it never} been used in mathematical physics up to our knowledge.  \\
We point out another difficulty existing because, in general, $ad({\cal{D}})$ is far from being involutive or even formally integrable whenever ${\cal{D}}$ is involutive. This is particularly true even for OD systems like the Kalman systems or the double pendulum as we saw. For this reason, we shall rather suppose that the coefficients of the operators or systems are in a differential field $K$ rather than in a ring $A$. In a word, one has to get used to a new language.   \\   \\

\noindent
{\bf 3) GENERAL RELATIVITY}  \\

From standard results in continuum mechanics and the preceding formulas, we have:  \\

\noindent
{\bf Proposition 3.1}: The Cauchy operator is the adjoint of the Killing operator.  \\

{\it Proof}: Let $X$ be a manifold of dimension $n$ with local coordinates $(x^1, ... , x^n)$, tangent bundle $T$ and cotangent bundle $T^*$. If $\omega \in S_2 T^*$ is a metric with $det(\omega)\neq 0$, we my introduce the standard Lie derivative in order to define the first order Killing operator:    \\
\[       {\cal{D}}: \xi \in T \rightarrow \Omega = \xi \rightarrow {\cal{L}}(\xi) \omega=\Omega\equiv ({\Omega}_{ij}= {\omega}_{rj}(x) {\partial}_i{\xi}^r + {\omega}_{ir}(x){\partial}_j{\xi}^r + 
{\xi}^r {\partial}_r {\omega}_{ij}(x) ) \in S_2T^*  \]
Here start the problems because, in our opinion at least, a systematic use of the adjoint operator has never been used in mathematical physics and even in continuum mechanics apart through a variational procedure. As we have seen, the purely intrinsic definition of the adjoint can only be done in the theory of differential modules by means of the so-called {\it side changing functor}. From a purely differential geometric point of view, the idea is to associate to any vector bundle $E$ over $X$ a new vector bundle $ad(E)= {\wedge}^nT^* \otimes E^*$ where $E^*$ is obtained from $E$ by patching local coordinates while inverting the transition matrices, exactly like $T^*$ is obtained from $T$. It follows that the stress $\sigma =({\sigma}^{ij}) \in ad(S_2T^*) =  {\wedge}^n T^* \otimes S_2 T$ is {\it not} a tensor but a tensor density, that is transforms like a tensor up to a certain power of the Jacobian matrix. When $n=4$,  the fact that such an object is called stress-energy tensor does not change anything as it cannot be related to the Einstein tensor which is a true {\it tensor} indeed. In any case, we may define as usual:  \\
\[  ad({\cal{D}}): {\wedge}^n T^* \otimes S_2T \rightarrow {\wedge}^n T^* \otimes T : \sigma \rightarrow \varphi \]
Multiplying ${\Omega}_{ij}$ by ${\sigma}^{ij}$ and integrating by parts, the factor of 
$ - 2 \, {\omega}_{kr}\,  {\xi}^r$ is easly seen to be:  \\
\[  \fbox{  $  {\nabla}_i {\sigma}^{ik} = {\partial}_i {\sigma}^{ik} +  {\gamma}^k_{ij} {\sigma}^{ij} = {\varphi}^k  $  } \] 
with well known Christoffel symbols 
$ {\gamma}^k_{ij} = \frac{1}{2} {\omega}^{kr} ({\partial}_i {\omega}_{rj} + {\partial}_j {\omega}_{ir} - 
{\partial}_r {\omega}_{ij}) $. \\
However, if the stress should be a tensor, we should get for the covariant derivative:  \\
\[ {\nabla}_r {\sigma}^{ij}= {\partial}_r{\sigma}^{ij} + {\gamma}^i_{rs}{\sigma}^{sj}+ {\gamma}^j_{rs} {\sigma}^{is} \Rightarrow 
{\nabla}_i {\sigma}^{ik} = {\partial}_i {\sigma}^{ik} + {\gamma}^r_{ri} {\sigma}^{ik} + 
{\gamma}^k_{ij} {\sigma}^{ij}    \]
The difficulty is to prove that we do not have a contradiction because $\sigma$ is a tensor density.  \\
If we have an invertible transformation like in Lemma $2.C.6$, we have successively:  \\
\[ {\tau}^{kl}(f(x))= \frac{1}{\Delta} {\partial}_i f^k(x) {\partial}_j f^l(x) {\sigma}^{ij}(x) \]
\[  \frac{\partial {\tau}^{kl}}{\partial y^k}= \frac{1}{\Delta}{\partial}_i f^k  \frac{\partial}{\partial y^k}({\partial}_j f^l) {\sigma}^{ij}+ 
\frac{1}{\Delta}{\partial}_i f^k {\partial}_j f^l \frac{\partial}{\partial y^k} {\sigma}^{ij}  
 = \frac{1}{\Delta} ({\partial}_{ij} f^l) {\sigma}^{ij}  + \frac{1}{\Delta} {\partial}_j f^l {\partial}_i {\sigma}^{ij}   \]
Now, we recall the transformation law of the Christoffel symbols, namely:   \\
\[    {\partial}_r f^u(x) {\gamma}^r_{ij} (x) =  {\partial}_{ij} f^u(x)  + {\partial}_i f^k(x) {\partial}_j f^l(x)  {\bar{\gamma}}^u_{kl} (f(x)) 
\Rightarrow   \frac{1}{\Delta}{\partial}_r f^u {\gamma}^r_{ij} {\sigma}^{ij}= \frac{1}{\Delta} {\partial}_{ij} f^u {\sigma}^{ij} + {\bar{\gamma}}^u_{kl}(y) {\tau}^{kl}   \]
Eliminating the second derivatives of $f$ we finally get:  \\
\[   {\psi}^u= \frac{\partial {\tau}^{ku}}{\partial y^k} + {\bar{ \gamma}}^u_{kl} {\tau}^{kl}= \frac{1}{\Delta} {\partial}_r f^u({\partial}_i {\sigma}^{ir} + {\gamma}^r_{ij} {\sigma}^{ij}) = \frac{1}{\Delta} {\partial}_r f^u {\varphi}^r  \]      
This tricky technical result, {\it which is not evident at all}, explains why the additional term we had is just disappearing in fact when $\sigma$ is a density. \\
One can prove, in a similar but even simpler fashion, that the two sets of Maxwell equations are invariant under any invertible transformation and that that the conformal group of spacetime is only the group of invariance of the Minkowski constitutive laws in vacuum (Pommaret, 2023 b).  \\
\hspace*{12cm}  $   \Box   $   \\

Linearizing the {\it Ricci} tensor ${\rho}_{ij}$ over the Minkowski metric $\omega$, we obtain the usual second order homogeneous {\it Ricci} operator $\Omega \rightarrow R$ with $4$ terms (This result can be found in {\it any} textbook on general relativity but (Pommaret, 2013) is a fine reference using the same notations):  \\
\[  2 R_{ij}= {\omega}^{rs}(d_{rs}{\Omega}_{ij}+ d_{ij}{\Omega}_{rs} - d_{ri}{\Omega}_{sj} - d_{sj}{\Omega}_{ri})= 2R_{ji}  \]
\[ tr(R)= {\omega}^{ij}R_{ij}={\omega}^{ij}d_{ij}tr(\Omega)-{\omega}^{ru}{\omega}^{sv}d_{rs}{\Omega}_{uv}  \]
We may define the $Einstein$ operator by setting $E_{ij}=R_{ij} - \frac{1}{2} {\omega}_{ij}tr(R)$ and obtain the $6$ terms:  \\
\[ 2E_{ij}= {\omega}^{rs}(d_{rs}{\Omega}_{ij} + d_{ij}{\Omega}_{rs} - d_{ri}{\Omega}_{sj} - d_{sj}{\Omega}_{ri})
- {\omega}_{ij}({\omega}^{rs}{\omega}^{uv}d_{rs}{\Omega}_{uv} - {\omega}^{ru}{\omega}^{sv}d_{rs}{\Omega}_{uv})  \]
We have the (locally exact) differential sequence of operators acting on sections of vector bundles where the order of an operator is written under its arrow:  \\
\[  \fbox{  $  \begin{array}{ccccccc}
T& \underset 1{\stackrel{Killing}{\longrightarrow}} & S_2T^* & \underset 2{\stackrel{Riemann}{\longrightarrow}} &  F_1 & \underset 1{\stackrel{Bianchi}{\longrightarrow}}  & F_2  \\
 n & \stackrel{{\cal{D}}}{ \longrightarrow} & n(n+1)/2 & \stackrel{{\cal{D}}_1}{\longrightarrow} & n^2(n^2-1)/12 & \stackrel{{\cal{D}}_2}{\longrightarrow} & 
 n^2(n^2-1)(n-2)/24  
\end{array}   $   }   \]
Our purpose is now first to study the differential sequence onto which its right part is projecting:  \\
\[   \fbox{  $ \begin{array}{ccccccc}
   S_2T^* & \underset 2 {\stackrel{Einstein}{\longrightarrow}} & S_2T^* & \underset 1{\stackrel{div}{\longrightarrow}} & T^* &  \rightarrow  & 0  \\
 n(n+1)/2 & \longrightarrow  & n(n+1)/2 &  \longrightarrow &  n & \rightarrow & 0  
 \end{array}   $  }  \]
 and then the following adjoint sequence where we have set (Pommaret, 2021 a) :    \\
\[ \fbox{  $  0  \longleftarrow  ad(T) \stackrel{Cauchy}{\longleftarrow} ad(S_2T^*)  \stackrel{Beltrami}{\longleftarrow} ad(F_1) \stackrel{Lanczos}{\longleftarrow} ad(F_2)   $  }   \]
In this sequence, if $E$ is a vector bundle over the ground manifold $X$ with dimension $n$, we may introduce, as we already said, the new vector bundle $ad(E)={\wedge}^nT^* \otimes E^*$ where $E^*$ is obtained from $E$ 
by inverting the transition rules exactly like $T^*$ is obtained from $T$. We have for example $ad(T)={\wedge}^nT^*\otimes T^*\simeq {\wedge}^nT^*\otimes T \simeq {\wedge}^{n-1}T^*$ because $T^*$ is isomorphic to $T$ by using the metric $\omega$.
The $10 \times 10 $ $ Einstein$ operator matrix is induced from the $10 \times 20$ $Riemann$ operator matrix and the $10 \times 4$ $div$ operator matrix is induced from the $20 \times 20$ $Bianchi$ operator matrix. We advise the reader not familiar with the formal theory of systems or operators to follow the computation in dimension $n=2$ with the $1 \times 3$ $Airy$ operator matrix, which is the formal adjoint of the $3 \times 1$ $Riemann$ operator matrix, and $n=3$ with the $6 \times 6$ $Beltrami$ operator matrix which is the formal adjoint of the $6 \times 6$ $Riemann$ operator matrix  which is easily seen to be self-adjoint up to a change of basis.\\ 
With more details, we have:  \\

\noindent
$\bullet \hspace{3mm}n=2$: The stress equations become $ d_1{\sigma}^{11}+ d_2{\sigma}^{12}=0, d_1{\sigma}^{21}+ d_2{\sigma}^{22}=0$. Their second order parametrization ${\sigma}^{11}= d_{22}\phi, {\sigma}^{12}={\sigma}^{21}= - d_{12}\phi, {\sigma}^{22}= d_{11}\phi$ has been provided by George Biddell Airy in 1863 
and is well known in plane elasticity. We get the second order system:  \\
\[ \left\{  \begin{array}{rll}
{\sigma}^{11} & \equiv d_{22}\phi =0 \\
-{\sigma}^{12} & \equiv d_{12}\phi =0 \\
{\sigma}^{22} & \equiv d_{11}\phi=0
\end{array}
\right. \fbox{ $ \begin{array}{ll}
1 & 2   \\
1 & \bullet \\  
1 & \bullet  
\end{array} $ } \]
which is involutive with one equation of class $2$, $2$ equations of class $1$ and it is easy to check that the $2$ corresponding first order CC are just the $Cauchy$  equations. Of course, the $Airy$ function ($1$ term) has absolutely nothing to do with the perturbation of the metric ($3$ terms). With more details, when $\omega$ is the Euclidean metric, we may consider the only component:   \\
\[  \begin{array}{rcl}
tr(R)  &  = &  (d_{11} + d_{22})({\Omega}_{11} + {\Omega}_{22}) - (d_{11}{\Omega}_{11} + 2 d_{12}{\Omega}_{12}+ d_{22}{\Omega}_{22})  \\
      &  =  &  d_{22}{\Omega}_{11} + d_{11}{\Omega}_{22} - 2 d_{12}{\Omega}_{12}
      \end{array}  \]
Multiplying by the Airy function $\phi$ and integrating by parts, we discover that:   \\
      \[  \fbox{ $  Airy=ad(Riemann) \,\,\,\,  \Leftrightarrow  \,\,\,\,  Riemann = ad(Airy)  $ } \]  \\
in the following adjoint differential sequences:
 \[ \fbox{  $ \begin{array}{rcccccccl}
    &  &  2  &  \underset 1 {\stackrel{Killing}{\longrightarrow}} & 3 &  \underset 2 {\stackrel{Riemann}{\longrightarrow}} &1 & \longrightarrow  &   0  \\
     &  &  &  &  &  &  &  &  \\
  0  & \longleftarrow   & 2 & \underset 1 {\stackrel{Cauchy}{\longleftarrow}}  &3  & \underset 2 {\stackrel{Airy}{\longleftarrow}}  & 1 & &   
      \end{array}  $  }  \]
      
\noindent
$\bullet \hspace{3mm} n=3$: It is quite more delicate to parametrize the $3$ PD equations: \\
\[ d_1{\sigma}^{11}+ d_2{\sigma}^{12}+ d_3{\sigma}^{13}=0,\hspace{3mm} d_1{\sigma}^{21}+ d_2{\sigma}^{22}+ d_3{\sigma}^{23}=0, \hspace{3mm} d_1{\sigma}^{31}+ d_2{\sigma}^{32}+ d_3{\sigma}^{33}=0 \]
As we explained in (Pommaret, 2017, 2021 b), a direct computational approach has been provided by Eugenio Beltrami in 1892, James Clerk Maxwell in 1870 and Giacinto Morera in 1892  by introducing the $6$ {\it stress functions} ${\phi}_{ij}={\phi}_{ji}$ in the {\it Beltrami parametrization}. The corresponding system:\\
\[   \left\{  \begin{array}{rll}
{\sigma}^{11} \equiv & d_{33}{\phi}_{22}+ d_{22}{\phi}_{33}-2 d_{23}{\phi}_{23}=0  \\
-{\sigma}^{12}\equiv & d_{33}{\phi}_{12}+ d_{12}{\phi}_{33}- d_{13}{\phi}_{23}- d_{23}{\phi}_{13}=0  \\
 {\sigma}^{22}\equiv & d_{33}{\phi}_{11}+ d_{11}{\phi}_{33}-2 d_{13}{\phi}_{13}=0  \\
{\sigma}^{13}\equiv & d_{23}{\phi}_{12}+ d_{12}{\phi}_{23}- d_{22}{\phi}_{13}- d_{13}{\phi}_{22} =0 \\
-{\sigma}^{23}\equiv & d_{23}{\phi}_{11}+ d_{11}{\phi}_{23}- d_{12}{\phi}_{13}- d_{13}{\phi}_{12} =0 \\
{\sigma}^{33}\equiv & d_{22}{\phi}_{11}+ d_{11}{\phi}_{22}-2 d_{12}{\phi}_{12}=0
\end{array}
\right. \fbox{ $ \begin{array}{lll}
1 & 2 & 3   \\
1 & 2 & 3  \\
1 & 2 & 3  \\
1 & 2 &  \bullet  \\
1 & 2 & \bullet  \\
1 & 2 & \bullet
\end{array} $ } \]
is involutive with $3$ equations of class $3$, $3$ equations of class $2$ and no equation of class $1$. We have $dim(g_2) = dim(S_2T^*\otimes S_2T^*) - dim(S_2T^*) = (6\times 6) - 6= 30$. The $3$ CC are describing the stress equations which admit therefore a parametrization ... but without any geometric framework, in particular without any possibility to imagine that the above second order operator is {\it nothing else but} the {\it formal adjoint} of the {\it Riemann operator}, namely the (linearized) Riemann tensor with $n^2(n^2-1)/2=6$ independent components when $n=3$ (Pommaret, 2023 b). \\
Breaking the canonical form of the six equations which is associated with the Janet tabular, we may rewrite the Beltrami parametrization of the Cauchy stress equations as follows, after exchanging the third row with the fourth row, keeping the ordering $\{(11)<(12)<(13)<(22)<(23)<(33)\}$:  \\
\[      \left(  \begin{array}{cccccc}
d_1& d_2 & d_3 &0 & 0 & 0 \\
 0 & d_1 &  0 & d_2 & d_3 & 0 \\
 0 & 0 & d_1 & 0 & d_2 & d_3 
\end{array}  \right)  
 \left(  \begin{array}{cccccc}
 0 & 0 & 0 & d_{33} & - 2d_{23} & d_{22} \\
 0 & - d_{33} & d_{23} & 0 & d_{13} & - d_{12}  \\
 0 & d_{23} & - d_{22} & - d_{13} & d_{12} & 0 \\
 d_{33}& 0 & - 2 d_{13} & 0 & 0 & d_{11}  \\
 - d_{23} & d_{13} & d_{12}& 0 & - d_{11} & 0 \\
 d_{22} & - 2 d_{12} & 0 & d_{11}& 0 & 0 
 \end{array} \right)  \equiv   0    \]
 as an identity where $0$ on the right denotes the zero operator. However, if  $\Omega$ is a perturbation of the metric $\omega$, the standard implicit summation used in continuum mechanics is, when $n=3$:  \\
 \[   \begin{array}{rcl}
{\sigma}^{ij}{\Omega}_{ij} & = & {\sigma}^{11}{\Omega}_{11} + 2 {\sigma}^{12}{\Omega}_{12} + 2 {\sigma}^{13}{\Omega}_{13} + {\sigma}^{22} {\Omega}_{22} + 2{\sigma}^{23}{\Omega}_{23} + {\sigma}^{33}{\Omega}_{33}  \\
   &  =  & {\Omega}_{22}d_{33}{\phi}_{11}+ {\Omega}_{33}d_{22}{\phi}_{11}- 2 {\Omega}_{23}d_{23}{\phi}_{11}+ ... \\
   &    & + {\Omega}_{23}d_{13}{\phi}_{12}+{\Omega}_{13}d_{23}{\phi}_{12}- {\Omega}_{12}d_{33}{\phi}_{12}- {\Omega}_{33}d_{12}{\phi}_{12} + ...
\end{array}  \]
because {\it the stress tensor density $\sigma$ is supposed to be symmetric}. Integrating by parts in order to construct the adjoint operator, we get:  \\
\[ \begin{array}{rcl}
 {\phi}_{11} &  \longrightarrow  &  d_{33}{\Omega}_{22} + d_{22}{\Omega}_{33} - 2 d_{23}{\Omega}_{23} \\
 {\phi}_{12} &  \longrightarrow   &  d_{13}{\Omega}_{23}+d_{23}{\Omega}_{13}-d_{33}{\Omega}_{12} - d_{12}{\Omega}_{33}
 \end{array}  \]
and so on. The identifications $ Beltrami = ad(Riemann), \,\, Lanczos = ad(Bianchi)$ in the diagram:  \\
   \[ \fbox{  $  \begin{array}{rcccccccl}  
   & 3  & \underset 1 {\stackrel{Killing}{\longrightarrow}} & 6  & \underset 2 {\stackrel{Riemann}{\longrightarrow}}& 6  & \underset 1 {\stackrel{Bianchi}{\longrightarrow }}  & 3 & \longrightarrow 0  \\
  0 \longleftarrow  & 3  & \underset 1 {\stackrel{Cauchy}{\longleftarrow}} & 6   & \underset 2 {\stackrel{Beltrami}{\longleftarrow}} 
  & 6 &  \underset 1 {\stackrel{Lanczos}{\longleftarrow}} & 3 &    
\end{array}. $   } \]
prove that the $Cauchy$ operator has nothing to do with the $Bianchi$ operator along with (Pommaret, 2017, 2021 b,2023 b). \\
When $\omega$ is the Euclidean metric, the link between the two sequences is established by means of the elastic constitutive relations $2 {\sigma}_{ij} = \lambda tr(\Omega) {\omega}_{ij} + 2 \mu {\Omega}_{ij}$ with the Lam\'{e} elastic constants $(\lambda, \mu) $ but mechanicians are usually setting $ {\Omega}_{ij} = 2 {\epsilon}_{ij} $. Using the standard Helmholtz decomposition ${\vec{\xi}} = {\vec{\nabla} }\varphi + {\vec{\nabla}} \wedge {\vec{\psi}} $ and substituting in the dynamical equation $ d_i {\sigma}^{ij} =  \rho d^2/dt^2 {\xi}^j $ where $\rho$ is the mass per unit volume, we get the longitudinal and transverse wave equations, namely $ \Delta \varphi - \frac{\rho}{\lambda + 2 \mu} \frac{d^2}{dt^2} \varphi = 0 $ and 
$ \Delta {\vec{\psi}} - \frac{\rho}{\mu} \frac{d^2}{dt^2} {\vec{\psi}} = 0 $, responsible for earthquakes !. \\

Then, taking into account the factor $2$ involved by multiplying the second, third and fifth row by $2$, we get the new $6\times 6$ operator matrix with rank $3$ which is clearly self-adjoint:  \\
\[ \fbox{  $   \left(  \begin{array}{cccccc}
 0 & 0 & 0 & d_{33} & - 2d_{23} & d_{22} \\
 0 & - 2d_{33} & 2d_{23} & 0 & 2d_{13} & - 2d_{12}  \\
 0 & 2d_{23} & - 2d_{22} & - 2d_{13} & 2d_{12} & 0 \\
 d_{33}& 0 & - 2 d_{13} & 0 & 0 & d_{11}  \\
 - 2d_{23} & 2d_{13} & 2d_{12}& 0 & - 2d_{11} & 0 \\
 d_{22} & - 2 d_{12} & 0 & d_{11}& 0 & 0 
 \end{array} \right)  \left( \begin{array}{c}
 {\Omega}_{11}  \\
 {\Omega}_{12}  \\
 {\Omega}_{13}  \\
 {\Omega}_{22}  \\
 {\Omega}_{23}  \\
 {\Omega}_{33}
 \end{array}  \right) =  - 1 \left( \begin{array}{r}
       E_{11}   \\
 2 \, E_{12}  \\
 2 \, E_{13}   \\
       E_{22}  \\
 2 \, E_{23}  \\
       E_{33}
 \end{array} \right)  $   }   \]

{\it Surprisingly}, the Maxwell parametrization is obtained by keeping ${\phi}_{11}=A, {\phi}_{22}=B, {\phi}_{33}=C$ while setting ${\phi}_{12}={\phi}_{23}={\phi}_{31}=0$ in order to obtain the system:\\
\[   \left\{  \begin{array}{rl}
{\sigma}^{11} \equiv& d_{33}B + d_{22}C=0  \\
{\sigma}^{22}\equiv & d_{33}A+ d_{11}C =0 \\
- {\sigma}^{23}\equiv & d_{23}A=0  \\
{\sigma}^{33}\equiv & d_{22}A+ d_{11}B=0  \\
- {\sigma}^{13}\equiv & d_{13}B=0 \\
- {\sigma}^{12}\equiv & d_{12}C=0
\end{array}
\right. \fbox{ $ \begin{array}{lll}
1 & 2 & 3   \\
1 & 2 & 3  \\
1 & 2 & \bullet  \\
1 & 2 &  \bullet  \\
1 & \bullet & \bullet  \\
1 & \bullet & \bullet
\end{array} $ } \]
{\it This system may not be involutive} and no CC can be found "{\it a priori} " because the coordinate system is surely not $\delta$-regular. Effecting the linear change of coordinates 
${\bar{x}}^1 = x^1, {\bar{x}}^2 = x^2, {\bar{x}}^3 = x^3 + x^2 + x^1 $ and taking out the bar for simplicity, we obtain the homogeneous involutive system with a quite tricky Pommaret basis:  \\
\[   \left\{  \begin{array}{l}
d_{33}C+ d_{13}C+ d_{23}C+ d_{12}C=0  \\
d_{33}B+ d_{13}B=0  \\
d_{33}A+ d_{23}A=0  \\
d_{23}C +d_{22}C - d_{13}C - d_{13}B - d_{12}C =0  \\
d_{23}A - d_{22}C + d_{13}B + 2 d_{12}C - d_{11}C=0  \\
d_{22}A + d_{22}C - 2 d_{12}C + d_{11}C + d_{11}B=0
\end{array} \right. 
\fbox{ $ \begin{array}{lll}
1 & 2 & 3   \\
1 & 2 & 3  \\
1 & 2 &  3  \\
1 & 2 &  \bullet  \\
1 &  2 & \bullet  \\
1 &  2 & \bullet
\end{array} $ } \]
It is easy to check that the $3$ CC obtained just amount to the desired $3$ stress equations when coming back to the original system of coordinates.
{\it We have thus a minimum parametrization}. \\

Again, {\it if there is a geometrical background, this change of local coordinates is hiding it totally}. The following new minimum involutive parametrization does not seem to be known or even used:  \\
\[   \left\{  \begin{array}{rll}
{\sigma}^{11} \equiv & d_{33}{\phi}_{22} = 0  \\
-{\sigma}^{12}\equiv & d_{33}{\phi}_{12} = 0  \\
 {\sigma}^{22}\equiv & d_{33}{\phi}_{11} = 0  \\
{\sigma}^{13}\equiv & d_{23}{\phi}_{12} - d_{13}{\phi}_{22} =0 \\
-{\sigma}^{23}\equiv & d_{23}{\phi}_{11} - d_{13}{\phi}_{12} =0 \\
{\sigma}^{33}\equiv & d_{22}{\phi}_{11}+ d_{11}{\phi}_{22}-2 d_{12}{\phi}_{12} = 0
\end{array}
\right. \fbox{ $ \begin{array}{lll}
1 & 2 & 3   \\
1 & 2 & 3  \\
1 & 2 & 3  \\
1 & 2 &  \bullet  \\
1 & 2 & \bullet  \\
1 & 2 & \bullet
\end{array} $ } \]

When $n=4$, the following crucial theorem is showing that the Einstein operator is useless contrary to the classical GR literature (Pommaret, 2024 a, 2024 b).  \\

\noindent
{\bf Theorem  3.2}: {\it The GW equations are deined by the adjoint of the Ricco operator which is not self-adjoint contrary to the Einstein operator 
which is self-adjoint}.  \\

\noindent
{\it Proof}: Multiplying the $Ricci$ operator by the Lagrange multipliers ${\lambda}^{ij}={\lambda}^{ji}$ used as test functions, setting $\Box = {\omega}^{rs} d_{rs}$ and integrating by parts, we get the adjoint operator $ad(Ricci): \lambda \rightarrow \sigma$:  \\
\[   \Box {\lambda}^{rs} + {\omega}^{rs} d_{ij} {\lambda}^{ij}  - {\omega}^{sj} d_{ij} {\lambda}^{ri}  - {\omega}^{ri} d_{ij} {\lambda}^{sj} = {\sigma}^{rs}  \]
that is, {\it exactly} but {\it backwards}, the operator defining GW in the literature [32]. We also obtain:   \\
\[ d_r {\sigma}^{rs} = {\omega}^{ij} d_{rij} {\lambda}^{rs}  + {\omega}^{rs} d_{rij} {\lambda}^{ij} - {\omega}^{sj} d_{rij} {\lambda}^{ri} - {\omega}^{ri} d_{rij} {\lambda}^{sj} = 0 \]
and finally the commutative diagram coherent with double differential duality: \\                                                                       
 \[ \fbox{  $ \begin{array}{rcccccccccl}
    &  &  4  &  \underset 1 {\stackrel{Killing}{\longrightarrow}} & 10 &  \underset 2 {\stackrel{Ricci}{\longrightarrow}} &10 & \longrightarrow  &   4 & \rightarrow & 0 \\
     &  &  &  &  &  &  &  &  \\
  0  & \longleftarrow   & 4 & \underset 1 {\stackrel{Cauchy}{\longleftarrow}}  &10 & \underset 2 {\stackrel{ad(Ricci)}{\longleftarrow}}  & 10 & &   &  &  
      \end{array}  $  }  \]
It follows that GW cannot exist as they cannot be considered as {\it ripples} of space-time because $\lambda$ has nothing to do with the deformation $\Omega$ of the metric $\omega$. \\
\hspace*{12cm}   $\Box$  \\

We finally prove that this result only depends on the second order jets of the conformal group of transformations of space-time, {\it a result highly not evident at first sight for sure and not known}. We need a few steps in order to show that:   \\
 {\it The mathematical foundations of conformal geometry must be entirely revisited}.   \\ \\  \\  \\

\noindent
{\bf 4) CONFORMAL GROUP}    \\

We start proving that the structure of the conformal with $(n+1)(n+2)/2$ parameters may not be related to a classification of Lie algebras ( Pommaret, 2021 c). \\
For $n=1$, the simplest such group of transformations of the real line with $3$ parameters is the projective group defined by the Schwarzian third order OD equation:
\[   \Phi(y, y_x, y_{xx}, y_{xxx}) \equiv \frac{y_{xxx}}{y_x} - \frac{3}{2} (\frac{y_{xx}}{y_x})^2 = \nu(x)\]
with linearization the only third order Medolaghi equation with symbol $g_3=0$ and no CC:
\[    L({\xi}_3)\nu \equiv  {\xi}_{xxx} + 2 \nu (x) {\xi}_x + \xi {\partial}_x \nu (x)= 0  \]
When $\nu = 0$, the general solution is simply $\xi = a x^2 + b x + c$ with $3$ parameters, namely $1$ {\it translation} + $1$ {\it dilatation} +$1$ {\it elation} with respective generators  $\{{\theta}_1 =  {\partial}_x, {\theta}_2 = x {\partial}_x, {\theta}_3 = \frac{1}{2} x^2 {\partial}_x\}$.  \\

For $n=2$, eliminating the conformal factor in the case of the Euclidean metric of the plane provides the two Cauchy-Riemann equations defining the infinitesimal complex transformations of the plane. The {\it only possibility} coherent with homogeneity is thus to consider the following system and to prove that it is defining a  system of 
infinitesimal Lie equations, leading to $6$  infinitesimal generators, namely: {\it 2 translations + 1 rotation + 1 dilatation + 2 elations}:  \\
\[  \left \{ \begin{array}{c}
  {\xi}^k_{ijr}=0   \\
   {\xi}^2_{22} - {\xi}^1_{12}=0, {\xi}^1_{22} + {\xi}^2_{12}=0, {\xi}^2_{12} - {\xi}^1_{11}=0, 
{\xi}^1_{12} + {\xi}^2_{11}=0   \\
  {\xi}^2_2 - {\xi}^1_1=0, {\xi}^1_2 + {\xi}^2_1=0   
  \end{array} \right.   \]
\[\{ {\theta}_1={\partial}_1, {\theta}_2={\partial}_2, {\theta}_3=x^1 {\partial}_2 - x^2 {\partial}_1, {\theta}_4=x^1 {\partial}_1 + x^2 {\partial}_2, {\theta}_5= - \frac{1}{2} ((x^1)^2 + (x^2)^2){\partial}_1  + x^1 (x^1 {\partial}_1 + x^2 {\partial}_2) , {\theta}_6 \}  \]
with the elation ${\theta}_6$ obtained from ${\theta}_5$ by exchanging $x^1$ with $x^2$. We have ${\hat{g}}_3=0$ when $n=1,2$.  \\

\noindent
{\bf Remark 4.1}: ({\it Special relativity}): Though surprising it may look like at first sight after replacing the Euclidean metric $d(x^1)^2 + d(x^2)^2$ by the Minkowski metric $d(x^1)^2 - d(x^2)^2$, the above example with $n=2$ perfectly fits with the original presentation of Lorentz transformations if one uses the "{\it hyperbolic}" notations $sh(\phi)=(e^{\phi} - e^{- \phi})/2, ch(\phi)= (e^{\phi}+ e^{-{\phi}})/2, th(\phi)= sh(\phi)/ch(\phi)$ with $ch^2(\phi) - sh^2(\phi) = 1$ instead of the classical $sin(\theta), cos (\theta), tan(\theta)= sin(\theta)/cos(\theta)$ with $cos^2(\theta) + sin^2(\theta) = 1$. Indeed, setting $x^1 = x, x^2 = c t$ and using the well defined  formula $th(\phi) = u/c$ among dimensionless quantities, the Lorentz transformation can be written:  
\[   {\bar{x}}^1 = \frac{x^1 - \frac{u}{c}x^2}{\sqrt{1- (\frac{u}{c})^2}}  ,  \,\,\,  {\bar{x}}^2 =  \frac{ - \frac{u}{c} x^1 + x^2}{\sqrt{1 - (\frac{u}{c})^2}}   \,\, \Leftrightarrow \,\,    
  {\bar{x}}^1 = ch(\phi) x^1 - sh(\phi) x^2, \,\,\,  {\bar{x}}^2 =  - sh(\phi) x^1  + ch(\phi) x^2        \]
Moreover, setting $th(\psi) = v/c$, we obtain easily for the composition of speeds:
\[ th(\phi + \psi) = \frac{(th(\phi) + th(\psi))}{( 1 + th(\phi) th(\psi))} \hspace{5mm} \Leftrightarrow  \hspace{5mm}
composition \,\, ( \frac{u}{c}, \frac{v}{c}) = \frac{(\frac{u}{c} + \frac{v}{c})}{(1 + \frac{u}{c}\frac{v}{c})}  \]  \\
without the need of any "{\it gedanken experiment}" on light signals.  \\

\noindent
{\bf Lemma 4.2}: We have as in (Pommaret, 2016 a, 2016 b):  \\
$\bullet$ ${\hat{g}}_1$ is finite type with ${\hat{g}}_3=0, \forall n \geq 3$.  \\
$\bullet$ ${\hat{g}}_2$ is $2$-acyclic when $n\geq 4$.  \\
$\bullet$ ${\hat{g}}_2$ is $3$-acyclic when $n\geq 5$.  \\

In order to convince the reader that both classical and conformal differential geometry must be revisited, let us prove that the analogue of the Weyl tensor is made by a third order {\it self-adjoint} operator when $n=3$, a result which is neither known nor acknowledged today (Pommaret, 2024 b). We shall proceed by diagram chasing as the local computation done by using computer algebra does not provide any geometric insight (See arXiv:1603.05030 and (Pommaret, 2016 b) for the details). We have $E=T$ and $dim({\hat{F}}_0)=5$ in the following commutative diagram providing 
${\hat{F}}_1$:  \\

\[   \begin{array}{rcccccccccl}
  & & 0  &  & 0  & &  0  &           \\
  & & \downarrow  &  &  \downarrow & & \downarrow &    \\
0  & \rightarrow & {\hat{g}}_4 & \rightarrow &S_4T^*\otimes T& \rightarrow &S_3 T^ *\otimes {\hat{F}}_0 & \rightarrow & {\hat{F}}_1 & \rightarrow  &  0 \\
& & \downarrow  &  &  \downarrow & & \parallel & &    \\
0 & \rightarrow & T^*\otimes {\hat{g}}_3 & \rightarrow &T^*\otimes S_3T^*\otimes T& \rightarrow & T^* \otimes S_2 T^ *\otimes {\hat{F}}_0 &  \rightarrow &0 & &   \\
& & \downarrow  &  &  \downarrow & & \downarrow & & &    \\
0 & \rightarrow &{\wedge}^2 T^*\otimes {\hat{g}}_2 & \rightarrow &{\wedge}^2T^* \otimes S_2T^* \otimes T& \rightarrow & {\wedge}^2 T^* \otimes T^* \otimes {\hat{F}}_0 &\rightarrow &    0   & &    \\
& & \downarrow  &  &  \downarrow & & \downarrow & & &   \\
0 & \rightarrow &{\wedge}^3 T^*\otimes {\hat{g}}_1 & \rightarrow &{\wedge}^3T^*\otimes T^*\otimes T& \rightarrow & {\wedge}^3 T^* \otimes {\hat{F}}_0 & \rightarrow  & 0  & \\
&   & \downarrow  &  &  \downarrow & & \downarrow & & &    \\
 &&  0  &  & 0  & &  0  &  &   &        
\end{array}  \]
\[   \begin{array}{rcccccccccl}
  & &   &  & 0  & &  0  &           \\
  & &   &  &\downarrow   & & \downarrow &    \\
  &  & 0 & \rightarrow &  45  & \rightarrow &  50 & \rightarrow & 5 & \rightarrow  &  0 \\
  & &   &  &  \downarrow & & \downarrow & &    \\
  &  & 0 & \rightarrow &  90 & \rightarrow &  90 &  \rightarrow &0 & &   \\
& & \downarrow  &  &  \downarrow & & \downarrow & & &    \\
0 & \rightarrow & 9 & \rightarrow & 54 & \rightarrow & 45 &\rightarrow & 0 & &    \\
& & \downarrow  &  &  \downarrow & & \downarrow &  &  \\
0 & \rightarrow & 4  & \rightarrow & 9 & \rightarrow &  5   &\rightarrow &0& \\
&   & \downarrow  &  &  \downarrow &  &  \downarrow & & &     \\
 &&  0  &  & 0  & & 0   &  &   &        
\end{array}  \]
A delicate double circular chase provides ${\hat{F}}_1=H^2_2({\hat{g}}_1)$ in the short exact sequence: \\
\[   0 \longrightarrow {\hat{F}}_1 \longrightarrow {\wedge}^2 T^* \otimes {\hat{g}}_2 \stackrel{\delta}{\longrightarrow} {\wedge}^3 T^* \otimes {\hat{g}}_1 \rightarrow 0  \hspace{2cm}
      0 \longrightarrow 5 \longrightarrow  9  \stackrel{\delta}{\longrightarrow}  4  \rightarrow 0  \]
but we have to prove that the map $\delta$ on the right is surjective, a result that it is almost impossible to find in local coordinates. Let us prove it by means of circular diagram chasing in the preceding commutative diagram as follows. Lift any $a \in {\wedge}^3 T^* \otimes {\hat{g}}_1 \subset {\wedge}^3 T^* \otimes T^* \otimes T$ to $b \in {\wedge}^2 T^* \otimes  S_2T^* \otimes T$ because the vertical $\delta$-sequence for $S_4 T^*$ is exact. Project it by the symbol map ${\sigma}_1({\hat{\Phi}})$ to $c \in {\wedge}^2 T^* \otimes  T^* \otimes {\hat{F}}_0$. Then, lift $c$ to $d \in T^* \otimes S_2 T^* \otimes {\hat{F}}_0$ that we may lift {\it backwards horizontally} to $e \in T^* \otimes S_2 T^* \otimes T$ to which we may apply $\delta$ to obtain $f \in {\wedge}^2 T^* \otimes S_2T^* \otimes T$. By commutativity, both $f $ {\it and} $b$ map to $c$ and the difference $f - b$ maps thus to zero. Finally, we may find $g \in {\wedge}^2 T^* \otimes {\hat{g}}_2 $ such that $ b = g + \delta (e)$ and we obtain thus $a = \delta (g) + {\delta}^2 (e) = \delta (g)$, proving therefore the desired surjectivity. We have $10$ parameters: {\it 3  translations +  3 rotations + 1 dilatation + 3 elations} and the totally unexpected formally exact sequences on the jet level are thus, showing in particular that second order CC do not exist:  \\
\[  0 \rightarrow {\hat{R}}_3 \rightarrow J_3(T)  \rightarrow J_2({\hat{F}}_0) \rightarrow 0 \,\, \Rightarrow \,\, 
0 \rightarrow 10 \rightarrow 60 \rightarrow 50 \rightarrow 0  \] 
\[     0  \rightarrow {\hat{R}}_4 \rightarrow J_4(T)  \rightarrow J_3({\hat{F}}_0) \rightarrow {\hat{F}}_1 \rightarrow 0 \,\,\, \Rightarrow \,\, 
0 \rightarrow 10 \rightarrow 105 \rightarrow 100 \rightarrow 5 \rightarrow 0 \]
\[     0  \rightarrow {\hat{R}}_5 \rightarrow J_5(T)  \rightarrow J_4({\hat{F}}_0) \rightarrow J_1({\hat{F}}_1) \rightarrow {\hat{F}}_2  \rightarrow 0 \,\,\, \Rightarrow \,\, 
0 \rightarrow 10 \rightarrow 168 \rightarrow 175 \rightarrow 20 \rightarrow   3    \rightarrow 0 \]
We obtain the minimum differential sequence, {\it which is nervertheless  not a Janet sequence}:
\[  \fbox{  $   0 \rightarrow {\hat{\Theta}} \rightarrow T \underset 1{\stackrel{\hat{\cal{D}}}{\rightarrow}} {\hat{F}}_0 \underset 3{\rightarrow} {\hat{F}}_1 \underset 1{\rightarrow}  {\hat{F}}_2  \rightarrow 0 \,\, \Rightarrow \,\, 
0 \rightarrow {\hat{\Theta}} \rightarrow 3 \underset 1{\stackrel{\hat{\cal{D}}}{\rightarrow}} 5 \underset 3{\rightarrow} 5 \underset 1{\rightarrow} 3 \rightarrow 0  $  } \]
with $\hat{\cal{D}}$ the conformal Killing operator and vanishing Euler-Poincar\'{e} characteristic  
$3 - 5 + 5 - 3 = 0$.   \\   
When $n=4$, we have {\it 4 translations + 6 rotations + 1 dilatation + 4 elations = 15 parameters}.  \\
Also, ${\hat{g}}_3=0  \Rightarrow {\hat{g}}_4 = 0 \Rightarrow  {\hat{g}}_5 = 0 $ in the conformal case, we have the commutative diagram with exact vertical long $\delta$-sequences {\it but the left one} and where the second row proves that there cannot exist first order Bianchi-like identities for the Weyl tensor, {\it contrary to what is still believed today}:

\[  \begin{array}{rcccccccccl}
  & &  0 & & 0 & & 0 &  & 0 &  & \\
  & & \downarrow & & \downarrow & & \downarrow & & \downarrow & & \\
0 & \rightarrow & {\hat{g}}_4 & \rightarrow &  S_4T^*\otimes T & \rightarrow & S_3T^*\otimes {\hat{F}}_0& \rightarrow & T^* \otimes {\hat{F}}_1 & \rightarrow & 0  \\
  & & \hspace{2mm}\downarrow  \delta  & & \hspace{2mm}\downarrow \delta & &\hspace{2mm} \downarrow \delta & & \parallel & & \\
0 & \rightarrow& T^*\otimes {\hat{g}}_3 &\rightarrow &T^*\otimes S_3T^*\otimes T & \rightarrow &T^*\otimes  S_2 T^*\otimes {\hat{F}}_0 &\rightarrow & T^* \otimes {\hat{F}}_1  & \rightarrow & 0 \\
  & &\hspace{2mm} \downarrow \delta &  &\hspace{2mm} \downarrow \delta & &\hspace{2mm}\downarrow \delta &  & \downarrow  & &  \\
0 & \rightarrow & {\wedge}^2T^*\otimes {\hat{g}}_2 & \rightarrow & {\wedge}^2T^*\otimes  S_2 T^*\otimes T & \rightarrow & {\wedge}^2T^*\otimes  T^* \otimes {\hat{F}}_0 & \rightarrow & 0 &&  \\
 &  &\hspace{2mm}\downarrow \delta  &  & \hspace{2mm} \downarrow \delta  &  & \hspace{2mm} \downarrow \delta & &  & & \\
0 & \rightarrow & {\wedge}^3T^*\otimes {\hat{g}}_1 & =  & {\wedge}^3T^*\otimes T^* \otimes  T  & \rightarrow   &  {\wedge}^3 T^* \otimes {\hat{F}}_0  & \rightarrow  & 0 &  & \\
 &   & \hspace{2mm} \downarrow \delta &  & \hspace{2mm}  \downarrow \delta  &  &  \downarrow    &  &  &  &\\
0  & \rightarrow  &  {\wedge}^4 T^* \otimes T  & =  & {\wedge}^4 T^* \otimes T & \rightarrow & 0 &  &  &&  \\
  &   & \downarrow &  &  \downarrow  &  &  &  &  &  &  \\
   & &  0  &  &  0  &  &  &  & &  &
\end{array}  \]

\[  \begin{array}{rcccccccccl}
  & &   & & 0 & & 0 &  & 0 & &  \\
  & & & & \downarrow  & & \downarrow & & \downarrow &  & \\
   & & 0 & \rightarrow &  140 & \rightarrow & 180 & \rightarrow & 40 & \rightarrow &  0  \\
  &  & & & \hspace{2mm} \downarrow \delta  & &\hspace{2mm} \downarrow \delta & & \parallel & &  \\
 & & 0 & \rightarrow & 320& \rightarrow & 360 &\rightarrow & 40  & \rightarrow & 0  \\
 &  & \downarrow  &  &\hspace{2mm} \downarrow \delta & &\hspace{2mm}\downarrow \delta &  & \downarrow  &  &  \\
0 & \rightarrow & 24 & \rightarrow & 240 & \rightarrow & 216 & \rightarrow & 0 & &  \\
  & &\hspace{2mm}\downarrow \delta  &  & \hspace{2mm} \downarrow \delta  &  & \hspace{2mm} \downarrow \delta  & & & & \\
0 & \rightarrow & 28 & \rightarrow  & 64 &\rightarrow   & 36    & \rightarrow  & 0 &  &  \\
  &  &  \hspace{2mm} \downarrow \delta  &  & \hspace{2mm} \downarrow \delta &  & \downarrow &  &  &  & \\
 0 & \rightarrow  &  4  & =  & 4  &\rightarrow  & 0 &  &  & &  \\
  &  &  \downarrow &  & \downarrow & &  &  &  &  &  \\
  &  &   0   &  &  0  &  &  &  &  &  &  
\end{array}  \]
A diagonal snake chase proves that ${\hat{F}}_1 \simeq H^2 ({\hat{g}}_1)$. However, we have the $\delta$-sequence:  \\
\[   0 \rightarrow  T^*\otimes  {\hat{g}}_2 \stackrel{\delta}{\rightarrow} {\wedge}^2 T^* \otimes {\hat{g}}_1 \stackrel{\delta}{\rightarrow } {\wedge}^3 T^*\otimes T \rightarrow 0  \]
We obtain $dim(B^2_2({\hat{g}}_1))= 4 \times 4 = 16$ and let the reader prove as before that the map $\delta$ on the right is surjective, a result leading to
$dim(Z^2_2({\hat{g}}_1))= (6 \times (6 + 1)) - (4 \times 4) = 42 - 16 = 26$. The Weyl tensor has thus $dim({\hat{F}}_1) = 26 - 16 = 10$ components, a way that must be compared to the standard one that can be found in the GR literature. We obtain the minimum differential sequence, {\it which is nervertheless  not a Janet sequence}:
\[  \fbox{  $  0 \rightarrow {\hat{\Theta}} \rightarrow T \underset 1{\stackrel{\hat{\cal{D}}}{\rightarrow}} {\hat{F}}_0 \underset 2{\rightarrow} {\hat{F}}_1 \underset 2{\rightarrow}  {\hat{F}}_2  \underset 1{\rightarrow}  {\hat{F}}_3  \rightarrow 0 \,\, \Rightarrow \,\, 
0 \rightarrow {\hat{\Theta}} \rightarrow 4  \underset 1{\stackrel{\hat{\cal{D}}}{\rightarrow}} 9 \underset 2{\rightarrow} 10 \underset 2{\rightarrow} 9  \underset 1{\rightarrow}  4  \rightarrow 0  $  } \]

\noindent
{\bf 5) POINCARE EQUATIONS}  \\

If $X$ is a manifold and $G$ is a lie group ({\it not acting necessarily on} $X$), let us consider maps $a:X\rightarrow G: (x)\rightarrow (a(x))$ or equivalently sections of the trivial (principal) bundle $X\times G$ over $X$. If $x+dx$ is a point of $X$ close to $x$, then $T(a)$ will provide a point $a+da=a+\frac{\partial a}{\partial x}dx$ close to $a$ on $G$. We may bring $a$ back to $e$ on $G$ by acting on $a$ with $a^{-1}$, {\it  either on the left or on the right}, getting therefore a $1$-form $a^{-1}da=A$ or $(da)a^{-1}=B$ with value in $\cal{G}$. As $aa^{-1}=e$ we also get $(da)a^{-1}=-ada^{-1}=-b^{-1}db$ if we set $b=a^{-1}$ as a way to link $A$ with $B$. When there is an action $y=ax$, we have $x=a^{-1}y=by$ and thus $dy=dax=(da)a^{-1}y$, a result leading through the first fundamental theorem of Lie and the Maurer-Cartan (MC) $1$-forms ${\omega}^{\tau} = {\omega}^{\tau}_{\sigma} (a) d a^{\sigma}$ to the equivalent formulas:\\
\[   a^{-1}da=A=({A}^{\tau}_i(x)dx^i=-{\omega}^{\tau}_{\sigma}(b(x)){\partial}_ib^{\sigma}(x)dx^i)  \]
\[   (da)a^{-1}=B=({B}^{\tau}_i(x)dx^i={\omega}^{\tau}_{\sigma}(a(x)){\partial}_ia^{\sigma}(x)dx^i)  \]
Introducing the induced bracket $[A,A](\xi,\eta)=[A(\xi),A(\eta)]\in {\cal{G}}, \forall \xi,\eta\in T$, we may define the {\it curvature} $2$-form $dA-[A,A]=F\in {\wedge}^2T^*\otimes {\cal{G}}$ by the local formula ({\it care again to the sign}):\\
\[     {\partial}_iA^{\tau}_j(x)-{\partial}_jA^{\tau}_i(x)-c^{\tau}_{\rho\sigma}A^{\rho}_i(x)A^{\sigma}_j(x)=F^{\tau}_{ij}(x)  \]
This definition can also be adapted to $B$ by using $dB + [B,B]$ and we obtain from the second fundamental theorem of Lie:\\

\noindent
{\bf Theorem 5.1}: There is a {\it nonlinear gauge sequence}:\\
\[ \fbox{  $   \begin{array}{ccccc}
X\times G & \longrightarrow & T^*\otimes {\cal{G}} &\stackrel{MC}{ \longrightarrow} & {\wedge}^2T^*\otimes {\cal{G}}  \\
a                & \longrightarrow  &    a^{-1}da=A         &    \longrightarrow & dA-[A,A]=F
\end{array}  $ }    \eqno{(1)}   \]

In 1956, at the birth of GT, the above notations were coming from the EM potential $A$ and EM field $dA=F$ of relativistic Maxwell theory. Accordingly, $G=U(1)$ (unit circle in the complex plane) $\longrightarrow dim ({\cal{G}})=1$) {\it was the only possibility} to get a $1$-form $A$ and a $2$-form $F$ with vanishing structure constants $c=0$ (Abelian group).  \\

Choosing now $a$ "close" to $e$, that is $a(x)=e+t\lambda(x)+...$ and linearizing as usual, we obtain the linear operator $d:{\wedge}^0T^*\otimes {\cal{G}}\rightarrow {\wedge}^1T^*\otimes {\cal{G}}:({\lambda}^{\tau}(x))\rightarrow ({\partial}_i{\lambda}^{\tau}(x))$ leading to (Pommaret, 1983 b, 1988, 1994):\\

\noindent
{\bf Corollary  5.2}: There is a {\it linear gauge sequence}:\\ 
\[  \fbox{ $ {\wedge}^0T^*\otimes {\cal{G}}\stackrel{d}{\longrightarrow} {\wedge}^1T^*\otimes {\cal{G}} \stackrel{d}{\longrightarrow} {\wedge}^2T^*\otimes{\cal{G}} \stackrel{d}{\longrightarrow} ... \stackrel{d}{\longrightarrow} {\wedge}^nT^*\otimes {\cal{G}}\longrightarrow  0  $  }     \eqno{(2)}  \]
which is the tensor product by $\cal{G}$ of the Poincar\'{e} sequence for the exterior derivative. \\

 In order to introduce the previous results into a variational framework along the rarely quoted paper of H. Poincar\'{e} in $1901$ (Poincar\'{e}, 1901), we may consider a Lagrangian on $T^*\otimes \cal{G}$, that is an {\it action} $W=\int w(A)dx$ where $dx=dx^1\wedge ...\wedge dx^n$ and to vary it. With $A=a^{-1}da= - (db)b^{-1}$ we may introduce $\lambda=a^{-1}\delta a= - (\delta b)b^{-1}\in {\cal{G}}={\wedge}^0T^*\otimes {\cal{G}}$ with local coordinates ${\lambda}^{\tau}(x)=-{\omega}^{\tau}_{\sigma}(b(x))\delta b^{\sigma}(x)$ and we obtain in local coordinates (Pommaret, 1994, p 180-185):
 \[ \fbox{ $   \delta A=d\lambda - [A,\lambda]  \Leftrightarrow  \delta A^{\tau}_i={\partial}_i\lambda^{\tau}-c^{\tau}_{\rho\sigma}A^{\rho}_i{\lambda}^{\sigma} $  }  \eqno{(3)}  \] 
 Then, setting $\partial w/\partial A={\cal{A}}=({\cal{A}}^i_{\tau})\in {\wedge}^{n-1}T^*\otimes {\cal{G}}^* $, we get:\\
\[  \delta W=\int {\cal{A}}\delta Adx=\int {\cal{A}}(d\lambda-[A,\lambda])dx  \]
and thus, after integration by part, the Poincar\'{e} equations, rarely quoted (Pommaret, 2014):\\
\[   \fbox{ $   {\partial}_i{\cal{A}}^i_{\tau}+c^{\sigma}_{\rho\tau}A^{\rho}_i{\cal{A}}^i_{\sigma}=0   $ }  \eqno{(4)} \]
Such a linear operator for $\cal{A}$ has non constant coefficients linearly depending on $A$ and is the adjoint of the previous operator (up to sign). \\
However, setting now $(\delta a)a^{-1}=\mu\in {\cal{G}}$, we get $\lambda=a^{-1}((\delta a)a^{-1})a=Ad(a)\mu$ while, setting $a'=ab$, we get the {\it gauge transformation} for any $b\in G$ (See (Pommaret, 1994), Proposition $14$, p $182$):
\[ \fbox{   $  A \rightarrow A'=(ab)^{-1}d(ab)=b^{-1}a^{-1}((da)b+adb)=Ad(b)A+b^{-1}db,   \Rightarrow F' = Ad(b) F$  }  \eqno{(5)}   \] 
Setting $b=e+t\lambda+...$ with $t\ll 1$, then $\delta A$ becomes an infinitesimal gauge transformation. Finally, $ a'=ba\Rightarrow A'=a^{-1}b^{-1}((db)a+a(db))=a^{-1}(b^{-1}db)a+A\Rightarrow \delta A=Ad(a)d\mu$ when $b=e+t\mu +...$ with $t\ll 1$.
Therefore, introducing $\cal{B}$ such that ${\cal{B}}\mu= {\cal{A}}\lambda$, we get the divergence-like equations:
\[  \fbox{  $  {\partial}_i{\cal{B}}^i_{\sigma}=0  $  }  \eqno{(6)}   \]

We provide some more details on the  {\it Adjoint representation} $Ad: G \rightarrow aut({\cal{G}}): a \rightarrow Ad(a)$. which is defined by a linear map $ M(a) = ({M}^{\tau}_{\rho}(a)) \in aut({\cal{G}})$ and we have the involutive system (Pommaret, 1994, Proposition 10, p 180):  
\[  \fbox{  $      \frac{\partial {M}^{\tau}_{\mu}}{\partial a^r} + c ^{\tau}_{\rho \sigma} {\omega}^{\rho}_r (a) {M}^{\sigma}_{\mu} = 0   $   }  \eqno{(7)} \] 
by using the fact that {\it any right invariant vector field on $G$ commutes with any left invariant vector field on $G$} (See (Bialynicki-Birula, 1962 and Pommaret, 1983 b, 2023 c) for applications of ({\it reciprocal distributions}) to Differential Galois Theory). In addition, as $a^{-1} \delta a = \lambda$, we obtain therefore successively:
\[    Ad(a^{-1})\lambda = a \lambda a^{-1} = a (a^{-1} \delta a ) a^{-1} = \delta a a^{-1} = \mu \, \,  \Rightarrow  \, \, d Ad(a^{-1}) \lambda = (d \delta a) a^{-1}- \delta a a^{-1} d a a^{-1}  \]
\[  \begin{array}{lcl}
 Ad(a) d Ad(a^{-1}) \lambda   &  =  &  a^{-1} (d \delta a a^{-1} - \delta a a^{-1} d a a^{-1})a   \\
        & =  &   a^{-1} d \delta a - (a^{-1} \delta a a^{-1}) d a   \\
        & =  &   a^{-1} \delta d a + (\delta a^{-1}) d a \\
        & =  &  \delta A  
        \end{array}   \]
As a byproduct, the operator $\nabla: \lambda \rightarrow d \lambda - [A, \lambda]$ only depends on $A$ and is the "{\it twist }" of the derivative $d$ under the action of $Ad(a)$ on 
${\cal{G}}$ whenever $A= a^{-1} d a$ in the gauge sequence. Setting:
\[   \nabla (\alpha \otimes \lambda) = d \alpha \otimes \lambda + (-1)^r \alpha \wedge \nabla \lambda, \forall \alpha \in {\wedge}^r T^* ; \lambda \in {\cal{G}} \]
an easy computation shows that $(\nabla \circ \nabla \lambda)^{\tau}_{ij} = c^{\tau}_{\rho \sigma} F^{\rho}_{ij} {\lambda}^{\sigma} = 0 $ and we obtain:  \\

\noindent
{\bf Corollary 5.3}: The following ${\nabla}$-sequence:
\[   \fbox{  $     {\wedge}^0 T^* \otimes {\cal{G}} \stackrel{\nabla} \longrightarrow {\wedge}^1T^* \otimes {\cal{G}} \stackrel{\nabla}{\longrightarrow} ... \stackrel{\nabla}{\longrightarrow} 
{\wedge}^n T^* \otimes {\cal{G}} \longrightarrow 0  $  }  \eqno(8)  \]
is {\it anothe}r locally exact linearization of the non-linear gauge sequence which is isomorphic to copies of the Poincar\'{e} sequence and describes infinitesimal gauge transformations.  \\   \\

{\it In a completely different local setting}, if $G$ acts on $X$ and $Y$ is a copy of $X$ with an action graph $X \times G \rightarrow X \times Y: (x, a) \rightarrow (x, y = a x = f(x,a))$, we may use the theorems of Lie in order to find a basis $\{ {\theta}_{\tau} \mid 1 \leq \tau \leq p=dim(G) \}$ of infinitesimal generators of the action. If $\mu = ({\mu}_1, ..., {\mu}_n)$ is a multi-index of length $\mid \mu \mid = {\mu}_1 + ... + {\mu}_n$ and ${\mu}+1_{i} = ({\mu}_1,..,{\mu}_{i-1},{\mu}_i + 1, {\mu}_{i+1} ,...,{\mu}_n)$, we may introduce the system of infinitesimal Lie equations or Lie algebroid $R_q \subset J_q(T)$ with sections defined by ${\xi}^k_{\mu} (x) = {\lambda}^{\tau} (x) {\partial}_{\mu} {\theta}^k_{\tau}(x)$ for an arbitrary section ${\lambda} \in {\wedge}^0 T^* \otimes {\cal{G}}$ and the trivially involutive operator $j_q: T \rightarrow J_q(T): \theta \rightarrow ({\partial}_{\mu} \theta , 0 \leq \mid \mu \mid \leq q)$ of order $q$. We finally obtain the {\it Spencer operator} through the chain rule for derivatives [9, 19, 38, 39]:  \\
\[   \fbox{  $  (d{\xi}_{q+1})^k_{\mu, i}(x) =   {\partial}_i{\xi}^k_{\mu}(x) -{\xi}^k_{\mu +1_i}(x) = {\partial}_i{\lambda}^{\tau}(x) {\partial}_{\mu} {\theta}^k_{\tau}(x)  $   }  \eqno(9) \]   \\

\noindent
{\bf Theorem 5.4}: When $q$ is large enough to have an isomorphism $R_{q+1} \simeq R_q \simeq {\wedge}^0 T^* \otimes {\cal{G}}$ and the following  {\it linear Spencer sequence} in which the operators $D_r$ are induced by $d$ as above:  \\
\[ \fbox{  $  0 \longrightarrow \Theta \stackrel{j_q}{\longrightarrow} R_q \stackrel{D_1}{\longrightarrow} T^* \otimes R_q \stackrel{D_2}{\longrightarrow} {\wedge}^2 T^* \otimes R_q \stackrel{D_3}{\longrightarrow} ... \stackrel{D_n}{\longrightarrow} {\wedge}^nT^*\otimes R_q \longrightarrow  0  $  }    \eqno(10)   \]
is isomorphic to the linear gauge sequence but with a completely different meaning because $G$ is now acting on $X$ and $\Theta \subset T$ is such that $[\Theta, \Theta ] \subset \Theta$.  \\

 The idea is to notice that the brothers are {\it always} dealing with the group of rigid motions, namely a group with $6$ parameters ($3$ {\it translations} + $3$ {\it rotations}). In 1909 it should have been strictly impossible for the two brothers to extend their approach to bigger groups, in particular to include the only additional {\it dilatation} of the Weyl group that will provide the virial theorem and, {\it a fortiori}, the {\it elations} of the conformal group considered later on by H.Weyl [40]. In order to emphasize the reason for using Lie equations, we now provide the explicit form of the highly nonlinear $n$ finite elations and their infinitesimal counterpart in a conformal Riemannian space of dimension $n$ with a non-degenerate metric 
 $\omega$, namely:\\
\[  \fbox{  $   y=\frac{x-x^2b}{1-2(bx)+b^2x^2}  \Rightarrow  {\theta}_s= - \frac{1}{2} x^2 {\delta}^r_s{\partial}_r+{\omega}_{st}x^tx^r{\partial}_r   \Rightarrow \  
{\partial}_r{\theta}^r_s=n{\omega}_{st}x^t, \hspace{2mm}   \forall 1\leq r,s,t \leq n   $   }  \]
where the underlying metric is used for the scalar products $x^2,bx,b^2$ involved.  \\

Our purpose is to exhibit {\it directly} the Cauchy, Cosserat,  Clausius, Maxwell, Weyl (CCCMW) equations by computing with full details the adjoint of the first Spencer operator $D_1: {\hat{R}}_3 \rightarrow T^* \otimes {\hat{R}}_3$ for the conformal involutive finite type third order system ${\hat{R}}_3 \subset J_3(T)$ for any dimension $n\geq 1$, in particular for $n\geq 1$ along with the results obtained in [30]. In general, one has $n$ {\it translations} $+$  $n(n-1)/2$ {\it rotations} +  $1$ {\it dilatation} + $n$ nonlinear {\it elations}, that is a total of  $(n+1)(n+2) /2$ parameters, thus $15$ when $n=4$. As a byproduct, the Cosserat couple-stress equations will be obtained for the Killing involutive finite type second order system ${\hat{R}}_2 \subset J_2(T)$. It must be noticed that not even a single comma must be changed when $n=3$ when our results are compared to the original formulas provided by the bothers Cosserat in $1909$. We only need recall the specific features of the standard first order Spencer operator $d: {\hat{R}}_3 \rightarrow T^* \otimes {\hat{R}}_2$ as follows by considering the multi-indices for the various parameters, separately as follows:   \\
\[   ({\xi}^k(x), {\xi}^k_i(x), {\xi}^k_{ij}(x), {\xi}^k_{ijr}(x) = 0) \rightarrow  ({\partial}_i {\xi}^k(x) - {\xi}^k_i(x), \,\,  {\partial}_i {\xi}^k_j(x) - {\xi}^k_{ij}(x) , \,\,  {\partial}_i {\xi}^r_r(x) - {\xi}^r_{ri}(x), {\partial}_r {\xi}^k_{ij}(x) )    \]
in the {\it duality summation}:  \\
\[  \fbox{  $   {\sigma}^i_k ({\partial}_i {\xi}^k(x) - {\xi}^k_i(x) )+ {\mu}^{ij}_k ({\partial}_i {\xi}^k_j(x) - {\xi}^k_{ij}(x)) + {\nu}^i ({\partial}_i {\xi}^r_r(x) - {\xi}^r_{ri}(x)) + 
 {\pi}^{ij,r}_k ({\partial}_r {\xi}^k_{ij}(x) )  $  }    \]
We have obtained a first simplification by noticing that the third order jets vanish, that is to say ${\xi}^k_{rij}=0$. Indeed, starting with the Euclidean or Minkowski metric 
$\omega$ with vanishing Christoffel symbols $\gamma=0$, the second order conformal equations can be provided in the parametric form:  \\
 \[  \fbox{  $  {\xi}^k_{ij} = {\delta}^k_i A_j(x) + {\delta}^k_j A_i(x) - {\omega}_{ij} {\omega}^{kr} A_r(x) \Leftrightarrow {\xi}^r_{ri} = n A_i(x)  $  }   \]
The desired result follows from the fact that this system is homogeneous and ${\hat{g}}_3=0, \forall n\geq 3$.  \\
A second simplification may be obtained by using the (constant) metric in order to raise or lower the indices in the implicit summations considered. In particular, we have successively:  \\
\[     {\omega}_{rj} {\xi}^r_i + {\omega}_{ir} {\xi}^r_i = 2 A(x) {\omega}_{ij}  \Rightarrow  A(x) = {\xi}^1_1(x) = {\xi}^2_2 (x) = ... = {\xi}^n_n (x) = \frac{1}{n} {\xi}^r_r (x)  \]
In this situation, ${\sigma}^{i,j} {\xi}_{i,j} = {\Sigma}_{i < j}({\sigma}^{i,j} - {\sigma}^{j,i}) {\xi}_{i,j} + \frac{1}{n} {\sigma}^r_r {\xi}^r_r $ and we may set 
${\mu}^{ij}_k {\xi}^k_{ij} = - {\mu}^i A_i$ where ${\mu}^i$ is a linear (tricky) function of the ${\mu}^{ij}_k$ with constant coefficients only depending on $\omega$. The new equivalent duality summation becomes:
\[   {\sigma}^{i,r} {\partial}_r {\xi}_i + {\Sigma}_{i<j} ( {\mu}^{ij,r} {\partial}_r{\xi}_{i,j}  - ({\sigma}^{i,j} - {\sigma}^{j,i} ){\xi}_{i,j} )  - {\sigma}^r_r A(x)  -  {\mu}^i A_i(x) + {\nu}^i ({\partial}_i A (x) - A_i(x)) + {\pi}^{i,r}({\partial}_r A_i(x))      \]
and is important to notice that we may have ${\sigma}^{i,j} \neq {\sigma}^{j,i}$ when $i < j$. Integrating by parts and changing the signs, we finally obtain 
the Poincar\'{e} equations in the following form that allows to avoid using the structure constants of the conformal Lie algebra:  \\

\[  \fbox{ $  {\xi}_i \longrightarrow {\partial}_r {\sigma}^{i,r} = f^i    \,\,\, (Cauchy \,\, stress \,\, equations)   $  }   \]

\[  \fbox{  $  {\xi}_{i,j}, i< j \longrightarrow  {\partial}_r {\mu}^{ij,r} + {\sigma}^{ij} - {\sigma}^{ji} = m^{ij}  \,\,\, (Cosserat \,\,  couple{-}stress \, \,equations) $ }  \]

\[  \fbox{  $  A(x) \longrightarrow {\partial}_r {\nu}^r   + {\sigma}^r_r =  u             \,\,\, (Clausius \,\, virial \,\, equation ) $  }  \]

\[  \fbox{  $  A_i(x) \longrightarrow         {\partial}_r {\pi}^{i,r}  + {\mu}^i + {\nu}^i =  v^i    \,\,\,( Maxwell/Weyl \,\, electromagnetic \,\, equations )    $   } \]   \\   \\
Transforming these equations into pure divergence-like equations as we already did for the Poincar\'{e} equations by using now the isomorphism 
$ R_q \simeq {\wedge}^0T^*\otimes {\cal{G}}$ is more difficult ( arXiv: 2401.14563 )  \\

\noindent
{\bf Example 5.5}: ({\it Projective group of the real line}): With $n=1$ and $K=\mathbb{Q}$, let us consider the Lie pseudogroup defined by the third order Schwarzian OD equation with standard jet notations:   \\
\[     \frac{y_{xxx}}{y_x} - \frac{3}{2} (\frac{y_{xx}}{y_x})^2 = 0   \,\,\,\,  \Rightarrow  \,\,\,\,   {\xi}_{xxx}=0  \]
A basis of infinitesimal generators is $\{ {\theta}_1 = {\partial}_x, \, {\theta}_2 = x \, {\partial}_x, \, {\theta}_3 = \frac{1}{2} x^2 \}$ and we have the following diagram in which the columns of the $3\times 3$ matrix describes the components of $j_2(\theta)$:  \\ 
\[ \fbox{  $   \left( \begin{array}{lcl}
{\xi} &  &   \\
{\xi}_x & =  &  A  \\
 {\xi}_{xx} &  =  & A_x
\end{array} \right)   =
\left( \begin{array}{cccccc}
1 & x & \frac{1}{2} x^2   \\
0 & 1 & x   \\
 0 & 0 & 1
\end{array} \right)
\left( \begin{array}{c}
{\lambda}^1 \\
{\lambda}^2  \\
{\lambda}^3  
\end{array} \right)   $  }  \]

\noindent
In order to construct the adjoint of the first Spencer operator $D_1: {\hat{R}}_3  \rightarrow T^* \otimes {\hat{R}}_3$ when there are only one translation, no rotation but only one dilatation and  only one elation, we have to consider the duality sum with ${\xi}_{xxx} = 0$:  \\
\[   \sigma \, ({\partial}_x \xi - {\xi}_x) + \nu \, ({\partial}_x {\xi}_x - {\xi}_{xx}) +  \pi \, ({\partial}_x {\xi}_{xx} - 0)  \]
Integrating by parts and changing the sign, we get the board of first order operators allowing to define $ad(D_1)$, namely the Cauchy stress equation, the Clausius virial equation and the Maxwell/Weyl equation successively:  \\
\[ \fbox{  $   \left\{  \begin{array}{lclcc}
\xi & \longrightarrow & {\partial}_x \sigma & = & f  \\
{\xi}_x & \longrightarrow & {\partial}_x \nu + \sigma & = & u  \\
{\xi}_{xx} & \longrightarrow & {\partial}_x \pi + \nu & = & v  
\end{array} \right.  $  }  \]
We may obtain therefore the pure divergence equations:  \\
\[   \fbox{  $  \left\{ \begin{array}{lcl}
{\partial}_x (\sigma) & = &  f  \\
{\partial}_x ( \nu + x \, \sigma ) & = &  x \, f  + u \\
{\partial}_x ( \pi + x \nu + \frac{1}{2} x^2 \sigma) & = & \frac{1}{2} x^2 f + x \, u +  v
\end{array}  \right.  $  }  \]
Coming back now to the Janet sequence in the following {\it Fundamental Diagram I} of [30]:  \\
\[ \fbox{  $  \begin{array}{rccccccccc}
  & && & &0 & & 0 &  &  \\
   & & & & & \downarrow & & \downarrow & &  \\
   & 0 &\longrightarrow &\Theta &\stackrel{j_3}{\longrightarrow} & 3 & \underset 1{\stackrel{D_1}{\longrightarrow} } & 3 &\longrightarrow 0 & \hspace{5mm} Spencer   \\
   & & & & & \downarrow & & \parallel  &  &  \\
    & 0 & \longrightarrow &1 &\stackrel{j_3}{\longrightarrow} & 4 & \underset 1{\stackrel{D_1}{\longrightarrow} }& 3 & \longrightarrow 0&   \\
    & & & \parallel & & \hspace{3mm}\downarrow \Phi & & \downarrow & &  \\
    0\longrightarrow & \Theta & \longrightarrow & 1 & \underset 3{\stackrel{{\cal{D}}}{\longrightarrow} } & 1 & \longrightarrow & 0 & & \hspace{5mm} Janet  \\
    & & & & & \downarrow & & & &  \\
     & & & & & 0 & & & & 
 \end{array}  $  }  \]
we notice that the central row splits because $j_3$ is an injective operator and the corresponding sequence of differential modules is the splitting sequence $ 0 \rightarrow D^3 \rightarrow D^4 \rightarrow D \rightarrow 0 $. It follow that the adjoint sequence also splits in the following commutative and exact adjoint diagram:  \\
\[ \fbox{  $   \begin{array}{ccccccccc}
    & & & & 0 & & 0 &  &  \\
    & &  & & \uparrow & & \uparrow &  \\
    & & 0 & \longleftarrow & 3 & \underset 1{\stackrel{ ad(D_1) }{\longleftarrow} } & 3 &  &    \\
    & & \uparrow & & \uparrow & & \parallel  &  &  \\
 0 & \longleftarrow & 1 &\stackrel{ ad(j_3 ) }{\longleftarrow} & 4 & \underset 1{\stackrel{ ad(D_1) }{\longleftarrow} }& 3 & \longleftarrow  & 0   \\
    & & \parallel & & \uparrow  & & \uparrow  & &  \\
 0 & \longleftarrow & 1 & \underset 3{\stackrel{ad({\cal{D}})}{\longleftarrow} } & 1 &  &  0  & &  \\
    & &\uparrow & & \uparrow & & & &  \\
    & & 0 & & 0 & & & & 
 \end{array} $  }   \]
In a more effective but quite surprising way, the kernel of $ad(D_1)$ in the adjoint Spencer sequence is defined by the successive differential conditions:  \\
\[  (f=0, \, u=0, \, v=0 ) \Rightarrow ({\partial}_x \, \sigma = 0, \, {\partial}_x \,  \nu + \sigma =0, \, {\partial }_x \, \pi + \nu = 0 ) \Rightarrow  \fbox{ $ {\partial}_{xxx} \, \pi =0  $ }  \]
It is thus isomorphic to the kernel of $ad({\cal{D}})$ in the adjoint of the Janet sequence. But ${\cal{D}}$ is a Lie operator in the sense that ${\cal{D}}\xi= 0, {\cal{D}} \eta =0 \Rightarrow {\cal{D}} [\xi, \eta] =0$ or, equivalently, $[ \Theta, \Theta ] \subset \Theta$. It follows that the " stress " appearing in the Cauchy operator which is the adjoint of the Lie operator ${\cal{D}}$ in the Janet sequence has {\it strictly nothing to do}  with the " stress " appearing in the Cosserat couple-stress equations provided by the adjoint of $D_1$ appearing in the Spencer sequence. This confusion, which is even worst than the controversy Cartan/Vessiot (Pommaret, 2022, b)) or the controversy Beltrami/Einstein, leads to revisit the mathematical foundations of both continuum mechanics and general relativity because of the well known field-matter couplings like piezzoelectricity or photoelasticity (Pommaret, 2019). \\   \\

\noindent
{\bf 6) ELECTROMAGNETISM AND GRAVITATION}  \\

When $n=4$, the comparison with the Maxwell equations of electromagnetism is easily obtained as follows. Indeed, writing a part of the dualizing summation in the form:\\
\[      {\cal{J}}^i ({\partial}_i A - A_i )+ \frac{1}{2} {\cal{F}}^{ij} ({\partial}_i A_j - {\partial}_j A_i) = -  {\cal{J}}^i A_i + {\sum}_{i\leq j} {\cal{F}}^{ij} ({\partial}_i A_j - {\partial}_j A_i) + ... \]
\[  = - {\cal{J}}^1A_1 + ... + {\cal{F}}^{12} ({\partial}_1 A_2 - {\partial}_2 A_1 + {\cal{F}}^{13} ({\partial}_1 A_3 - {\partial}_3 A_1) + {\cal{F}}^{14} ({\partial}_1 A_4 - {\partial}_4 A_1) + ...  \]
\[  =  - ({\cal{J}}^1 A_1+ ... + ({\cal{F}}^{12}{\partial}_2 A_1+  {\cal{F}}^{13}{\partial}_3 A_1 +  {\cal{F}}^{14}{\partial}_4 A_1) + ...) \]
\[   = div(...) + ( - {\cal{J}}^1 + {\partial}_2 {\cal{F}}^{12}+  {\partial}_3 {\cal{F}}^{13} +  {\partial}_4 {\cal{F}}^{14} ) A_1 + ... \]
Integrating by parts and changing the sign as usual, we obtain as usual the second set of Maxwell equations for the induction ${\cal{F}}$:  \\
\[    \fbox{ $  {\partial}_r {\cal{F}}^{ir} - {\cal{J}}^i =0  \,\,\, \Rightarrow  \,\,\, {\partial}_i {\cal{J}}^i = {\partial}_{ij}{\cal{F}}^{ij} = 0  $  }  \]
Such a result is coherent with the virial equation on the condition to have ${\sigma}^r_r = 0$ in a coherent way with the classical Maxwell (impulsion-energy) stress tensor density: \\
\[ \fbox{  $    {\sigma}^i_j =  {\cal{F}}^{ir} F_{rj} + \frac{1}{4} {\delta}^i_j {\cal{F}}^{rs} F_{rs}  \Rightarrow {\sigma}^r_r = 0   $  }  \]
which is traceless with a divergence producing the Lorentz force as we have indeed when $n=4$:
\[   \fbox{  $ d F = 0 \,\, \Rightarrow \,\, {\partial}_i {\sigma}^i_j = {\cal{J}}^r F_{rj} + \frac{1}{2} {\cal{F}}^{rs} ( {\partial}_r F_{sj} + {\partial}_s F_{jr} + {\partial}_j F_{rs}) =
    {\cal{J}}^r F_{rj}  $  }  \]
The mathematical foundations of EM, that is both the first and second Maxwell equations, thus only depend on the group structure of the conformal group of space-time, a fact that can only be understood by using the Spencer operator as we saw and which is therefore not even known.  \\

As we have explained in the recent (Pommaret, 2022), studying the mathematical structure of gravitation is much more delicate as it involves third order jets. Our purpose at the end of this paper is to consider only the linearized framework. The crucial idea is to notice that the Poisson equation has only to do with the trace of the stress tensor density, contrary to the EM situation as we just saw. \\

Defining the vector bundle ${\hat{F}}_0 = J_1(T)/{\hat{R}}_1 \simeq T^* \otimes {\hat{g}}_1$ when $n\geq 4$, another difficulty can be discovered in the following commutative and exact diagrams obtained by applying the Spencer $\delta$-map to the symbol sequence with $dim({\hat{g}}_1) = dim(g_1) + 1 = (n(n-1)/2 ) + 1$:
\[     0  \rightarrow  {\hat{g}}_1  \rightarrow   T^*\otimes T  \rightarrow   {\hat{F}}_0  \rightarrow  0   \]    
then to its first prolongation with $dim({\hat{g}}_2)= n$:   \\
\[  0   \rightarrow  {\hat{g}}_2 \rightarrow S_2T^*\otimes T  \rightarrow T^*\otimes {\hat{F}}_0 \rightarrow  0  \]
and finally to its second prolongation in which ${\hat{g}}_3= 0$:  \\
\[  \fbox{  $  \begin{array}{rcccccccccl}
  & &  0 & & 0 & & 0 &  &  &  & \\
  & & \downarrow & & \downarrow & & \downarrow & & & & \\
0 & \rightarrow & {\hat{g}}_3 & \rightarrow &  S_3T^*\otimes T & \rightarrow & S_2T^*\otimes {\hat{F}}_0& \rightarrow & {\hat{F}}_1 & \rightarrow & 0  \\
  & & \hspace{2mm}\downarrow  \delta  & & \hspace{2mm}\downarrow \delta & &\hspace{2mm} \downarrow \delta & & & & \\
0 & \rightarrow& T^*\otimes {\hat{g}}_2&\rightarrow &T^*\otimes S_2T^*\otimes T & \rightarrow &T^*\otimes T^*\otimes {\hat{F}}_0 &\rightarrow & 0 &&  \\
  & &\hspace{2mm} \downarrow \delta &  &\hspace{2mm} \downarrow \delta & &\hspace{2mm}\downarrow \delta &  &  & &  \\
0 & \rightarrow & {\wedge}^2T^*\otimes {\hat{g}}_1 & \rightarrow & \underline{ {\wedge}^2T^*\otimes T^*\otimes T }& \rightarrow & {\wedge}^2T^*\otimes {\hat{F}}_0 & \rightarrow & 0 &&  \\
 &  &\hspace{2mm}\downarrow \delta  &  & \hspace{2mm} \downarrow \delta  &  & \downarrow  & &  & & \\
0 & \rightarrow & {\wedge}^3T^*\otimes T & =  & {\wedge}^3T^*\otimes T  &\rightarrow   & 0  &  &  &  & \\
 &   &  \downarrow  &  &  \downarrow  &  &  &  &  &  &\\
  &  &  0  &   & 0  & &  &  &  &&
\end{array}  $  }  \]
A snake chase allows to introduce the {\it Weyl} bundle ${\hat{F}}_1$ defined by the short exact sequence:  \\
\[      0  \longrightarrow T^* \otimes  {\hat{g}}_2 \stackrel{\delta}{\longrightarrow}  Z^2_1({\hat{g}}_1) \longrightarrow  {\hat{F}}_1 \longrightarrow 0 \]
in which the cocycle bundle $Z^2_1({\hat{g}}_1)$ is defined by the short exact sequence: \\
\[   0 \longrightarrow Z^2_1({\hat{g}}_1) \longrightarrow  {\wedge}^2 T^* \otimes {\hat{g}}_1 \stackrel{\delta}{\longrightarrow} {\wedge}^3 T^* \otimes T \longrightarrow 0  \]
We have of course $dim({\hat{F}}_1)= 10$ when $n=4$ but more generally: \\
\[   \begin{array}{ccl}
dim({\hat{F}}_1) &  =  &  (n (n+1)/2)(n(n+1)/2 - 1) - n^2(n + 1)(n + 2)/6 \\
                           &  =  &  ((n(n-1)/2)(n(n-1)/2 + 1) - n^2 (n-1)(n-2)/6) - n^2  \\ 
                           &  =  &  n(n+1)(n+2)(n-3)/12
\end{array}   \]
In the purely Riemannian case, as $g_2=0$, we have $ F_1 \simeq Z^2_1(g_1)$ and thus:  \\
\[   \begin{array}{ccl}
dim(F_1) &  =  &  (n (n+1)/2)(n(n+1)/2 - 1) - n^2(n + 1)(n + 2)/6 \\
                           &  =  &  n^2 (n-1)/2)^2- n^2 (n-1)(n-2)/6    \\
                           &  =  &   n^2 (n^2 - 1)/12 
\end{array}   \]
a result leading to the unexpected formula \fbox{  $  dim(F_1) - dim({\hat{F}}_1) = n(n + 1) / 2$  }. \\

Needless to say that no classical method can produce such results which are summarized in the following {\it Fundamental Diagram II} provided as early as in $1983$ (Pommaret, 1983 a):  \\ 
\[  \fbox{  $   \begin{array}{rcccccccccl}
&  &  &  &  &  &  &  &  &  & \\
  &  &  &  &  &  &   &  &  0  & & \\
  &  &  &  &  &  &   &  & \downarrow &  & \\
  &  &  &  &  &  &  0  &  &  Ricci & &  \\
  &  &  &  &  &  &  \downarrow &  & \downarrow &  &  \\
  &  &  &  &  0 & \longrightarrow & Z^2_1(g_1) & \longrightarrow & Riemann & \longrightarrow & 0  \\
  &  &  &  &   \downarrow &  & \downarrow &  & \downarrow &  &  \\
  &  & 0 & \longrightarrow & T^* \otimes {\hat{g}}_2 & \stackrel{\delta}{\longrightarrow} & Z^2_1({\hat{g}}_1) & \longrightarrow & Weyl  & \longrightarrow & 0  \\
  &  &  &  &  \downarrow &  &  \downarrow &  &  \downarrow &  &  \\
 0 &  \longrightarrow & S_2T^* &  \stackrel{\delta}{\longrightarrow} &  T^* \otimes T^* & \stackrel{\delta}{\longrightarrow} & {\wedge}^2 T^* &  \longrightarrow & 0 &  &  \\      
  &  &  &  & \downarrow &  &  \downarrow &  &  &  &  \\
  &  &  &  &  0 &  &  0  &  &  &  &\\
  &  &  &  &  &  &  &  &  &  & 
  \end{array}  $  }  \]   \\

\noindent
{\bf Theorem 6.1}: This commutative and exact diagram splits and a diagonal snake chase proves that $Ricci \simeq S_2T^*$.  \\

\noindent
{\it Proof}: The monomorphism $\delta: S_2 T^* \rightarrow T^* \otimes T^*$ splits with $\frac{1}{2} (A_{i,j} + A_{j,i}) \leftarrow A_{i,j}$ while the epimorphism $\delta: T^* \otimes T^* \rightarrow {\wedge}^2 T^*:A_{i,j} \rightarrow A_{i,j} - A_{j,i}$ splits with $ \frac{1}{2} F_{ij} \leftarrow F_{ij}= - F_{ji}$. We explain how the well known result $T^* \otimes T^* \simeq S_2T^* \oplus {\wedge}^2 T^*$, which is coming from the elementary formula $n^2 = n(n+1)/2 + n(n-1)/2$ may be related to the Spencer $\delta$-cohomology interpretation of the Riemann and Weyl bundles. For this, we have to give details on the "{\it snake} " chase:  \\
\[  S_2T^* \rightarrow T^* \otimes T^* \rightarrow T^* \otimes {\hat{g}}_2 \rightarrow Z^2_1({\hat{g}}_1) \rightarrow Z^2_1(g_1) \rightarrow Riemann \rightarrow Ricci  \]
Starting with $(A_{ij} = A_{i,j} = A_{j,i} = A_{ji} ) \in S_2T^*\subset T^* \otimes T^* $, we may define:  \\
\[  {\xi}^r_{ri,j} = n A_{i,j}=n A_{ij} = nA_{ji} = {\xi}^r_{rj,i} \Rightarrow ({\xi}^k_{lj,i} ={\delta}^k_l A_{j,i} + {\delta}^k_j A_{l,i} - {\omega}_{ij} {\omega}^{kr} A_{r,i}) \in T^* \otimes {\hat{g}}_2  \]           
\[  \Rightarrow  (R^k_{l,ij} = {\xi}^k_{li,j} - {\xi}^k_{lj,i} ) \in Z^2_1({\hat{g}}_1) \in {\wedge}^2 T^* \otimes {\hat{g}}_1 \in {\wedge}^2 T^* \otimes T^* \otimes T \]
\[ \Rightarrow   R^r_{r,ij} = {\xi}^r_{zi,j} - {\xi}^r_{rj,i} = n (A_{i,j} - A_{j,i}) = 0 \Rightarrow (R^k_{l,ij}) \in   Z^2_1 (g_1)\]
\[  \Rightarrow  n R^k_{l,ij} = ({\delta}^k_l {\xi}^r_{ri,j} +{\delta}^k_i {\xi}^r_{rl,j}  - {\omega}_{li} {\omega}^{ks} {\xi}^r_{rs,i}) - ( {\delta}^k_l{\xi}^r_{rj,i} + {\delta}^k_j {\xi}^r_{rl,i} - 
{\omega}_{lj} {\omega}^{ks} {\xi} ^r_{rs,i})  \]                                                                                                                                                                                                                                                  
\[  \Rightarrow  R^k_{l,ij} = ({\delta}^k_i A_{lj} - {\delta}^k_jA_{li}) - {\omega}^{ks} ( {\omega}_{li} A_{sj} - {\omega}_{ lj} A_{si} )  \]
Introducing $tr(A) = {\omega}^{ij}A_{ij} $ and  $R_{ij} = R^r_{ i,rj} = (n A_{ij} - A_{ij} ) - (A_{ij} - {\omega}_{lj} tr(A))  $, we get:  \\                                                                                                                                                                                                                                                                                                                                                                                                                                                           
\[  \fbox{  $  R_{ij} = (n - 2) A_{ij}  + {\omega}_{ij} tr(A) = R_{ji}  \Rightarrow  tr(R) = {\omega}^{ij} R_{ij} = 2 ( n - 1) tr(A) $  } \]
Substituting, we finally obtain $A_{ij} = \frac{1}{(n-2)} R_{ij} - \frac{1}{ 2 (n-1)(n-2)} {\omega}_{ij} tr (R)$ and the tricky formula:    \\                                   
\[ \fbox{  $   R^k_{l,ij} = \frac{1}{(n-2)} ({\delta}^k_i R_{lj} - {\delta}^k_j R_{li}  - {\omega}^{ks} ({\omega}_{li} R_{sj}- {\omega}_{lj} R_{si}) - 
\frac{1}{(n-1)(n-2)} ({\delta}^k_i {\omega}_{lj} - {\delta}^k_j {\omega}_{li} ) tr(R)  $  }   \]                                           
totally independently from the standard elimination of the derivatives of a conformal factor.  \\
Contracting in $k$ and $i$, we obtain indeed the lift:  \\
\[    Riemann = H^2_1(g_1) \rightarrow S_2 T^* \simeq Ricci: R^k_{l,ij} \rightarrow R^r_{i,rj} = R_{ij} = R_{ji}  \]
in a coherent way. Using a standard result of homological algebra [16, 17, 37], we obtain therefore a splitting
  $Weyl = H^2_1({\hat{g}}_1) \rightarrow H^2_1(g_1) = Riemann $:  \\
\[ \fbox{  $  W^k_{l,ij} = R^k_{l,ij}  -  ( \frac{1}{(n-2)} ({\delta}^k_i R_{lj} - {\delta}^k_j R_{li}  - {\omega}^{ks} ({\omega}_{li} R_{sj}- {\omega}_{lj} R_{si}) - 
\frac{1}{(n-1)(n-2)} ({\delta}^k_i {\omega}_{lj} - {\delta}^k_j {\omega}_{li} ) tr(R))  $  }   \]   
in such a way that $W^r_{i,rj} = 0$, a result leading to the isomorphism $Riemann \simeq Ricci \oplus Weyl$.   \\
\hspace*{12cm}  $\Box$  \\

We are now ready to apply the previous diagrams by proving the following crucial Theorem:  \\

\noindent
{\bf Theorem 6.2}: When $n=4$, the linear Spencer sequence for the Lie algebra ${\hat{\Theta}}$  of infinitesimal conformal group of transformations projects onto a part of the Poincar\'{e} sequence for the exterior derivative {\it with a shift by one step} according to the following commutative and locally exact diagram:  
\[ \fbox{  $   \begin{array}{rcccccccl}
0 \longrightarrow & \hat{\Theta} & \stackrel{j_2}{\longrightarrow} & {\hat{R}}_2 & \stackrel{D_1}{\longrightarrow} & T^* \otimes  {\hat{R}}_2 & \stackrel{D_2}{\longrightarrow} &  {\wedge}^2T^*\otimes {\hat{R}}_2   \\
&  &  &  \downarrow & \swarrow &  \downarrow  &  &  \downarrow   \\
&  &  &  T^* & \stackrel{d}{\longrightarrow}  & {\wedge}^2T^* & \stackrel{d}{\longrightarrow} &  {\wedge}^3T^* \\
&  &  &  A &  &  d A=F &  &d F=0   
\end{array}   $  }  \]
{\it This purely mathematical result also contradicts classical gauge theory} because it proves that EM only depends on the structure of the conformal group of space-time but not on $U(1)$.  \\

\noindent
{\it Proof}: Restricting our study to the linear framework, we introduce a new system ${\tilde{R}}_1\subset J_1(T)$ of infinitesimal Lie equations defined by 
$L({\xi}_1)\omega = 2 A \omega$ with prolongation defined by  by $L({\xi}_2)\gamma = 0$  in such a way that                                                                                                                                                                                                                                                                                                                                                                                                                                                                                                                                                                                                                                           
$R_1 \subset   {\tilde{R}}_1 = {\hat{R}}_1$ with a strict inclusion and the strict inclusions $R_2 \subset {\tilde{R}}_2 \subset {\hat{R}}_2$. \\
Let us prove that there is an isomorphism ${\hat{R}}_2 / {\tilde{R}}_2 \simeq {\hat{g}}_2$.  \\
Indeed, from the definitions, we obtain the following commutative and exact diagram:  \\
\[       \begin{array}{ccccccccc}
  &  &  &  &  0 &  &  &  & \\
  &  &  &  & \downarrow & &  &  &  \\
  &  &  0 & \rightarrow & {\hat{g}}_2  &  &  &  &  \\  
  &  &  \downarrow &  & \downarrow & \searrow &  &  &  \\
  0 & \rightarrow & {\tilde{R}}_2 & \rightarrow & {\hat{R}}_2 & \rightarrow& {\hat{R}}_2/{\tilde{R}}_2 & \rightarrow & 0 \\
  &  &  \downarrow & & \downarrow & & \downarrow &   &   \\
  0 & \rightarrow & {\tilde{R}}_1 & =  & {\hat{R}}_1&  \rightarrow &  0 &  &  \\
  &  &  \downarrow  &  &  \downarrow &  &  &  &  \\
  &  &  0  &  &  0  &  &  &  & 
  \end{array}   \]
The  south-east arrow is an isomorphism as it is both a monomorphism and an epimorphism by using a snake chase showing that ${\hat{R}}_2 = {\tilde{R}}_2 \oplus {\hat{g}}_2$.  \\
                                                                                                                                                                                                                                                                                                                     A first problem to solve is to construct vector bundles from the components of the image of $D_1$. Using the corresponding capital letter for denoting the linearization, let us introduce:   \\
\[   {\partial}_i {\xi}^k_{\mu} - {\xi}^k_{\mu + 1_i} = X^k_{\mu, i} \,\,  \Rightarrow \,\,  B^k_{\mu, i} \,\, (tensors) \]
\[ (B^k_{l,i}=X^k_{l,i}+{\gamma}^k_{ls}X^s_{,i}) \in T^*\otimes T^*\otimes T \Rightarrow  (B^r_{r,i}=B_i)\in T^*\]
\[  (B^k_{lj,i}=X^k_{lj,i}+{\gamma}^k_{sj}X^s_{l,i}+{\gamma}^k_{ls}X^s_{j,i}-{\gamma}^s_{lj}X^k_{s,i}+X^r_{,i}{\partial}_r{\gamma}^k_{lj}) \in T^*\otimes S_2T^*\otimes T \Rightarrow (B^r_{ri,j}-B^r_{rj,i}=F_{ij})\in {\wedge}^2T^*  \]
We obtain from the relations ${\partial}_i{\gamma}^r_{rj}={\partial}_j{\gamma}^r_{ri}$ and the previous results:  \\
\[ \begin{array}{rcl}
F_{ij}=B^r_{ri,j}-B^r_{rj,i} & = & X^r_{ri,j}-X^r_{rj,i}+{\gamma}^r_{rs}X^s_{i,j}-{\gamma}^r_{rs}X^s_{j,i}+X^r_{,j}{\partial}_r{\gamma}^s_{si}-X^r_{,i}{\partial}_r{\gamma}^s_{sj}  \\
  &  =  & {\partial}_iX^r_{r,j}-{\partial}_jX^r_{r,i}+{\gamma}^r_{rs}(X^s_{i,j}-X^s_{j,i})+X^r_{,j}{\partial}_i{\gamma}^s_{sr}-X^r_{,i}{\partial}_j{\gamma}^s_{sr} \\
    &  =  & {\partial}_i(X^r_{r,j}+{\gamma}^r_{rs}X^s_{,j})-{\partial}_j(X^r_{r,i}+{\gamma}^r_{rs}X^s_{s,i})  \\
      &  =  &  {\partial}_iB_j-{\partial}_jB_i
      \end{array}   \]
Now, using the contracted formula ${\xi}^r_{ri}+ {\gamma}^r_{rs}{\xi}^s_i + {\xi}^s{\partial}_s{\gamma}^r_{ri}=nA_i$, we obtain:  \\
\[ \begin{array}{rcl}
 B_i & =  & ({\partial}_i{\xi}^r_r - {\xi}^r_{ri})+{\gamma}^r_{rs}({\partial}_i{\xi}^s - {\xi}^s_i)\\
    &  =  &{\partial}_i{\xi}^r_r + {\gamma}^r_{rs}{\partial}_i{\xi}^s+
{\xi}^s {\partial}_s{\gamma}^r_{ri} - nA_i \\
  &  =  &{\partial}_i({\xi}^r_r + {\gamma}^r_{rs}{\xi}^s) - nA_i \\
    &  =  &n ({\partial}_iA - A_i) 
\end{array}   \]  
and we finally get $F_{ij}=n({\partial}_jA_i-{\partial}_iA_j)$ {\it which is no longer depending on} $A$, a result fully solving the dream of Weyl. Of course, when $n=4$ and $\omega$ is the Minkowski metric, then we have $\gamma=0$ in actual practice and the previous formulas become particularly simple. \\     

It follows that $d B=F \Leftrightarrow - ndA=F$ in ${\wedge}^2T^*$ and thus $d F=0$, that is $ {\partial}_i{F}_{jk} + {\partial}_j{F}_{ki} + {\partial}_k{F}_{ij}=0 $, has an intrinsic meaning in ${\wedge}^3T^*$. It is finally important to notice that the usual EM Lagrangian is defined on sections of ${\hat{C}}_1$ killed by $D_2$ but {\it not} on ${\hat{C}}_2$. Finally, the south west arrow in the left square is the composition:   \\
\[  {\xi}_2 \in {\hat{R}}_2 \stackrel{D_1}{\longrightarrow} X_2 \in T^* \otimes {\hat{R}}_2 \stackrel{{\pi}^2_1}{\longrightarrow } X_1 \in T^*\otimes {\hat{R}}_1 \stackrel{(\gamma)}{\longrightarrow} (B_i) \in  T^*  \Leftrightarrow {\xi}_2 \in {\hat{R}}_2 \rightarrow (nA_i)\in T^*  \]
More generally, using the Lemma, we have the composition of epimorphisms:  \\
\[  {\hat{C}}_r \rightarrow {\hat{C}}_r/ {\tilde{C}}_r \simeq {\wedge}^r T^* \otimes ({\hat{R}}_2 / {\tilde{R}}_2) \simeq {\wedge}^r T^* \otimes {\hat{g}}_2 \simeq {\wedge}^r T^* \otimes T^* \stackrel{\delta}{\rightarrow} {\wedge}^{r+1} T^* \]
Accordingly, though $A$ and $B$ are potentials for $F$, then $B$ can also be considered as a part of the {\it field} but the important fact is that the first set of ({\it linear}) Maxwell equations $d F=0$ is induced by the ({\it linear}) operator $D_2$ because we are only dealing with involutive and thus formally integrable operators, a {\it fact} justifying the commutativity of the square on the left of the diagram.   \\
\hspace*{12cm}     $\Box $     \\

If we introduce the gravitational potential $\phi= \frac{GM}{r}$ where $r$ is the distance at the central attractive mass $M$ and $G$ is the gravitational constant, then we have $\frac{\phi}{c^2}\ll 1$ as a dimensionless number and $\Theta=1$ when there is no gravity. When there is static gravity, the conformal factor $\Theta$ must be therefore close to $1$ with vanishing Laplacian and $\frac{\partial \Theta}{\partial y}<0$. We have proved in [Pommaret ...] that the only coherent possibility is to set $\Theta = 1 + \frac{\phi}{c^2}$ in order to correct the value $\Theta= 1- \frac{\phi}{c^2}$ we found in (Pommaret,  1994,                                    , p 450). Hence, {\it gravity in vacuum only depends on the conformal isotropy groupoid through the conformal factor} but this new approach is quite different from that of [Weyl (1940)]. \\
The {\it large infinitesimal equivalence principle} initiated by the Cosserat brothers becomes natural in this framework, namely an observer cannot measure sections of $R_q$ but can only measure their images by $D_1$ or, equivalently, can only measure sections of $C_1$ killed by $D_2$. Accordingly, for a free falling particle in a constant gravitational field, we have successively:    \\
 \[  \fbox{  $  {\partial}_4 {\xi}^k - {\xi}^k_4=0, \,\,\, {\partial}_4 {\xi}^k_4 - {\xi}^k_{44}=0, \,\,\,{\partial}_i{\xi}^k_{44} - 0=0, \,\,\,1 \leq i,k\leq 3  $  }  \]
This result explains why the elations are sometimes called "{\it accelerations} " by physicists . Indeed, we have 
${\xi}^k_{44}= - {\omega}_{44} {\omega}^{kr} A_r \sim A_k$ whenever $1\leq k \leq 3$ because the ${\xi}^k_{44}$ are transformed like ${\partial}_{44} {\xi}^k$ according to jet theory, exactly like the ${\xi}^k_4$ are transformed like the "{\it speed} " ${\partial}_4 {\xi}^k$ for the same reason.  \\

Finally, when $n=4$ {\it only}, we have the following Euler-Lagrange equations:  \\
\[   \left\{\begin{array}{lcl}
{\xi}^r_{ri}  & \rightarrow & \exists  \,\, gravitational \,\, potential  \\
{\xi}^r_r    & \rightarrow & \exists \,\, Poisson \,\, equation  \\
{\xi}^r  & \rightarrow & \exists \,\, Newton \,\, law
\end{array}   \right.\]   \\  
according to (Pommaret, 1884).  \\   \\

\noindent
{\bf 7) CONCLUSION}   \\

Summarizing the results obtained in the preceding sections, we can only refer to the Zentralblatt review   Zbl 1079.93001 
for comments on the new mathematical methods that can be found in the corresponding book (1000 pages !) and have successively:  \\

7.1) The Einstein operator has been written down 25 years before A. Einstein by 
the Italian mechanician E. Beltrami in dimension n=3 for parametrizing the Cauchy stress equations by 
the Beltrami stress functions used as potentials through the Beltrami = ad(Riemann) operator.
The explicit comparison, {\it that has never been done}, needs no comment ( See (4.1) in (Pommaret, 2023)).   \\

7.2) The Einstein operator is self-adjoint ({\it who knows} !) and Einstein made {\it two dual confusions}:  \\
\noindent
*  between Beltrami stress functions and the deformation of the metric (n(n+1)/2 components !).\\
\noindent
** {\it Last but not least}, between the Cauchy = ad(Killing) operator and the "div" operator induced from the Bianchi operator, 
{\it by far the worst confusion} disappearing of course when n=2.  \\

7.3) These two confusions can only be understood through homological algebra, because:  \\
\noindent
*  The Einstein operator goes from the variation of the metric to another symmetric tensor 
having {\it strictly nothing to do with stress}.   \\
\noindent
** The adjoint of the Einstein (or Ricci) operator goes from stress functions 
having also {\it strictly nothing to do with the metric}, to the stress tensor density.   \\

7.4) As a byproduct:   \\
\noindent
*  Einstein equations are not coherent with differential duality, {\it contrary to Maxwell equations}.  \\
Indeed, according to Poincar\'{e}, as the ({\it geometrical}) left member is a tensor, the ({\it physical}) right member {\it must} be a tensor density. \\
\noindent
**  GW cannot be ripples of space-time, and {\it cannot thus exist for a purely mathematical reason}.
This is why Einstein hesitated so many times all along his life as he could not quote Beltrami !.   \\

7.5) These results could have been found since 20 years because the double pendulum and the impossibility to parametrize the Einstein operator 
are already in (Pommaret, 2005, p. 201).  \\

7.6) {\it Electromagnetism and gravitation only depend on the elations of the conformal group of space-time} 
by chasing in the {\it Fundamental diagram} II (Pommaret, 1983 a).  Such a result is {\it not} coherent with classical gauge theory because $U(1)$ is {\it not } acting on space-time contrary to the conformal group and also because the EM field is a section of the first Spencer bundle and {\it not} of the curvature bundle described by the second Spencer bundle.   \\

\newpage

\noindent
{\bf REFERENCES}  \\

\noindent
[1] Bialynicki-Birula, A. (1962) On Galois Theory of Fields with Operators, Am. J. Math., 84, 89-109.  \\
\noindent
[2] Cosserat, E., \& Cosserat, F. (1909) Th\'{e}orie des Corps D\'{e}formables, Hermann, Paris.\\
\noindent
[3] Janet, M. (1920) Sur les Syst\`{e}mes aux D\'{e}riv\'{e}es Partielles, Journal de Math., 8, 65-151. \\
\noindent 
[4] Kashiwara, M. (1995) Algebraic Study of Systems of Partial Differential Equations, M\'{e}moires de la Soci\'{e}t\'{e} Math\'{e}matique de France, 63 
(Translation from Japanese of his $1970$ Master Thesis).  \\
\noindent
[5] Macaulay, F.S. (1916) The Algebraic Theory of Modular Systems, Cambridge Tract 19, Cambridge University Press, London (Reprinted by Stechert-Hafner Service Agency, New York, 1964).  \\
\noindent
[6] Oberst, U. (1990) Multidimensional Constant Linear Systems, Acta Applicandae Mathematica, 20, 1-175.    \\
https://doi.org/10.1007/BF00046908    \\
\noindent
[7] Poincar\'{e}, H. (1901) Sur une Forme Nouvelle des Equations de la M\'{e}canique, C. R. Acad. Sc. Paris, 132, 7, 369-371.  \\
\noindent
[8] Pommaret, J.-F. (1978) Systems of Partial Differential Equations and Lie Pseudogroups, Gordon and Breach, New York 
(Russian translation: MIR, Moscow,1983).\\
\noindent
[9] Pommaret, J.-F. (1983 a) La Structure de l'Electromagnetisme et de la Gravitation, C. R. Acad. Sc. Paris, 297, 493-496.  \\
\noindent
[10] Pommaret, J.-F. (1983 b) Differential Galois Theory, Gordon and Breach, New York.\\
\noindent
[11] Pommaret, J.-F.( (1988) Lie Pseudogroups and Mechanics, Gordon and Breach, New York.\\
\noindent
[12] Pommaret, J.-F. (1994) Partial Differential Equations and Group Theory, Kluwer.\\
http://dx.doi.org/10.1007/978-94-017-2539-2.    \\
\noindent
[13] Pommaret, J.-F. (1995) Dualit\'{e} Diff\'{e}rentielle et Applications. Comptes Rendus Acad. Sciences Paris, S\'{e}rie I, 320, 1225-1230.  \\
\noindent
[14] Pommaret, J.-F. (1997) Fran\c{c}ois Cosserat and the Secret of the Mathematical Theory of Elasticity, Annales des Ponts et Chauss\'ees, 82, 59-66 
(Translation by D.H. Delphenich).  \\
\noindent
[15] Pommaret, J.-F. (2001) Partial Differential Control Theory, Kluwer, Dordrecht (Zbl 1079.93001).   \\
\noindent
[16] Pommaret, J.-F. (2005) Algebraic Analysis of Control Systems Defined by Partial Differential Equations, in "Advanced Topics in Control Systems Theory", Springer, Lecture Notes in Control and Information Sciences 311, Chapter 5, pp. 155-223.\\
\noindent
[17] Pommaret, J.-F. (2010) Parametrization of Cosserat Equations, Acta Mechanica, 215, 43-55.   \\
http://dx.doi.org/10.1007/s00707-010-0292-y.  \\
\noindent
[18] Pommaret, J.-F. (2012) Spencer Operator and Applications: From Continuum Mechanics to Mathematical Physics, in "Continuum Mechanics-Progress in Fundamentals and Engineering Applications", Dr. Yong Gan (Ed.), ISBN: 978-953-51-0447--6, InTech, Available from: \\
http://dx.doi.org/10.5772/35607    \\
\noindent
[19] Pommaret, J.-F. (2013) The Mathematical Foundations of General Relativity Revisited, Journal of Modern Physics, 4, 223-239.  \\
 https://dx.doi.org/10.4236/jmp.2013.48A022    \\
 \noindent
[20] Pommaret, J.-F. (2014) The Mathematical Foundations of Gauge Theory Revisited, Journal of Modern Physics, 5, 157-170.  \\
https://dx.doi.org/10.4236/jmp.2014.55026   \\
 \noindent
[21] Pommaret, J.-F.: (2015) Relative Parametrization of Linear Multidimensional Systems, Multidim. Syst. Sign. Process., 26, 405-437. \\ 
https://doi.org/10.1007/s11045-013-0265-0.  \\
\noindent
[22] Pommaret, J.-F. (2016 a) From Thermodynamics to Gauge Theory: The Virial Theorem Revisited, in L. Bailey Editor: " Gauge Theories and Differential Geometry", Nova Science Publishers, New York, (ISBN 978-1-63483-546-6). Chapter I, p 1-44.  \\
\noindent
[23] Pommaret, J.-F. (2016 b) Deformation Theory of Algebraic and Geometric Structures, Lambert Academic Publisher (LAP), Saarbrucken, Germany (2016). A short summary can be found in "Topics in Invariant Theory ", S\'{e}minaire P. Dubreil/M.-P. Malliavin, Springer 
Lecture Notes in Mathematics, 1478 (1990) 244-254. (http://arxiv.org/abs/1207.1964). \\
\noindent
[24] Pommaret, J.-F.: (2017) Why Gravitational Waves Cannot Exist, Journal of Modern Physics, 8, 2122-2158.  \\
https://doi.org/104236/jmp.2017.813130.    \\   \\
\noindent
[25] Pommaret, J.-F. (1918) New Mathematical Methods for Physics, Mathematical physics Books, Nova Science publishers, New York, 150 pp.  \\
\noindent
[26]  Pommaret, J.-F.: (2019) The Mathematical Foundations of Elasticity and Electromagnetism Revisited, Journal of Modern Physics, 10 (2019) 1566-1595.     \\
 https://doi.org/10.4236/jmp.2019.1013104   \\
\noindent
[27] Pommaret, J.-F.: (2021 a) Homological Solution of the Lanczos Problems in Arbitrary Dimension, Journal of Modern Physics, 12 , 829-858.  \\
https://doi.org/10.4236/jmp.2020.1110104  \\
\noindent
[28] Pommaret, J.-F. (2021 b) Minimum Parametrization of the Cauchy Stress Operator, Journal of modern Physics, 12, 453-482. 
https://doi.org/10.4236/jmp.2021.124032   \\
\noindent
[29] Pommaret, J.-F. (2021 c) The Conformal Group Revisited,   \\
\noindent
https://doi.org/10.4236/jmp.2021.1213106  \\
\noindent  
[30] Pommaret, J.-F. (2021 d) Differential Correspondences and Control Theory, Advances in Pure Mathematics, 30, 835-882.  \\
https://doi.org/10.4236:apm.2021.111056      \\
\noindent
[31]  Pommaret, J.-F.: (2022 a) Nonlinear Conformal Electromagnetism, Journal of Modern Physics, 13, 442-494. \\
\noindent
https://doi.org/10.4236/jmp.2022.134031    \\
\noindent
[32] Pommaret, J.-F.( 2022 b) How Many Structure Constants do Exist in Riemannian Geometry ?, Mathematics in Computer Science, 16:23, 
https://doi.org/10.1007/s11786-022-00546-3.  \\
\noindent
[33] Pommaret, J.-F. (2023 a) Killing Operator for the Kerr Metric, Journal of Modern Physics, 14, 31-59. https://arxiv.org/abs/2203.11694 , 
https://doi.org/10.4236/jmp.2023.141003.     \\
\noindent
[34] Pommaret, J.-F. (2023 b) Gravitational Waves and Parametrizations of Linear Differential Operators, in Gravitational Waves:Theory and Observations, 
Edited by C. Frajuca, IntechOpen.  \\ 
https://doi.org/10.5772/intechopen.1000851 (https://doi.org/10.5772/intechopen.1000226).  \\
\noindent
[35] Pommaret, J.-F. (2023 c) Differential Galois Theory and Hopf Algebras for Lie Pseudogroups, \\
https://arxiv.org/abs/2308.03759.  \\  
\noindent
[36] Pommaret, J.-F. (2023 d) Nonlinear Conformal Geometry, Journal of Modern Pgysics, 14, 1464-1496.   \
htts://doi.org/104236:jmp.2023.1411086    \\
\noindent
[37] Pommaret, J.-F. (2024 a) From Control Theory to Gravitational Waves, Advances in Pure Mathematics, 14, 49 -100.   \\
https://doi.org/10.4236/apm.2024.142004    \\
\noindent
[38] Pommaret, J.-F. (2024 b) Gravitational Waves and the Foundations of Riemannian Geometry, Advances in Mathematical Research, Volume 35, 95-161, 
Nova Science Publisher, ISBN: 979-8-89113-607-6  \\
\noindent
[39] Rotman, J.J.: (1979) An Introduction to Homological Algebra, Pure and Applied Mathematics, Academic Press.  \\
\noindent
[40] Spencer, D.C. (1965) Overdetermined Systems of Partial Differential Equations, Bull. Am. Math. Soc., 75, 1-114.\\
\noindent
[41] Spencer, D.C. and Kumpera, A. (1972) Lie Equations, Princeton University Press, Princeton.  \\
\noindent
[42] Weyl, H. (1918) Space, Time, Matter, New York: Dover; 1952.   \\

\end{document}